\documentclass[]{aa} 
\usepackage[varg]{txfonts}
\usepackage{time}
\usepackage{color}
\usepackage{natbib}
\usepackage[percent]{overpic}
\usepackage{amstext,amsmath,mathtools}
\usepackage{graphicx}
\usepackage{textcomp}
\usepackage{txfonts}
\usepackage{url}
\usepackage{booktabs} 
\begin{document}
\titlerunning{Performance of the revamped SST}
\title{Is the sky the limit?}
\subtitle{Performance of the revamped Swedish 1-m Solar Telescope\\ and its blue- and red-beam re-imaging systems.}

\author{G.B. Scharmer\inst{1,2,3}
\and
M.G. L\"ofdahl\inst{1,2}
\and
G. Sliepen\inst{1,2}
\and
J. de la Cruz Rodr\'iguez\inst{1,2}}

\institute{Institute for Solar Physics, Stockholm University,
AlbaNova University Center, SE 106\,91 Stockholm, Sweden \and
Stockholm Observatory, Dept. of Astronomy, Stockholm University,
AlbaNova University Center, SE 106\,91 Stockholm, Sweden \and
Royal Swedish Academy of Sciences, Box 50005, SE 104\,05 Stockholm, Sweden}
\date{Draft: \now\ \today}
\frenchspacing

\abstract{We discuss the use of measurements of the solar granulation contrast as a measure of optical quality. We demonstrate that for data recorded with a telescope that uses adaptive optics and/or post-processing to compensate for many low- and high-order aberrations, the RMS granulation contrast is directly proportional to the Strehl ratio calculated from the residual (small-scale) wavefront error (static and/or from seeing). We demonstrate that the wings of the high-order compensated PSF for SST are likely to extend to a radius of not more than about 2\arcsec, consistent with earlier conclusions drawn from straylight compensation of sunspot images. We report on simultaneous measurements of seeing and solar granulation contrast averaged over 2~sec time intervals at several wavelengths from 525~nm to 853.6~nm on the red-beam (CRISP beam) and wavelengths from 395~nm to 484~nm on the blue-beam (CHROMIS beam). These data were recorded with the Swedish 1-m Solar Telescope (SST) that has been revamped with an 85-electrode adaptive mirror and a new tip-tilt mirror, both of which were polished to exceptionally high optical quality. Compared to similar data obtained with the previous 37-electrode adaptive mirror in 2009 and 2011, there is a significant improvement in image contrast. The highest 2-sec average image contrast measured in April 2015 through 0.3-0.9~nm interference filters at 525~nm, 557~nm, 630~nm and 853.5~nm with compensation only for the diffraction limited point spread function of SST is 11.8\%, 11.8\%, 10.2\% and 7.2\% respectively. Similarly, the highest 2-sec contrast measured at 395~nm, 400~nm and 484~nm in May 2016 through 0.37-1.3~nm filters is 16\%, 16\% and 12.5\% respectively. The granulation contrast observed with SST compares favorably to measured values with SOT on Hinode and with Sunrise as well as major ground-based solar telescopes. Simultaneously with the above wideband red-beam data, we also recorded narrow-band continuum images with the CRISP imaging spectropolarimeter. We find that contrasts measured with CRISP are entirely consistent with the corresponding wide-band contrasts, demonstrating that any additional image degradation by the CRISP etalons and telecentric optical system is marginal or even insignificant. Finally, we discuss the origin of the 48~nm RMS wavefront error needed to bring consistency between the measured granulation contrast and that obtained from 3D simulations of convection.
}


\maketitle

\section{Introduction}
In the present era of solar physics, many complex dynamic processes in the solar atmosphere can only be explained through sophisticated time dependent 3D numerical simulations. Some of these simulations are already sufficiently realistic to provide \emph{quantitative} predictions of (spatial variations of) physical quantities such as temperature, velocities and the magnetic field. By combining observations of Stokes spectra obtained at high spatial resolution with modern inversion techniques \citep[see reviews by][]{2016LRSP...13....4D,2017SSRv..210..109D}, we can in principle confront theoretical simulations with experimental data and decide whether the simulations can be validated, need to be improved, or should be refuted. However, most of the dynamics in the solar atmosphere occurs at spatial scales that are close to the diffraction limit of even the largest solar telescopes, and inferences drawn from observations are therefore more often than not severely compromised by uncertainties in the estimated spatial point spread function (PSF) of the recorded data.

A classical example that illustrates the difficulties of confronting simulations with observations is the longstanding controversy over the continuum granulation contrast, which gives a measure of the temperature fluctuations of the solar atmosphere at the visible surface. Disturbingly, measurements of the granulation contrast stretching over more than 4 decades vary wildly. For example, from a compilation by \citet{2000ApJ...538..940S} we note that in 1969, \citet{1969SoPh....9...39B} inferred a contrast equivalent to 5.12\% when translated to a wavelength of 500~nm, while only 2 years later \citet{1971A&A....14...15L} obtained the equivalence of 18.55\% at the same wavelength. Much later, between 1991 and 1997, several measurements were published that resulted in values between 9.8\% and 14.85\%. 

Thanks to accurate measurements of the PSF  of the Solar Optical Telescope (SOT) on Hinode \citep{2008SoPh..249..167T}, there is now a satisfying congruence of granulation contrasts obtained from simulations and observations \citep{2008A&A...484L..17D,
  2009A&A...503..225W, 2009A&A...501L..19M}, demonstrating that we can indeed have confidence in the predictions made from convection simulations\footnote{This also means that the observed granulation contrast with any space- or ground-based solar telescope can be used to constrain the PSF of that telescope \citep{2011Sci...333..316S}.}. For example, it can be concluded that
the true value of the granulation contrast is approximately 21--23\% at 500~nm, based on the compilation of results from numerical
simulations \citep[][their Table 3]{2009A&A...503..225W} and assuming that the RMS contrast is inversely proportional to the wavelength.

Calibrating the PSF even in the case of a seeing-free space-based telescope is far from a trivial task. \citet{2008A&A...484L..17D} demonstrated that the diffraction limited PSF of SOT, with its large central obscuration, spider and CCD degrades the RMS contrast of spectra at 630~nm with the SOT spectropolarimeter \citep{2001ASPC..236...33L} from its theoretically established value of 14.4\% to 8.5\%, which is still significantly above the observed value of 7.0\%. These authors suggest that the remaining discrepancy can be explained by a combination of straylight and imperfections of the instrument, such as a focus error and/or low-order aberrations. A more detailed determination of the PSF of SOT was made possible by observing the planet Mercury against the solar disk \citep{2009A&A...501L..19M} and using a combination of solar eclipse and Mercury transit data \citep{2008A&A...487..399W}. Similar methods have been used to characterize the PSF's of several ground-based solar telescopes, but as discussed above there is a disappointing lack of consistency as regards the measured granulation contrast. 

Quite significant efforts to understand the PSF and performance of  an AO
system  were undertaken by \citet{2007PhDT........41M,  Marino:10}, discussed also by \citet{2011LRSP....8....2R}. These authors modified the wavefront sensor of the Dunn Solar Telescope to allow its AO system to lock on bright stars such as Sirius. Through long
exposures, they determined the core as well as the far wings of the PSF, and performed a comparison with the theoretical PSF
expected from the AO telemetry simultaneously obtained with the science exposures. These tests validated the method of obtaining the PSF from AO telemetry when an object with zero angular extent (a star) is the wavefront sensor target but does not provide a critical test of relevance to the performance of solar AO, which suffers from anisoplanatic effects when solar fine structure is used as wavefront sensor target. 

Analysis of data from the Swedish 1-m Solar Telescope \citep[SST;][]{scharmer03new} shows that the main source of straylight is from small-angle scattering over at most a few arcsec. This conclusion follows from analysis of images containing both a sunspot and granulation, recorded in 2010. These data were restored with multi-object multi-frame blind deconvolution methods \citep[MFBD and MOMFBD;][]{lofdahl02multi-frame, 2005SoPh..228..191V}. The measured granulation contrast of the MFBD processed images was  8.9\%, which is only 53\% of the expected 16.9\% contrast, whereas the measured minimum umbral intensity of a sunspot was as low as 15.8\%. This  constrained the corresponding stray-light PSF to have a FWHM of at most a few arcsec (else the restored umbra intensity would be negative), leading to the conclusion that \emph{the main source of stray-light must be small-scale aberrations} \citep[][their Supporting Online Material, SOM\footnote{{\small \url{http://www.sciencemag.org/content/333/6040/316/suppl/DC1}}}]{2011Sci...333..316S}. This conclusion was reinforced by later work \citep{2012A&A...537A..80L} indicating that a significant fraction of the stray-light comes from the re-imaging optics (including the tip-tilt and adaptive mirrors). Subsequent work by \citet{2016A&A...585A.140L} confirms that the \emph{far-wing} scattered light (such as from scattering in the Earth's atmosphere or from dust on the telescope optics) in SST data is insignificant in the context of explaining the observed granulation contrast.

In this paper, we discuss co-temporal measurements of seeing and granulation contrast recorded with the SST, after replacing both the tip-tilt and adaptive mirrors, and the theoretical explanation by means of numerical calculations. The paper is organized as follows: In Sect.~2, we discuss measurements of the seeing and granulation contrast, in Sect.~3 the interpretation, including theoretical calculations, and demonstrate that the measured granulation contrast gives an optical quality measure similar to the Strehl ratio. In Sect.~4 we summarize the results.

\section{Observations and seeing measurements}
\subsection{Telescope and adaptive optics}
The SST is a 1-meter evacuated solar telescope consisting of a primary and secondary optical system. The primary system consists of a 1.1~m singlet lens of fused silica with a clear aperture of 0.98~m and two 1.4~m flat Zerodur mirrors used to deflect the beam into a 17~m high tower. The secondary optical system consists of a 60~mm field mirror, a 250~mm clear aperture Schupmann corrector that is used to remove the chromatic aberrations of the singlet lens, a 50~mm field lens, a 42~mm tip-tilt mirror, a 34 mm clear aperture adaptive mirror, and a 40~mm re-imaging triplet lens. 

The present AO system (Scharmer et al., in prep.) was installed in 2013 and uses an 85-electrode deformable monomorph mirror from CILAS together with an 85-subaperture hexagonal microlens array from SMOS. The system is characterized by a very large wavefront sensor (WFS) field-of-view (FOV) of about 17\arcsec$\times$17\arcsec, an image scale of 0\farcs48 per pixel and an update frequency of 2~kHz.

\subsubsection{Seeing measurements}

\begin{figure}
\center
\includegraphics[angle=0, width=0.4\textwidth,clip]{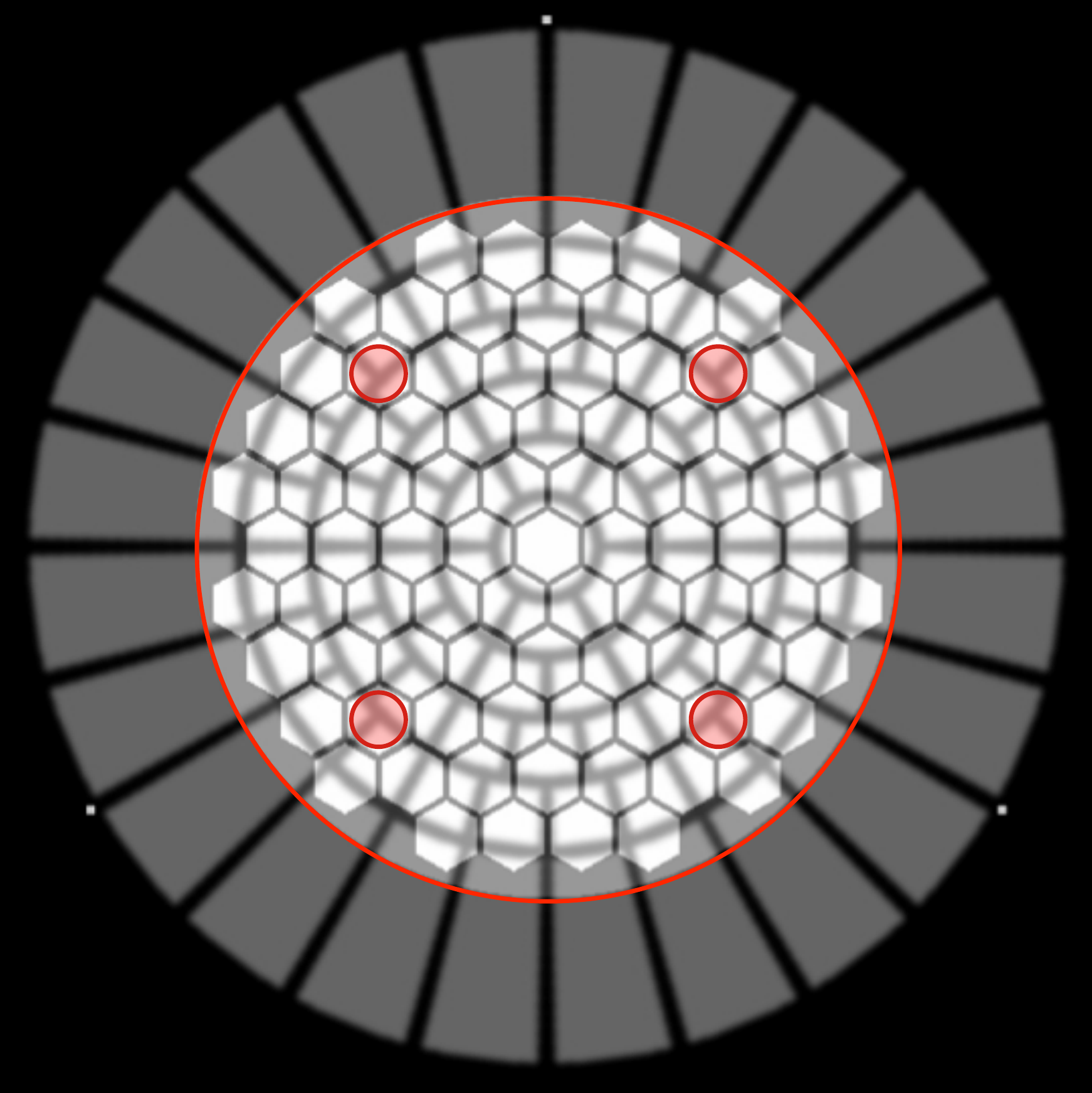}
 \caption{
Layout of the present 85-electrode monomorph deformable mirror and 85 subaperture wavefront sensor of the present AO system of the SST. The grey or white radial/azimuthal structures correspond to the electrodes of the deformable mirror, and the hexagonal structures correspond to the 85 lenslets. The four lenslets indicated with small red circles correspond to the subapertures used for seeing measurements. The large red circle corresponds to the pupil diameter. For details, see text.
}
\label{fig:AO_WFS}
\end{figure}

Since 2013, real-time seeing measurements have been routinely made at the SST using the WFS of the new AO system. The electrode and wavefront sensor layouts are shown in Fig.~\ref{fig:AO_WFS}. Of particular importance in the present context are the four lenslets highlighted with red circles. These are separated by approximately 47~cm and 48.8~cm in the horizontal and vertical directions respectively, when projected on the 98 cm pupil diameter. We use the measured longitudinal and transverse differential image motions between these four lenslets to provide measurements of the seeing quality in terms of Fried's parameter $r_0$. To estimate $r_0$, we use the approximate equations given by \citet{1990A&A...227..294S}, giving the expected variance of differential image displacements $x$ and $y$, 

 \begin{equation}
\langle (x(s)-x(0))^2\rangle  = 0.358 \lambda^2 r_0^{-5/3} D^{-1/3} (1 - 0.541 (s/D)^{-1/3}) ,
\end{equation}
for longitudinal image displacements (i.e., image displacements along the direction connecting the centers of the two subapertures involved), and
 \begin{equation}
\langle (y(s)-y(0))^2\rangle  = 0.358 \lambda^2 r_0^{-5/3} D^{-1/3} (1 - 0.811 (s/D)^{-1/3}) ,
\end{equation}
for transverse image displacements. Here, $D$ is the subaperture diameter, $s$ is the separation between the centers of the subapertures and $\lambda$ the wavelength. In order to compensate for the closed-loop actions of the DM on the measured differential image motions we multiply the vector of voltages with the inverse of a matrix that is obtained during the calibration of the AO system and that gives the $x$,$y$ image displacements
with unit voltage on each electrode for each subaperture. We have verified, by operating the closed loop of the AO system in on-off mode at 10 sec. intervals, that this compensation gives reliable $r_0$ estimates.

The AO system, wavefront sensor and software used at the SST for seeing measurements have a few notable properties: 

\begin{enumerate}
\item The optical surface of the adaptive mirror is likely of exceptional quality. The manufacturer (CILAS) estimates that the residual wavefront error is only 6~nm RMS (mechanical 3~nm RMS) after flattening with optimum voltages on the electrodes. In addition, simulations show that the combination of the WFS and electrode layout, which was designed by one of the authors (GS) allows 84 modes to be controlled, the only uncontrollable mode being piston. Finally, we recently replaced also the 40~mm diameter tip-tilt mirror with a 42~mm mirror polished by IC Optical Systems (ICOS) to Fabry--P\'erot quality (approaching 1/100 wave PV). We attribute the overall optical quality of the adaptive mirror and tip-tilt mirror as a major reason for the high granulation contrast.   
 \item We use relatively large subapertures, approximately 9.4~cm in diameter. A sufficiently large lenslet diameter is needed in order to allow image shift measurements with low noise, when using low-contrast granulation as WFS target. At the same time, a too large lenslet diameter may not allow wavefront slopes measurements when $r_0$ is small. Based on experience with this system and a wide-field WFS \citep{2010A&A...513A..25S} installed at the SST, the 9.4~cm lenslet diameter gives reasonably accurate measurements of $r_0$ when $r_0 > 6$~cm whereas the AO system can lock in as bad seeing as corresponding to $r_0\approx5$~cm. Note, that \citet{2010A&A...524A..90L} demonstrated that cross-correlation techniques using granulation as target results in measured image displacements that are linear in the wavefront slopes even when $r_0$ is significantly smaller than the subaperture diameter.
 \item The AO WFS uses a very large FOV for the cross-correlations, 24$\times$24 pixels or 12\arcsec$\times$12\arcsec, in order to average out as much as possible of the high-altitude seeing and thus to minimize the image quality degradation \emph{outside} the AO lock point. By splitting each 24$\times$24 pixel image into 3$\times$3 subfields of 8$\times$8 pixels, we also measure and average $r_0$ from 9 smaller subfields of only 4\arcsec$\times$4\arcsec. This is possible by measuring image shifts \emph{relative} to those of the 12\arcsec$\times$12\arcsec{} and only allowing $\pm1\arcsec$ relative shifts. Whereas the large FOV essentially only corresponds to low altitude seeing, the small FOV measurement gives an $r_0$ that is a combination of both low- and high-altitude seeing, roughly according to $r_0^{-5/3} \approx r_{0~\text{low}}^{-5/3} + r_{0~\text{high}}^{-5/3} $. 
 \item We  measure and compensate for the WFS noise of these measurements. This is done by continuously monitoring the variance of the difference in image motion between consecutive wavefront samples taken 500~ns apart and, assuming that such variations must be from noise, dividing this variance by a factor of 2. We verified for a few data sets that this gives estimates of the noise level that are similar to what is obtained from power spectra of the measured differential image motions.
 \item We calculate average variances of the relative image displacements over only 2~sec time intervals. However, the variances are \emph{not} calculated relative to an average differential image motion taken over the \emph{same} 2~sec. Instead, we calculate the variances \emph{relative to an average taken over 20--30~sec}\footnote{The algorithms used are the following: the variance of image motion in the x-direction is obtained as $\sigma_x^2=\langle(x-x_0)^2\rangle$, where the average is taken over 2 seconds and $x_0$ is calculated iteratively as ${x_0}^{(n)}=c{x_0}^{(n-1)}+(1-c)x$. The constant $c$ equals $1-1/(20f)$, where $f$ is the update frequency, which is close to 2 kHz, so $c\approx~0.999975$. If $x$ changes step-wise, then $x_0$ reaches 63\% of that step after 20~sec., 78\% after 30~sec. and 95\% after 1~min.}. In this way \emph{the variances calculated during 2~sec include both the fast (high temporal frequencies) and the slow (low temporal frequencies) of the seeing}, in spite of the short time averaging interval. The reason for this procedure is the strong intermittency of the (day-time) seeing, which is clearly evidenced by the strong time variability of the quality of the SST science images. This intermittency makes it meaningless to define a single average quality measure, such as $r_0$ or the granulation contrast, during time intervals of significantly longer duration than a few seconds \citep[as found also in our previous analysis;][]{2010A&A...521A..68S}.
 \item The $r_0$ measurements are averages over 2~sec. intervals as described above, but the measurements are made with overlapping time intervals such that we get one $r_0$ measurement per second.
\end{enumerate}

\begin{figure}
\center
\includegraphics[viewport=55 70 720 530, angle=0, width=0.45\textwidth,clip]{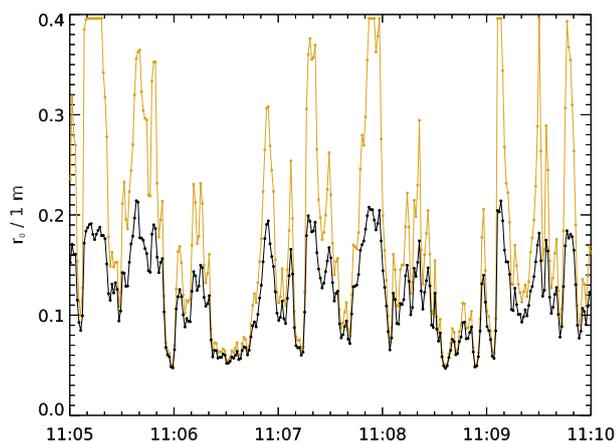}
 \caption{
The yellow curve, which is saturated in some peaks, shows the time variation of the SST seeing quality (Fried's parameter $r_0$) as
measured with the 12\farcs0$\times$12\farcs0 wavefront sensor of the SST AO system, and the black curve shows the corresponding variations measured with a much smaller field-of-view of 4\farcs0$\times$4\farcs0. The yellow curve corresponds roughly to ground-layer seeing and the black curve to a combination of ground- and high-altitude seeing (see text). Note the strong intermittency that requires high time resolution. The data were recorded on 13 May 2016. 
}
\label{fig:r0}
\end{figure}

Figure~\ref{fig:r0} shows an example of measurements of $r_0$ during a time span of only 5 minutes, with the yellow curve showing (essentially) near-ground seeing and the black curve the contribution of (essentially) the entire atmosphere. It is evident that the day-time seeing is highly intermittent in this case, as confirmed by the strongly variable quality of the science images, and this constitutes a clear rationale for dividing the analysis into 2~sec. intervals. It is also evident that when the seeing is relatively good ($r_0$ reaching 0.2 m or more), there is a large difference between the seeing inferred from the yellow and black curves. This is because the measurements corresponding to the yellow curve are only sensitive to ground-layer seeing whereas the data of the black curve also includes contributions from high-layer seeing. When the seeing temporarily becomes very poor, it is normally a consequence of ground heating and near-ground turbulence, which impacts equally on the black and yellow curves. However, when the ground-layer seeing becomes excellent for a brief moment, as revealed by the yellow curve, there is still the (independent) contribution from high-layer seeing. During the early mornings of good days, when the Sun is at very low elevation, it frequently happens that $r_0$ measured with 24$\times$24~pixels reaches values of over 50~cm, even though the live solar image shows strong and small-scale warping. These large $r_0$ values clearly are not reasonable. The values obtained with 8$\times$8~pixels, on the other hand, obviously respond to the small-scale differential morning seeing and overall appear much more reasonable.

From the above discussion and Fig.~\ref{fig:r0} it is evident that the seeing measurements with 24$\times$24~pixels are insensitive to high-altitude seeing and therefore unrealistic. In the following we only discuss seeing measurements made with the AO system using 8$\times$8~pixels, or 4\arcsec$\times$4\arcsec.

\subsection{Granulation contrast measurements and calculations}
\subsubsection{CRISP and CHROMIS re-imaging systems}

\begin{figure}
\center

\includegraphics[angle=0, width=0.49\textwidth,clip]{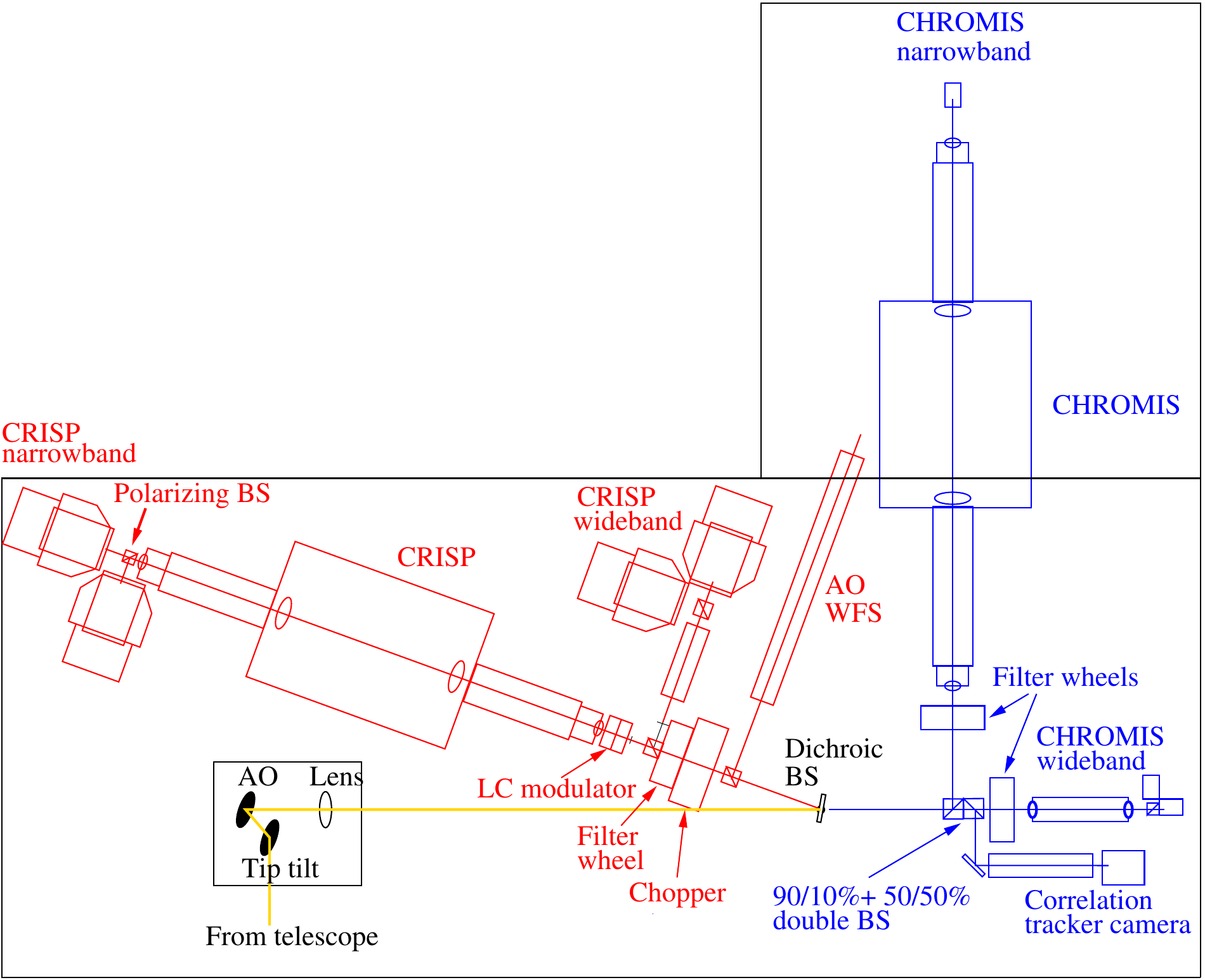}
 \caption{
Layout of the present optical setup at SST, including the adaptive optics and wavefront sensor, and the CRISP and CHROMIS narrowband and wideband re-imaging systems. At the time of the tests, CHROMIS was not yet installed and its wideband re-imaging together with the pre-filters of CHROMIS was used at the present location of CHROMIS. For details, see text.
}
\label{fig:Re-imaging}
\end{figure}

Figure~\ref{fig:Re-imaging} shows the optical setup of the main optical tables at SST, including the CRISP and CHROMIS dual Fabry--P\'erot filter-based narrow-band re-imaging systems, their wideband correspondences, and the AO system and its wavefront sensor. The beam enters the tables via a tip-tilt mirror, the 85-electrode adaptive mirror and a re-imaging triplet lens. The beam is then divided by a dichroic beam splitter into the red (CRISP) and blue (CHROMIS) beams. SST uses a separate tip-tilt mirror (controlled by the correlation tracker camera and its computer) to compensate image motion and the AO system a deformable mirror to compensate higher-order aberrations. The correlation tracker camera is located on the blue beam and the AO wavefront sensor is on the red beam. Each beam contains beam splitters to split light into the auxiliary wide-band systems of CRISP and CHROMIS, and to the AO wavefront sensor and the correlation tracker camera. 

CRISP has two ferro-electric liquid crystal modulators and a polarizing beam splitter to allow measurements of both circular and
linear polarization at any wavelength covered by the CRISP pre-filters. The CRISP beam allows for a so-called phase diversity
setup with one focused and another intentionally defocused camera \citep{lofdahl94wavefront} at the wide-band beam, but this feature was not in use at the time of recording the present data. CHROMIS also allows for a phase-diversity setup at the wide-band beam but does not yet allow for polarization measurements. Both CRISP and CHROMIS use telecentric re-imaging systems, with two anti-reflection (AR) coated doublet lenses and a pupil stop on both the input and output sides. The wideband re-imaging systems are also telecentric, and use two cemented doublet lenses with designs that are identical to those of the first and last lenses of the corresponding narrowband re-imaging systems. All etalons are wedged and have AR coatings to minimize light losses and spurious reflections. The low-resolution etalons (see also the next paragraph) of CRISP and CHROMIS are slightly tilted so that the pupil stops on the output side can completely eliminate ghost images from inter-etalon reflections. CRISP and CHROMIS have been designed to be as compact as possible with the constraint that the Strehl ratio must be at least 0.95 at all wavelengths and all field points within a 1\arcmin$\times$1\arcmin{} FOV. The minimum Strehl ratios are actually a bit higher than 0.95, according to the optical design. The overall length of CRISP from focal plane to focal plane is 1.5~m, that of CHROMIS 1.6~m.

Both CRISP and CHROMIS share an important design feature proposed by \citet{scharmer06comments}, namely to use a combination of high reflectivity for the high-resolution etalon and low reflectivity for the low-resolution etalon. This is fundamental in ensuring that the low-resolution etalon has a spectral passband that is wide enough to accomodate the wavelength shifts from cavity errors of the
high-resolution etalon. This in turn leads to high throughput and a spectral transmission profile that shows small variations, apart from
wavelength shifts, across the FOV. All in all, both CRISP and CHROMIS are highly efficient in terms of throughput and deliver exceptional image quality. The image scales for CRISP and CHROMIS are 0\farcs06 and 0\farcs04 per pixel, respectively, and the exposure times were 17.5 and 1--2~msec, respectively. The short exposure times are necessary to ``freeze'' the seeing and allow image reconstruction with compensation for residual aberrations.

\subsubsection{Observational data and contrast measurements}

\begin{table*}
   \centering
   \caption{Properties of data collected.}
   \begin{tabular}{ l r r r r r r r r}
   \hline\hline\noalign{\smallskip}
    Wavelength (nm) & 395 & 400 & 485 & 525 & 558 & 630 & 854 & 854  \\
    \hline\hline\noalign{\smallskip}
    Comments & H+K wings & H cont. & H$\beta$ cont. & Incl. Fe I & Incl. Fe I & Incl. Fe I & Incl. Ca II 854 & Ca II 854 wing \\
    Instrument & CHROMIS & CHROMIS & CHROMIS & CRISP & CRISP & CRISP & CRISP & CRISP \\
    WB/NB & WB & WB & WB & WB & WB & WB & WB & NB \\
    CWL (nm) & 395.0 & 399.04 & 484.55 & 525.055 & 557.80 & 630.26 & 854.16 & $\approx$~853.8 \\
    FWHM (nm) & 1.32 & 0.42 & 0.65 & 0.33 & 0.30 & 0.44 & 0.93 & 0.01\\
    Exp. time (ms) & 1--2 & 1--2 & 1--2 & 17.5 & 17.5 & 17.5 & 17.5 & 17.5 \\
    N:o frames & 100 & 100 & 100 & $\approx$~74 & $\approx$~74 & $\approx$~74 & $\approx$~74 & $\approx$74 \\
    \hline
    \label{obs_data}
   \end{tabular}
\tablefoot{Summary of data collected. WB and NB refers to the wideband
  and narrowband re-imaging systems, respectively, and CWL and FWHM to
  the center wavelength and full width at half maximum of the filters
  used. }
\end{table*}

The data discussed in this paper were obtained with SST on 5 April 2015 (CRISP narrow- and wideband) and 13 May 2016 (CHROMIS wideband), the latter observations taking place when pre-filters and the wideband re-imaging system for CHROMIS were available but not yet CHROMIS itself. Table~\ref{obs_data} gives a summary of the center wavelengths and passbands of the filters used with CHROMIS and CRISP.

On the red beam, the $r_0$ measurements in the log file of the AO system, accurately timed at 1~sec. intervals, were used to create
groups of images recorded during the time intervals precisely corresponding to the $r_0$ measurements. On the blue beam, images were recorded in fixed bursts of 100~images, collected during 2~sec., and instead $r_0$ was mapped to the time of the image bursts by interpolation of adjacent $r_0$ measurements. The bursts of images were then processed with multi-frame blind deconvolution (MFBD) \citep{lofdahl02multi-frame} using different number of corrected aberration modes: 2 modes (tip-tilt only), 36 modes and 100 modes. Here we shall only discuss the restorations using 2 and 100 modes.

For each of the images restored as described above, we calculated the RMS granulation contrast over quadratic subfields having different dimensions. We found, as expected, that using a small sub-field gives a slightly higher average contrast but also larger scatter when plotting granulation contrast against $r_0$.  We finally opted for a fairly large FOV of 18\arcsec$\times$18\arcsec, corresponding to 300$\times$300~pixels with CRISP and 450$\times$450 pixels with CHROMIS.

\begin{figure*}
\center
\includegraphics[angle=90, width=0.9\textwidth,clip]{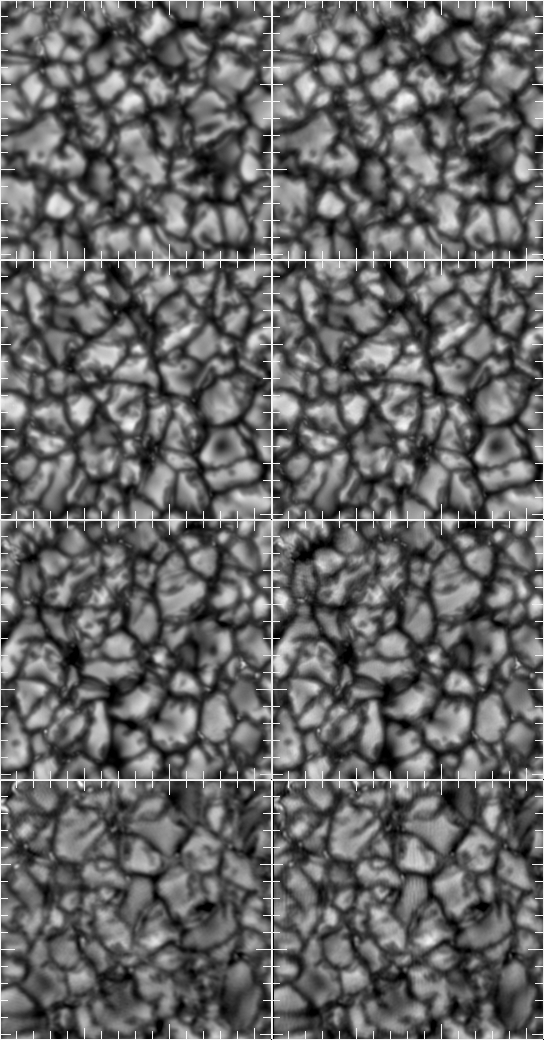}\\[2mm]
\includegraphics[angle=90, width=0.9\textwidth,clip]{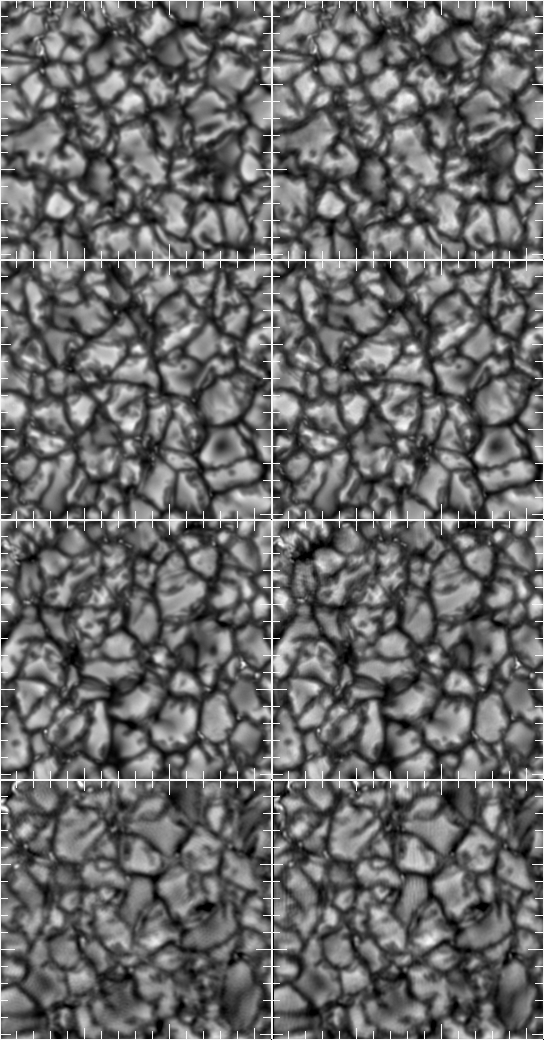}
 \caption{
The top 8 panels show the resulting WB (lower 4 panels) and CRISP NB (upper 4 panels) images after adding up the 74 observed images during 2~sec. and only compensating for relative image motion (tip and tilt) between the individual exposures and the PSF of the diffraction limited telescope. The bottom 8 panels show the resulting image of the same observed images but after applying MFBD image reconstruction to compensate for the largest 100 Karhunen--Lo\`eve modes. The wavelengths are (left to right) 525~nm, 557~nm, 630~nm and 853~nm.
}
\label{fig:CRISP1}
\end{figure*}

\begin{figure*}
\center
\includegraphics[angle=0, width=0.9\textwidth,clip]{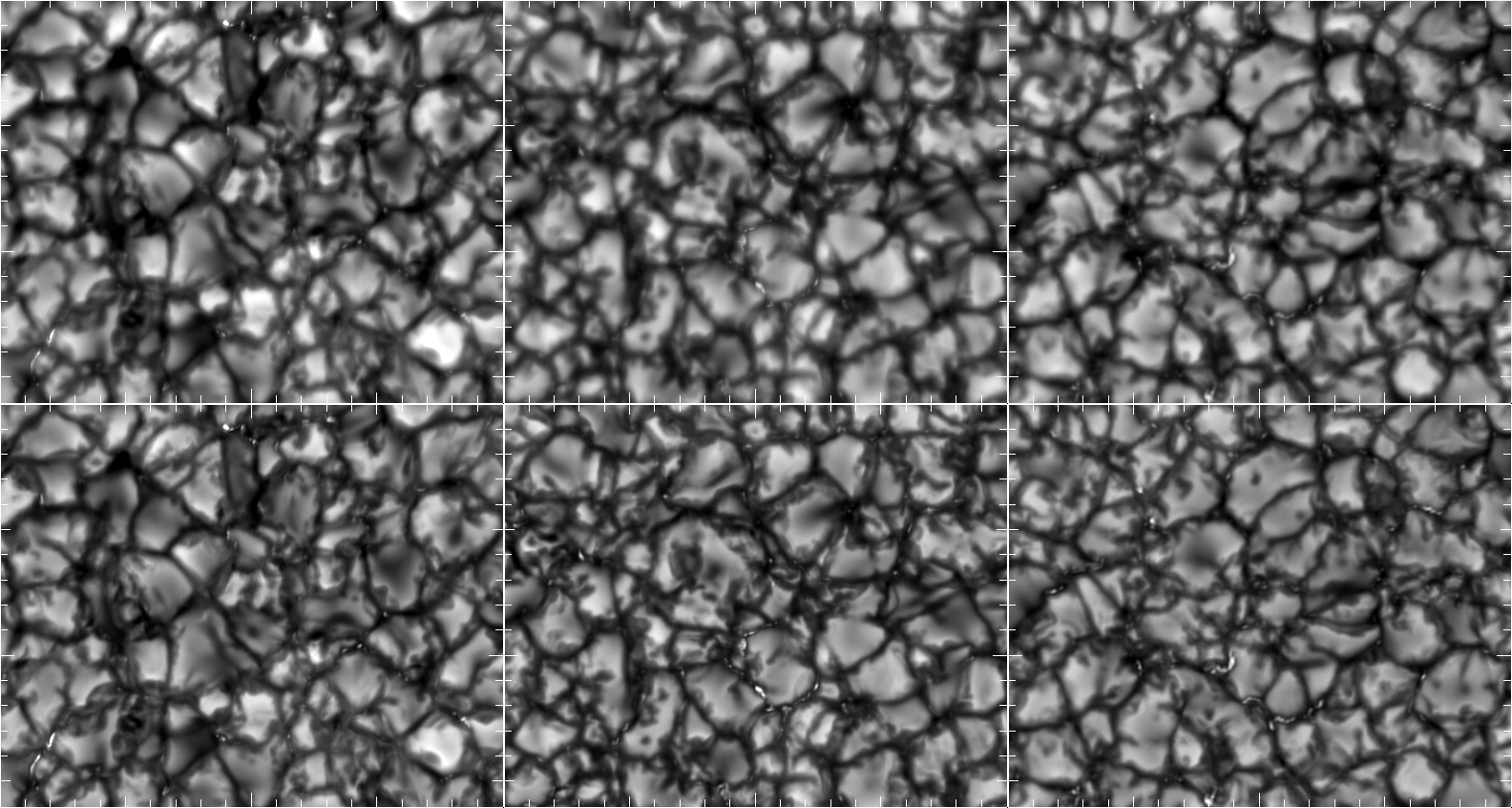}
 \caption{
The top 3 panels show the resulting CHROMIS WB images  after adding up the 100 observed images during 2~sec. and only compensating for relative image motion (tip and tilt) between the individual exposures. The bottom 3 panels show the resulting images of the same observed images but after applying MFBD image reconstruction to compensate for the largest 100 Karhunen--Lo\`eve modes. The wavelengths are (left to right) 395~nm, 400~nm, and 485~nm.
}
\label{fig:CHROMIS1}
\end{figure*}

\begin{table*}
  \centering
  \caption{Observed granulation contrasts with SST/CRISP}
  \begin{tabular}{ l r r r r r r r r r}
    \hline\hline\noalign{\smallskip}
   Wavelength (nm)& \multicolumn{2} {c} {525} &\multicolumn{2} {c} {558} & \multicolumn{2} {c} {630} &\multicolumn{2} {c} {853.5}  \\
   \cmidrule{2-3} \cmidrule{4-5} \cmidrule{6-7} \cmidrule{8-9} 
    & NB & WB & NB & WB & NB & WB & NB & WB \\
    \hline\noalign{\smallskip}
    No corr. & 10.9 &10.5 &10.7 &10.6 &9.2 &9.2 &6.3 &5.6\\
    MTF corr. & 11.8 &11.5 &11.8 &11.6 &10.2 &10.2 &7.2 &6.2 \\
    MFBD corr. & 13.9 &13.7 &13.4 &13.1 &11.7 &11.5 &8.2 &7.2 \\
    Num. simulations & 20.7 & 19.6 & 18.4 & 18.0 & 15.1 & 14.7 & 10.3 & 9.0 \\
        \hline\noalign{\smallskip}
    $r_0$ (m) & \multicolumn{2} {c} {0.164} &\multicolumn{2} {c} {0.239} & \multicolumn{2} {c} {0.238} &\multicolumn{2} {c} {0.270}  \\   
    \hline
  \end{tabular}
  \tablefoot{Summary of observed RMS granulation contrast, given in percent of the mean intensity, for the SST/CRISP wideband (WB) and narrowband (NB) wavelengths indicated. The contrasts given correspond to the best seeing conditions for each data set
    (wavelength) separately. That seeing quality is defined by the value of Fried's parameter $r_0$, also given in the Table, which
    is scaled from the measured value at 500~nm to that of the actual wavelength. Here and in the following, $r_0$ refers to measurements made over a 4\arcsec~$\times$~4\arcsec~field-of-view.  ``No corr.'' corresponds to image data that have not been post-processed beyond that of dark and gain correction, ``MTF corr.'' to images compensated also for tip-tilt and the
    diffraction limited point spread function, and ``MFBD corr.'' to images that have been processed with the multi-frame blind
    deconvolution technique to (partially) compensate also for residual seeing and telescope aberrations using 100 KL modes.}
\label{tab:table_obsRMS}
\end{table*}

In Figs.~\ref{fig:CRISP1} and \ref{fig:CHROMIS1} are shown images recorded through the CRISP and CHROMIS re-imaging systems during moments of excellent seeing during the two campaigns. Each image shown represents an average over 2~sec. and we show images obtained with only tip-tilt correction and compensation for the diffraction limited PSF of the telescope, as well as images that have been restored with the MFBD method to compensate for the 100 most significant KL modes. It is evident that the image quality through these re-imaging systems is outstanding during excellent seeing conditions. As regards CRISP, it also seems evident that the image quality with CRISP itself, which involves two Fabry--P\'erot etalons, is comparable with that obtained through its wideband system. Table \ref{tab:table_obsRMS} summarizes the highest measured contrast values, obtained during excellent seeing conditions, with the different CRISP wideband filters and CRISP itself. 

In the following we quantify and investigate the image quality of CRISP and CHROMIS in variable seeing conditions.  

\begin{figure*}
\center
\includegraphics[viewport=73 46 690 530, width=0.32\textwidth,clip]{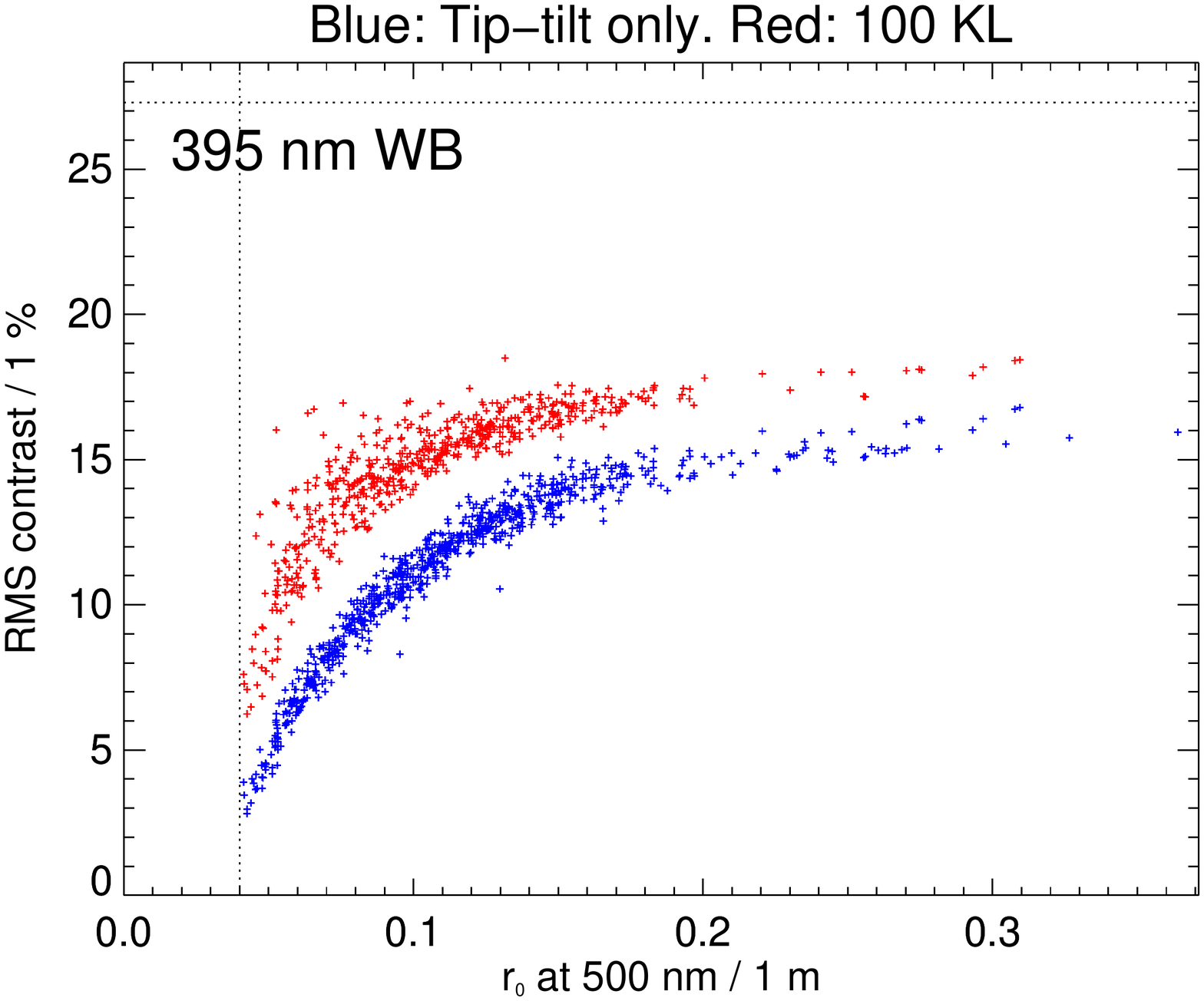}\quad
\includegraphics[viewport=73 46 690 530, width=0.32\textwidth,clip]{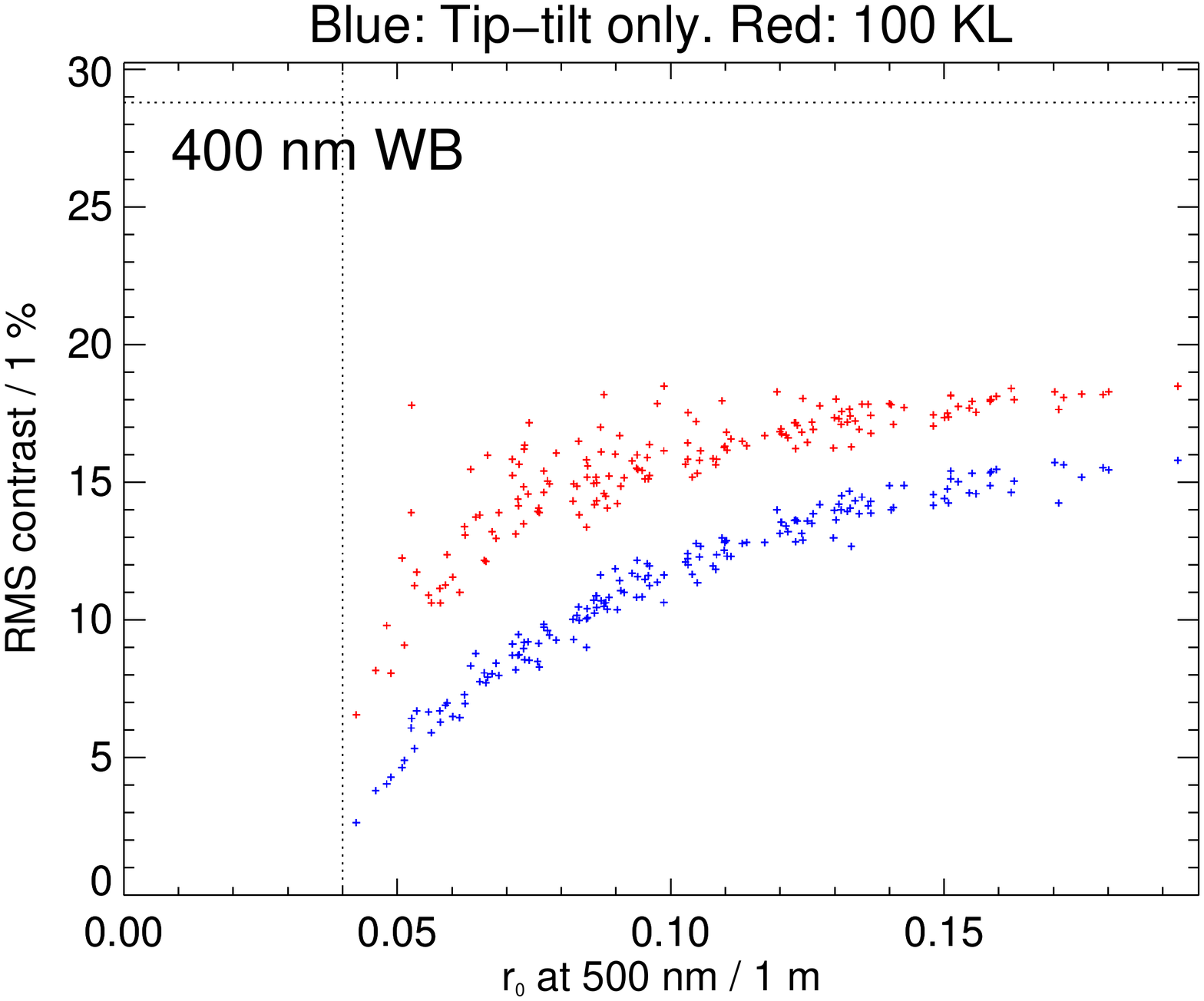}\quad
\includegraphics[viewport=73 46 690 530, width=0.32\textwidth,clip]{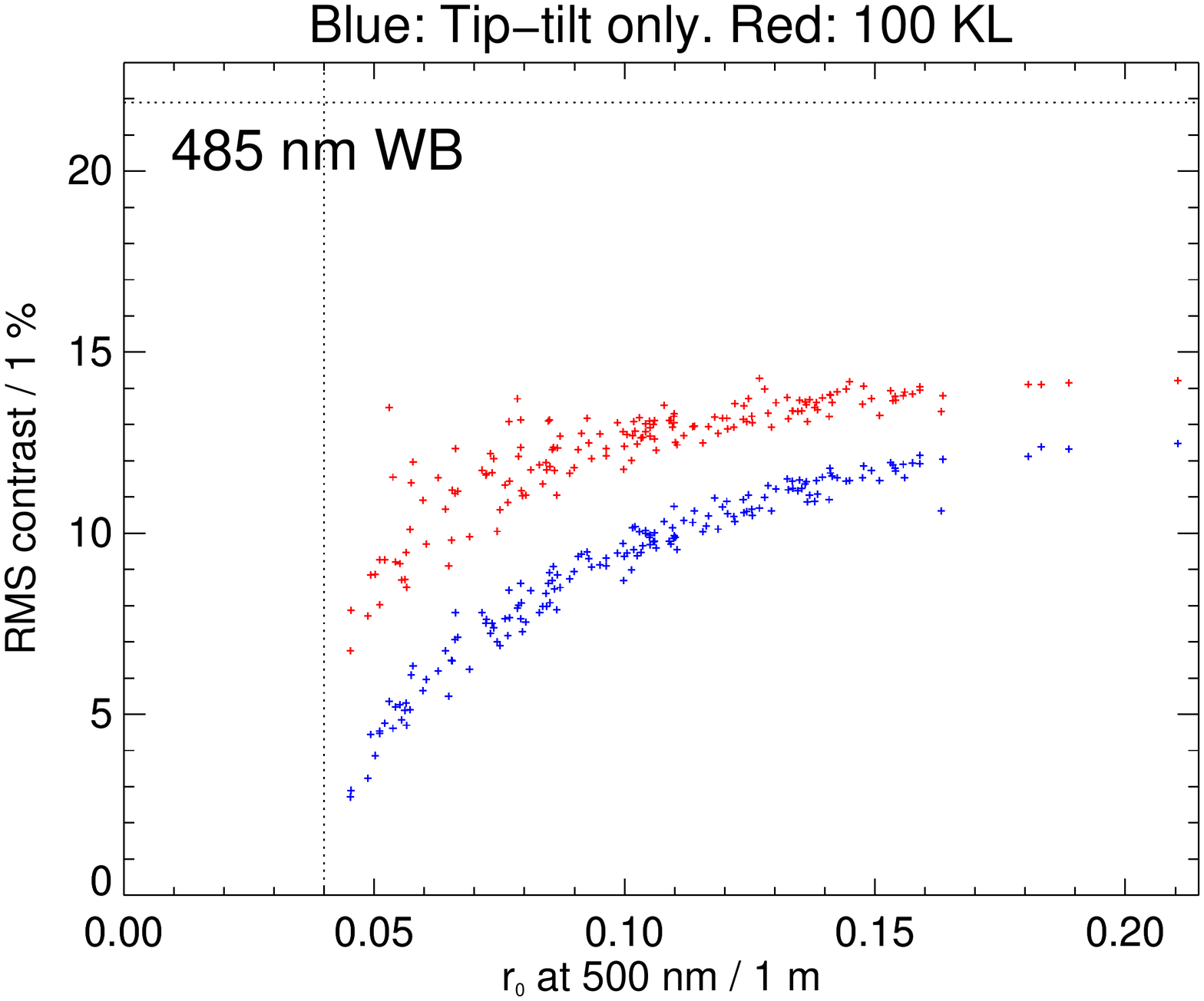}\\[1.5mm]
\includegraphics[viewport=73 46 690 530, width=0.32\textwidth,clip]{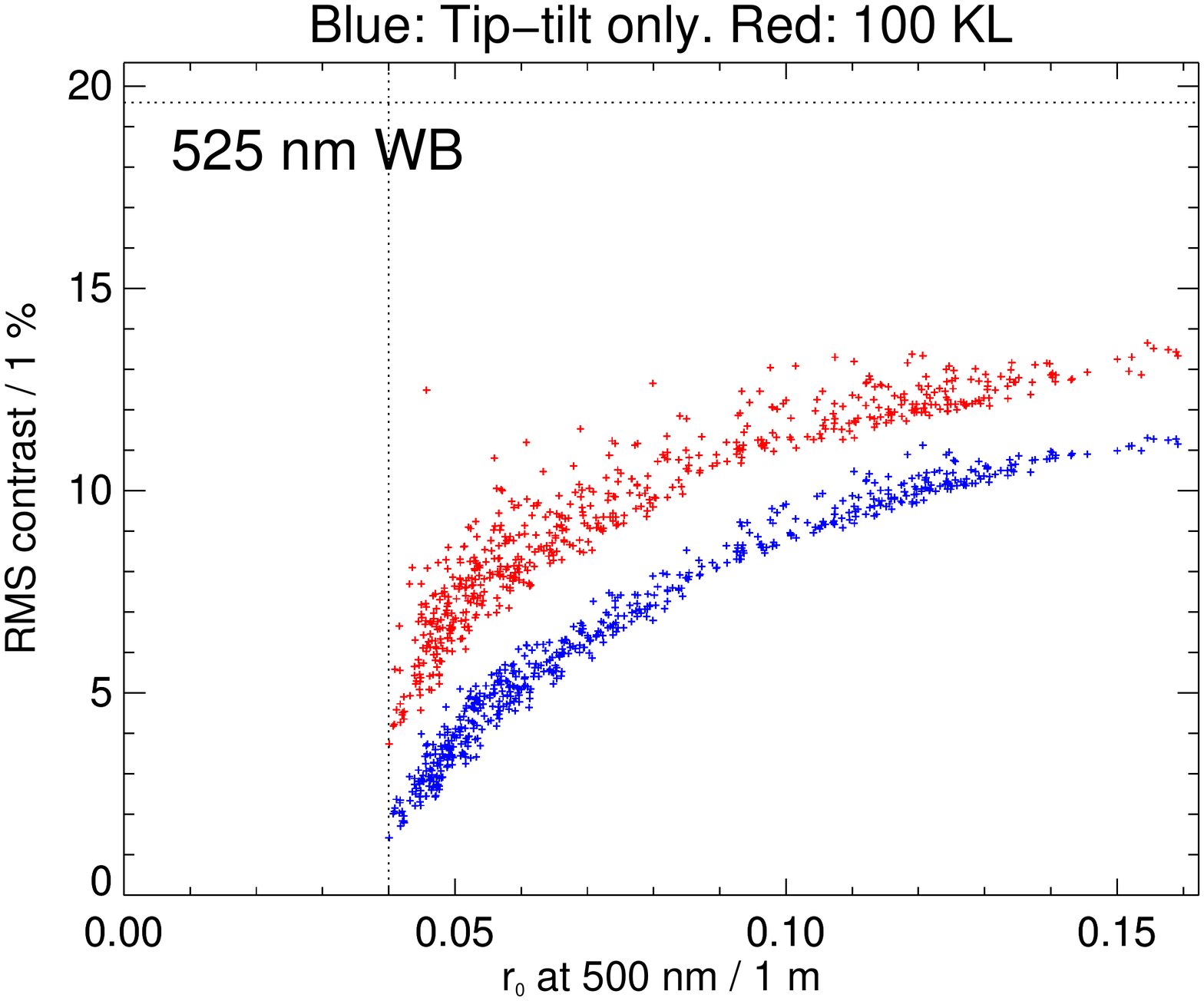}\quad
\includegraphics[viewport=73 46 690 530, width=0.32\textwidth,clip]{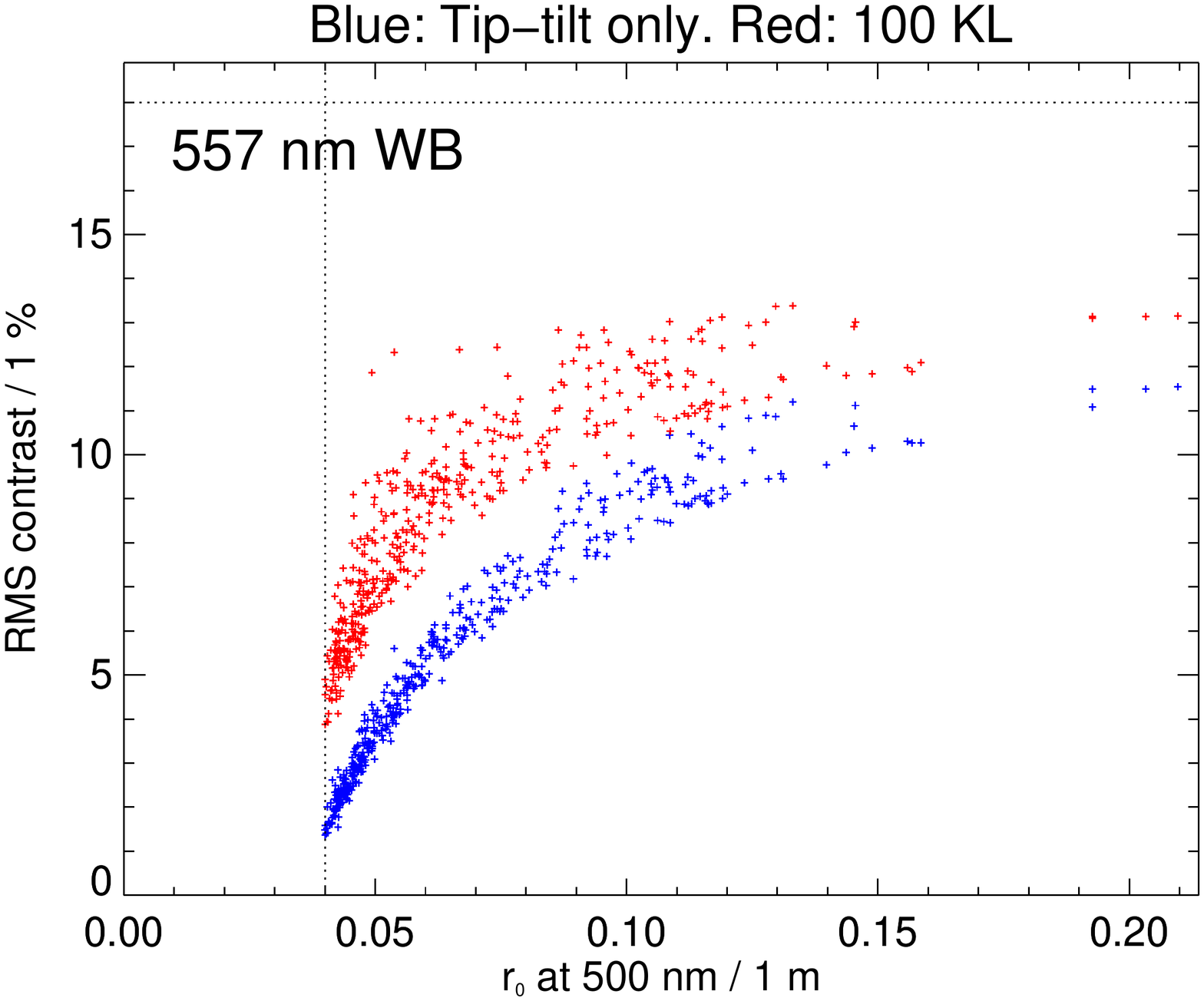}\quad
\includegraphics[viewport=73 46 690 530, width=0.32\textwidth,clip]{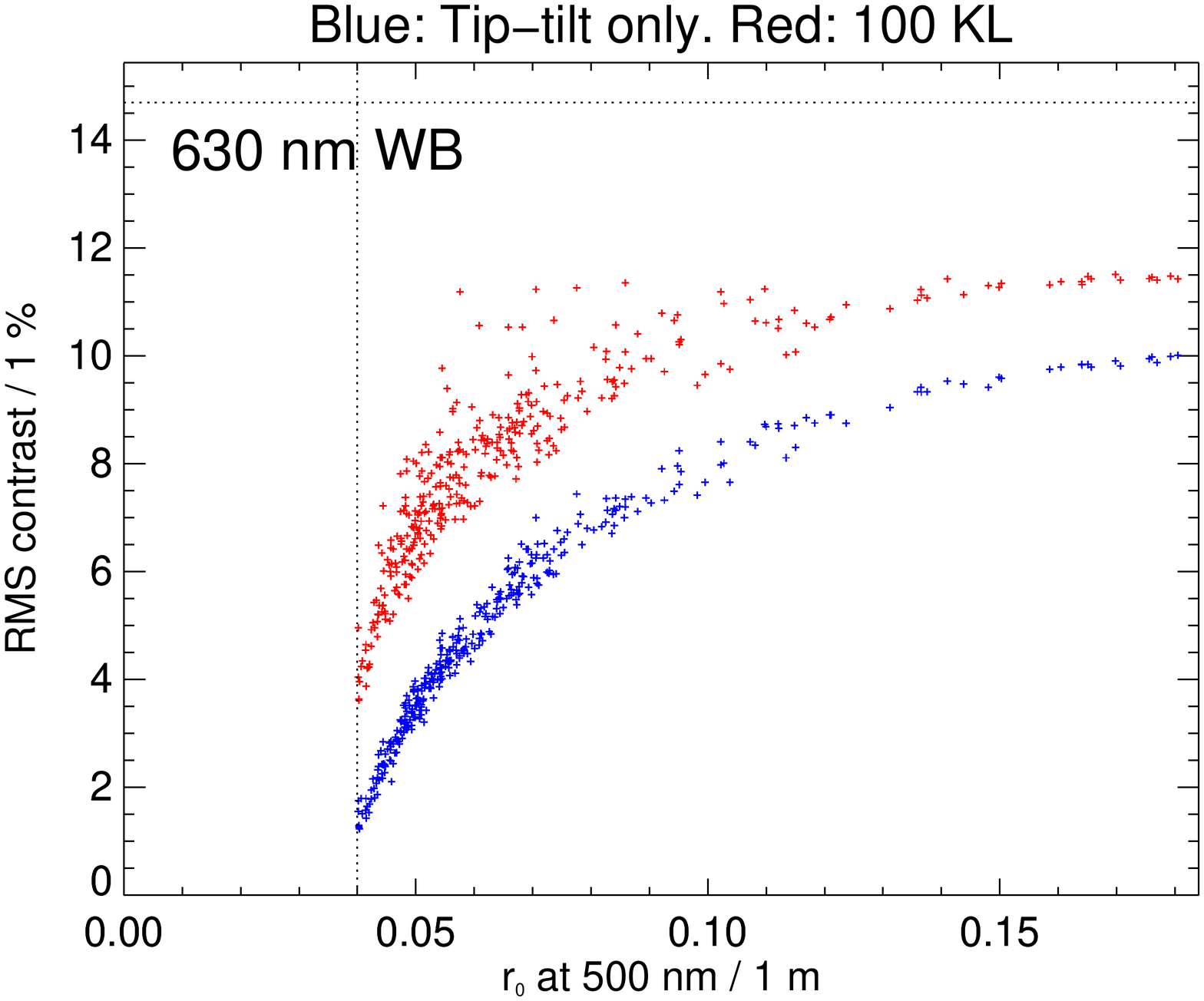}\\[1.5mm]
\includegraphics[viewport=73 46 690 530, width=0.32\textwidth,clip]{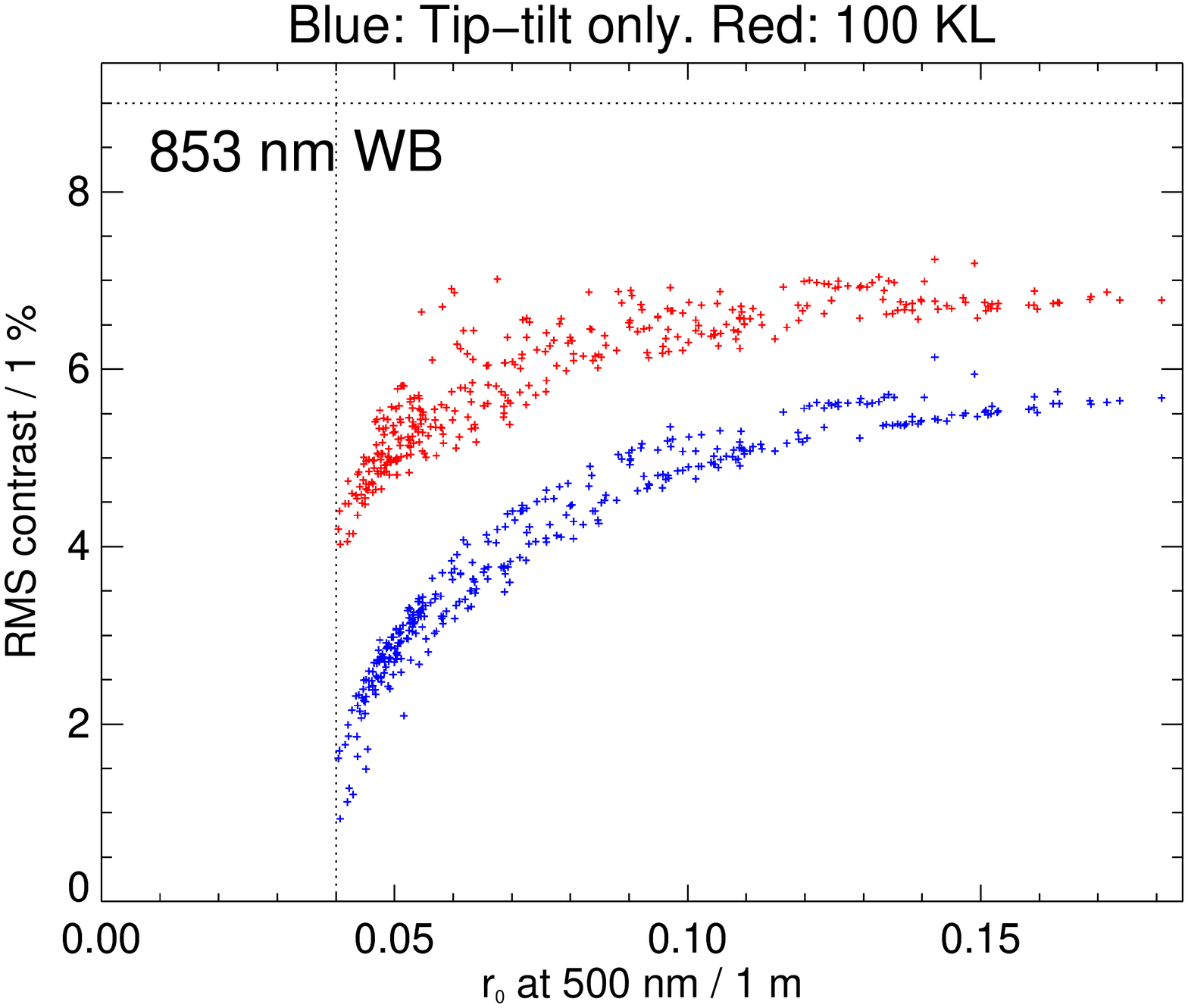}\quad
\includegraphics[viewport=73 46 690 530, width=0.32\textwidth,clip]{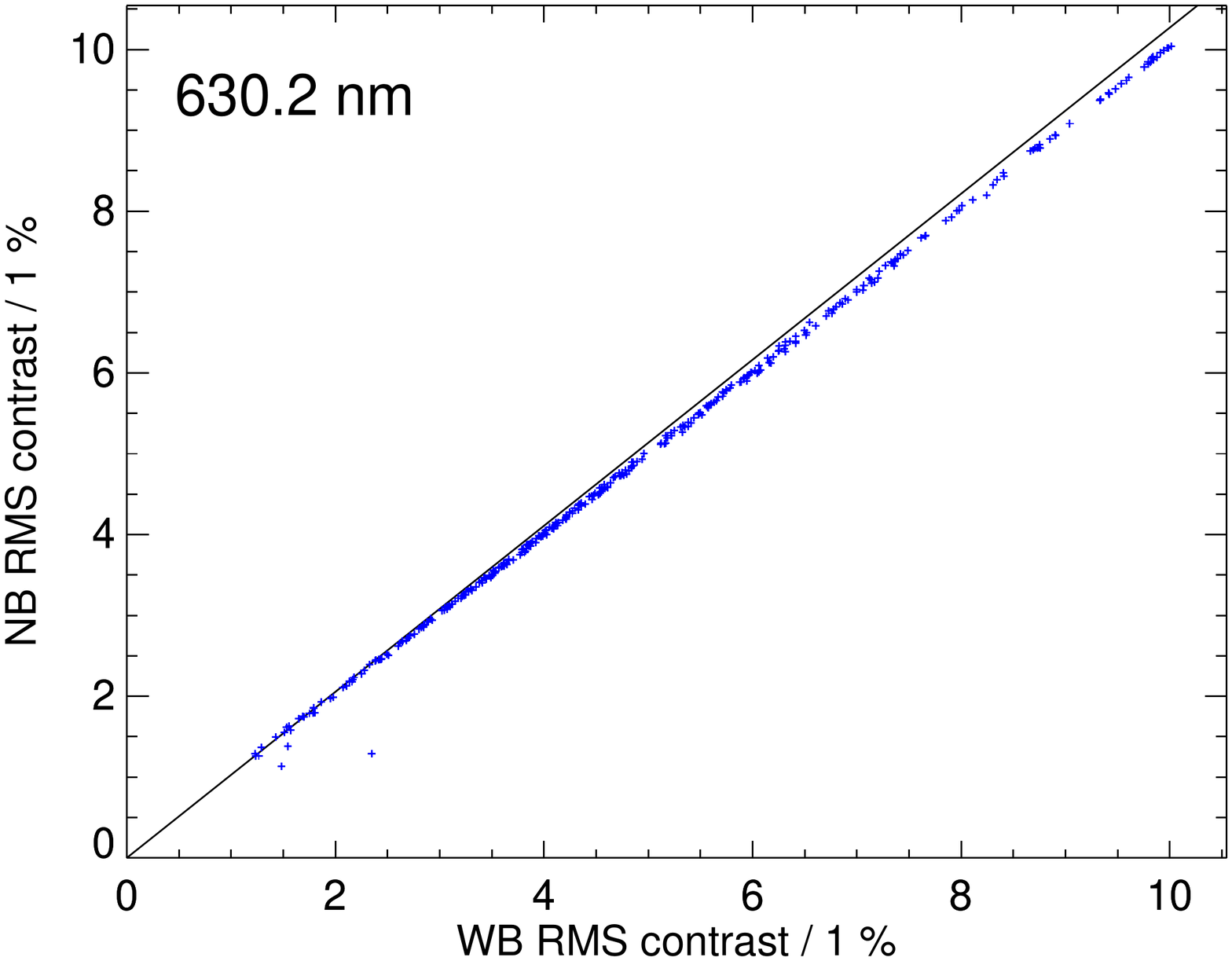}\quad
\includegraphics[viewport=73 46 690 530, width=0.32\textwidth,clip]{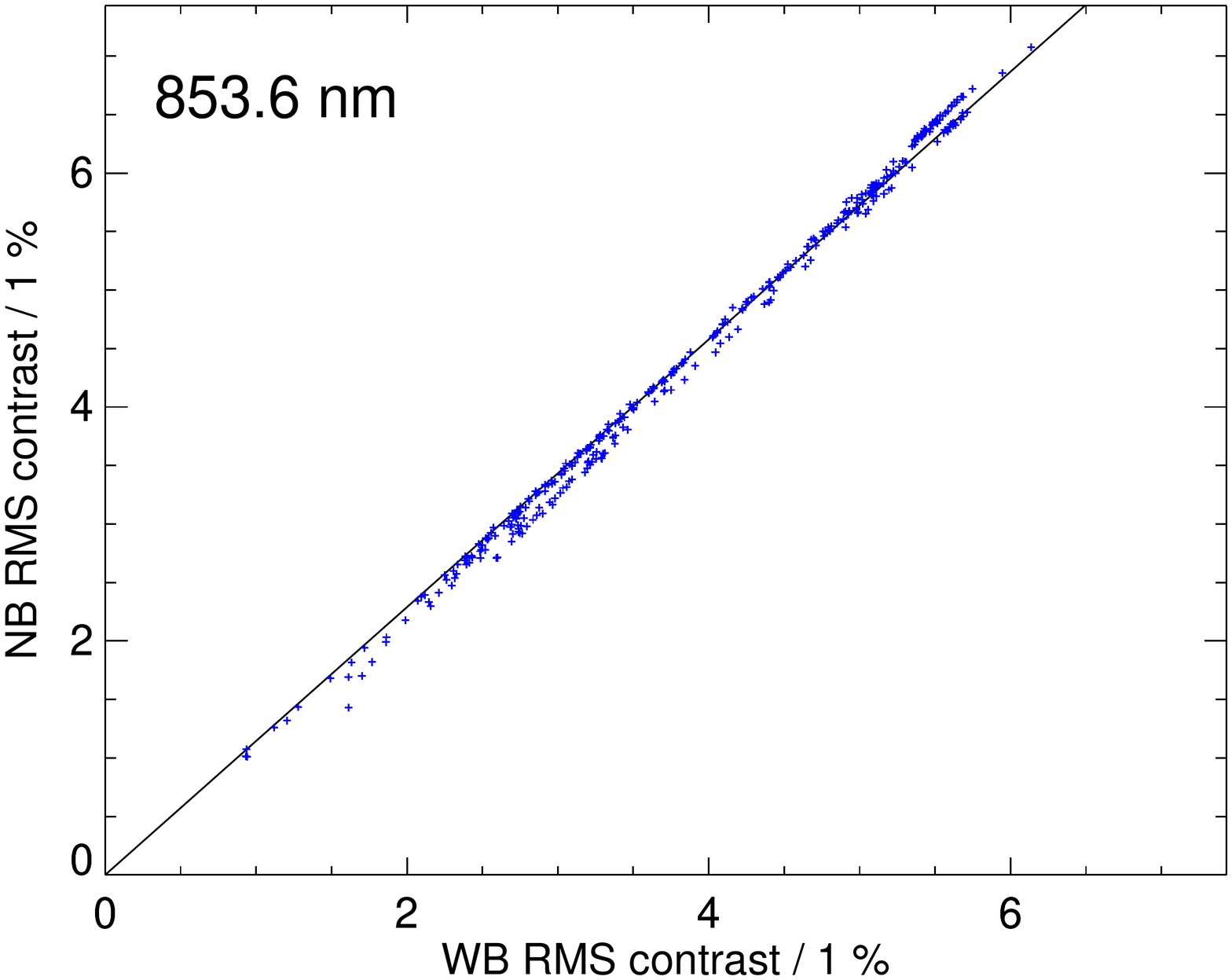}

 \caption{
RMS contrast of granulation measured from SST images recorded at different wavelengths and in different seeing conditions characterized by the Fried $r_0$ parameter. Both $r_0$ and the granulation contrast have been averaged over 2 sec. The blue dots refer to RMS contrasts of images recorded with the AO system and that are compensated only for the diffraction limited point spread function, the red dots to RMS contrasts of images further compensated for aberrations with the multi-frame blind deconvolution method \protect\citep{lofdahl02multi-frame}, and the horizontal dotted lines show the contrasts of 3D MHD simulations. Images were recorded with wide-band (WB) filters on both the ``red'' (CRISP) and ``blue'' (CHROMIS) beams. Note that the plot limits for $r_0$ vary from one plot to another and that data for $r_0$ values smaller than 0.04~m, shown as vertical dotted lines, are excluded. Two of the panels in the lowermost row show a comparison between RMS contrast obtained from images recorded simultaneously through wideband pre-filters (labeled WB) and the corresponding narrowband (NB) CRISP spectropolarimeter filter system on the ``red'' beam of the SST (see text). The significantly higher contrast of the NB beam at 853.6~nm is because of the strong influence of the Ca 854.2~nm line within the passband of the WB filter.
}
\label{fig:CRISP2}
\end{figure*}

Figure~\ref{fig:CRISP2} shows the correlation between $r_0$ and the measured granulation contrast. The blue dots show the correlation obtained with MFBD restorations when compensating for the tip-tilt modes only, and the red dots the results with 100 modes. The dotted horizontal line shows the contrast expected from theoretical simulations, as described in Sect.~\ref{MHD_simulations}.

The blue dots show an excellent correlation between the measured seeing quality ($r_0$), obtained from the AO system, and the granulation contrast measured from science cameras on the blue and red beams. This demonstrates the high relevance of our seeing measurements as an indicator of data quality. For the MFBD processed images with 100 aberration modes (the red dots), the contrast is systematically higher but with a larger spread.

The relations between $r_0$ and the image contrast consistently is highly non-linear and with clear evidence of showing asymptotic convergence to values that are well below those expected from theoretical MHD simulations. The origin of this behaviour is discussed in the following section.

Finally, the last two plots in Fig.~\ref{fig:CRISP2} show the relation between the contrasts measured with the wideband and narrowband beams of CRISP for two wavelengths (630.2~nm and 853.6~nm), along with the relation expected from theoretical simulations (shown as a straight line). It is evident that the contrast in the narrowband CRISP data is only marginally less than expected from MHD simulations, and the contrast in the corresponding wideband images.

\section{Theoretical interpretation}
To explain the observed relation between $r_0$ and the measured granulation contrast, we rely on the (by now) well established theory of the the PSF in the presence of turbulence-induced seeing and the effects of partial compensation of the PSF for such seeing by an AO system - for early reviews see, e.g., \citet{1992ESOC...42..475C,1998aoat.book.....H,2006PASP..118..885B}. Such partial compensation results in a PSF that can be roughly described as the linear combination of a diffraction limited PSF and another much wider PSF, usually referred to as the ``halo''
\begin{equation}
 P = S P_d + (1-S) P_h
\end{equation}
where $S$ is the Strehl ratio, $P_d$ the diffraction limited PSF and $P_h$ the PSF corresponding to the halo. Since the full width at half maximum (FWHM) of $P_h$ with a well-functioning high-order AO system on a meter-class telescope will typically be one order of magnitude larger than the FWHM of  $P_d$, it will be comparable to the scale of granulation, which is about 1\farcs5. We therefore expect that primarily $P_d$ will contribute to the observed contrast, which will then tend to equal that of the diffraction limited telescope reduced by a factor $S$. 

\begin{figure*}
\center
\includegraphics[viewport=73 46 690 530, width=0.32\textwidth,clip]{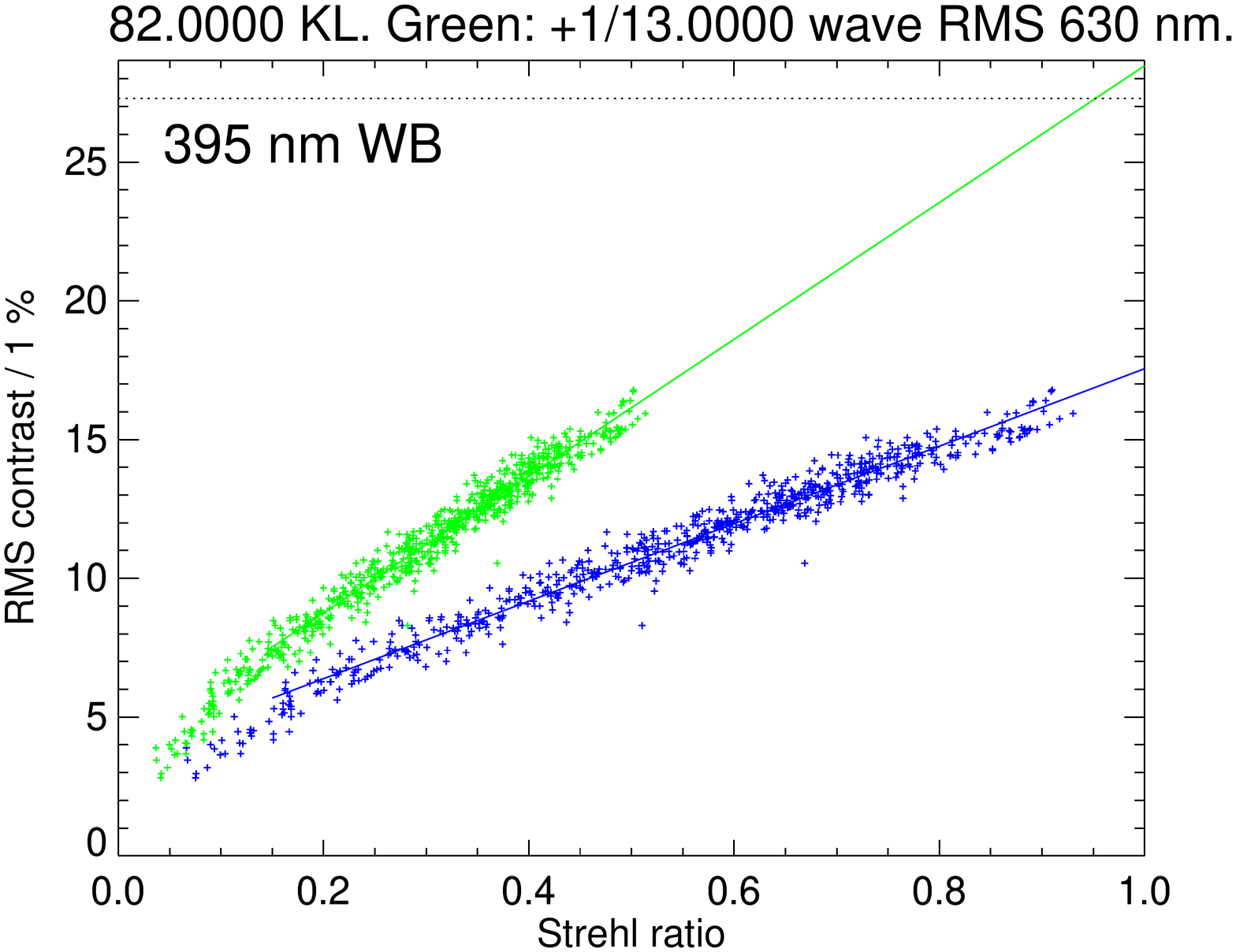}\quad
\includegraphics[viewport=73 46 690 530, width=0.32\textwidth,clip]{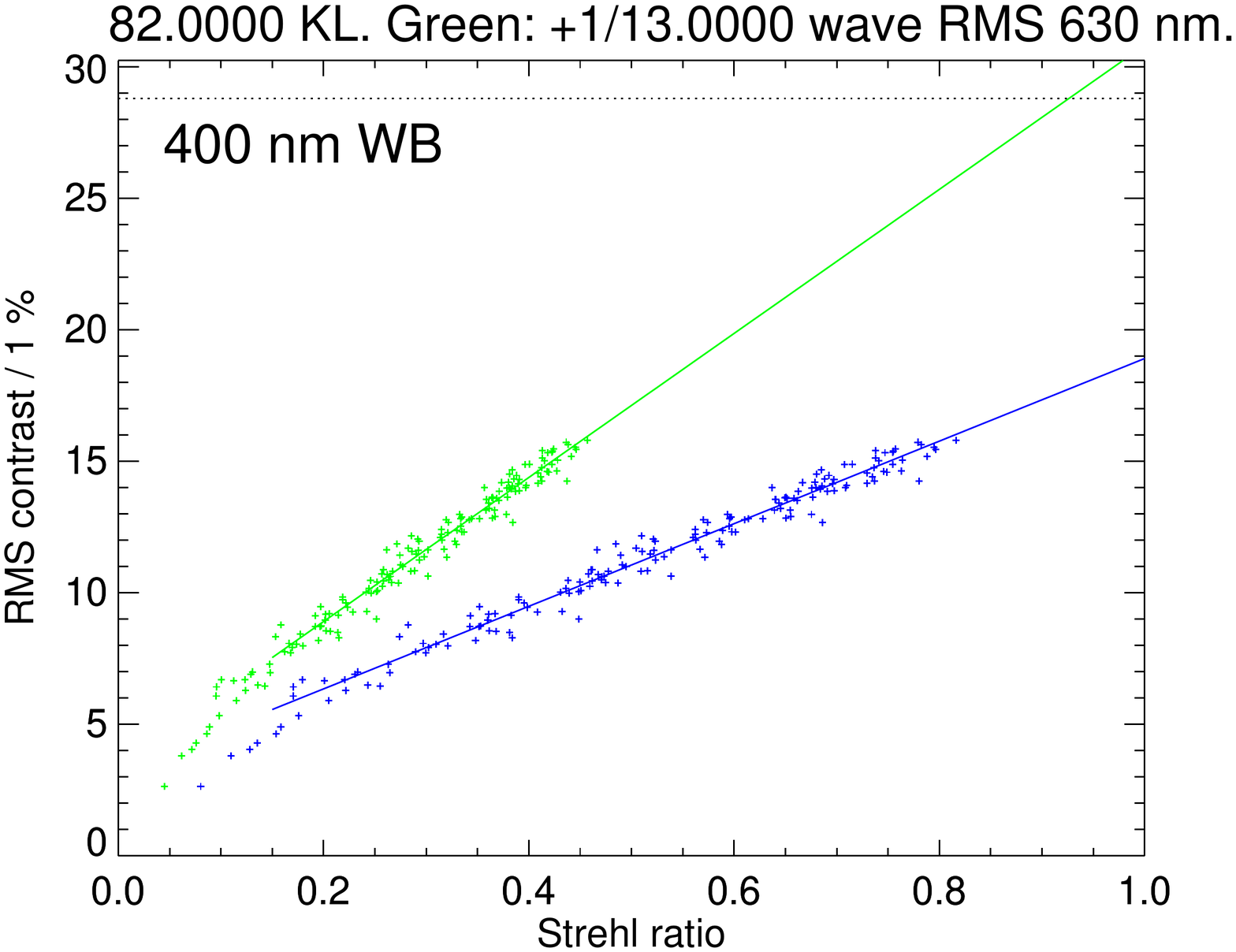}\quad
\includegraphics[viewport=73 46 690 530, width=0.32\textwidth,clip]{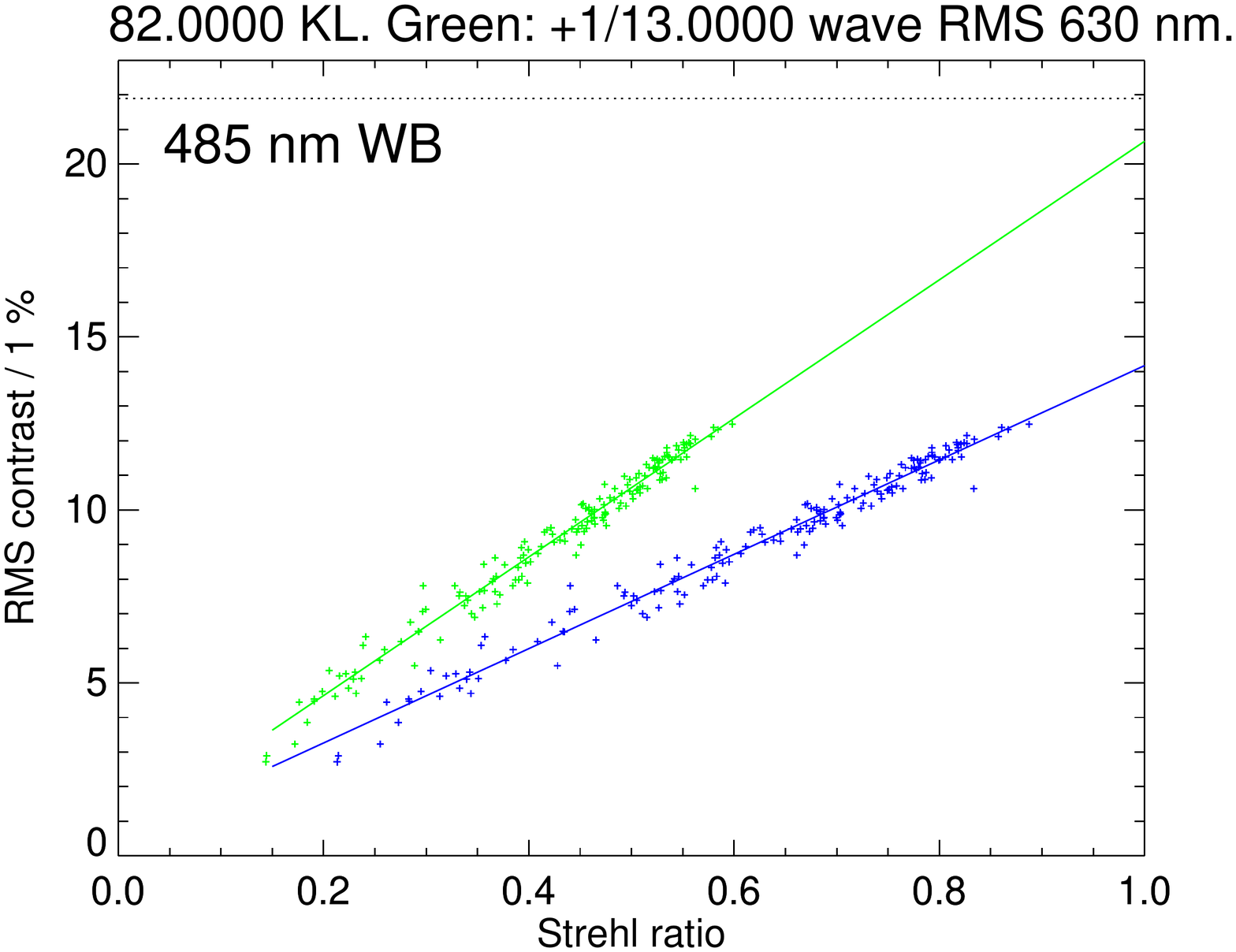}\\[1.5mm]
\includegraphics[viewport=73 46 690 530, width=0.32\textwidth,clip]{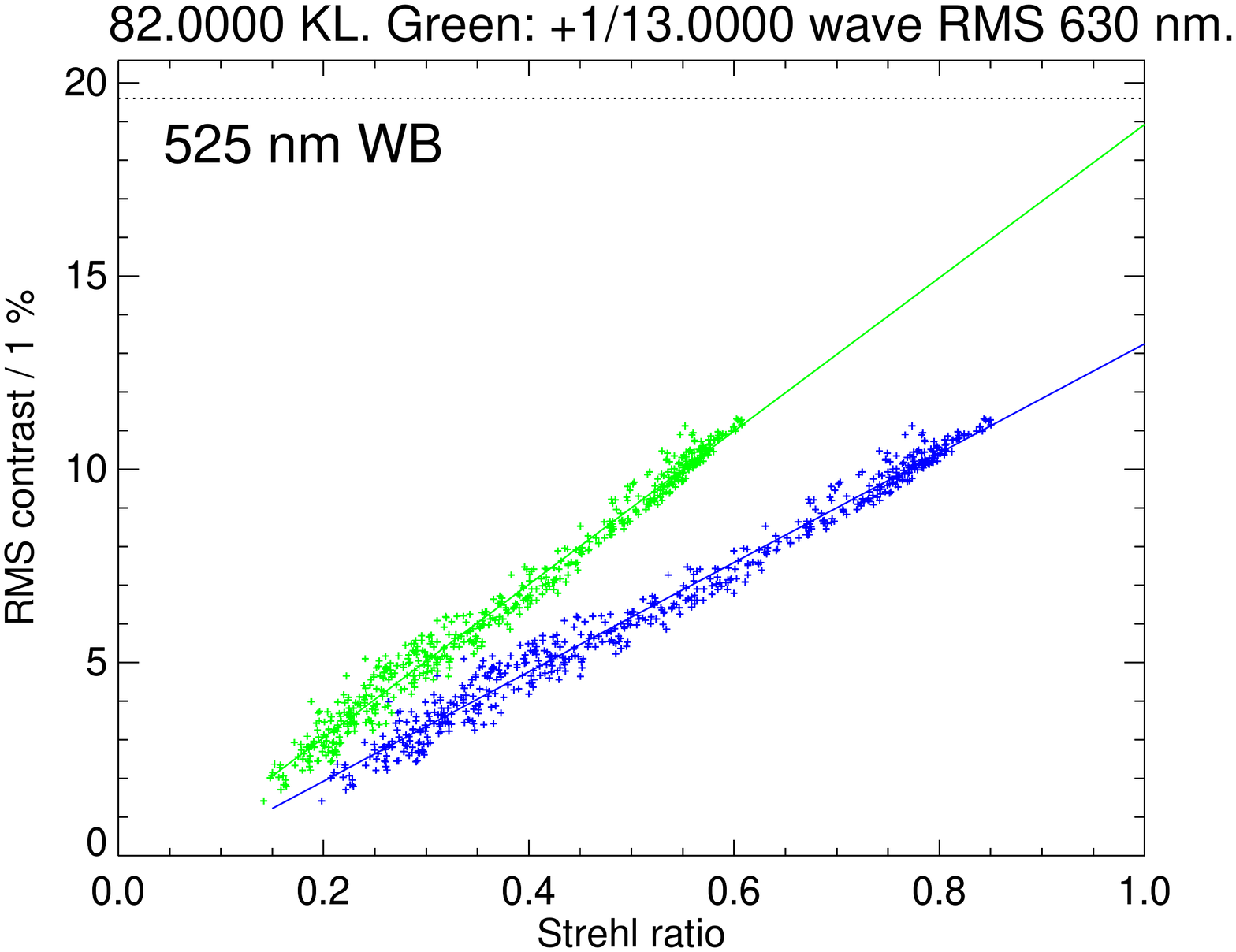}\quad
\includegraphics[viewport=73 46 690 530, width=0.32\textwidth,clip]{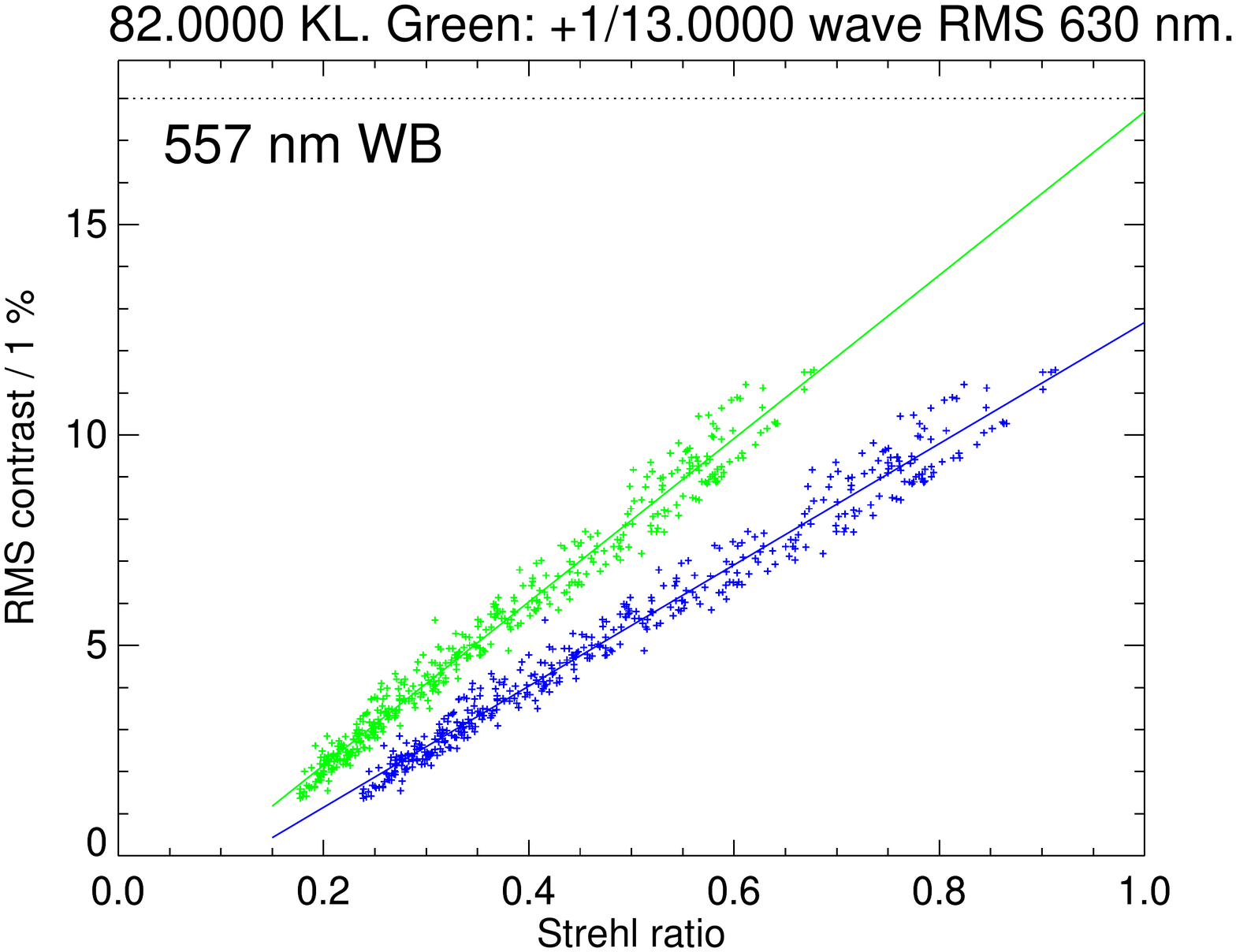}\quad
\includegraphics[viewport=73 46 690 530, width=0.32\textwidth,clip]{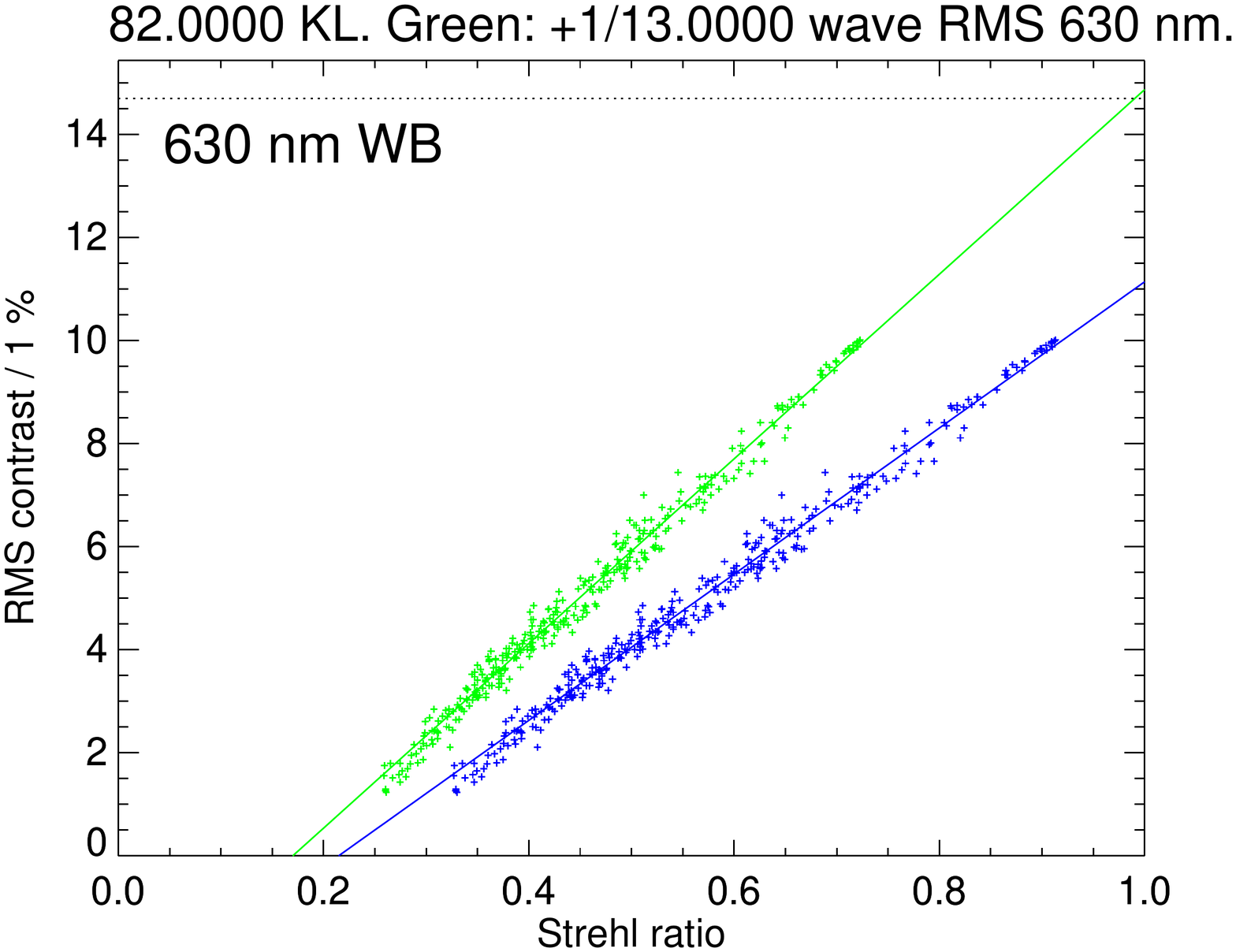}\\[1.5mm]
\includegraphics[viewport=73 46 690 530, width=0.32\textwidth,clip]{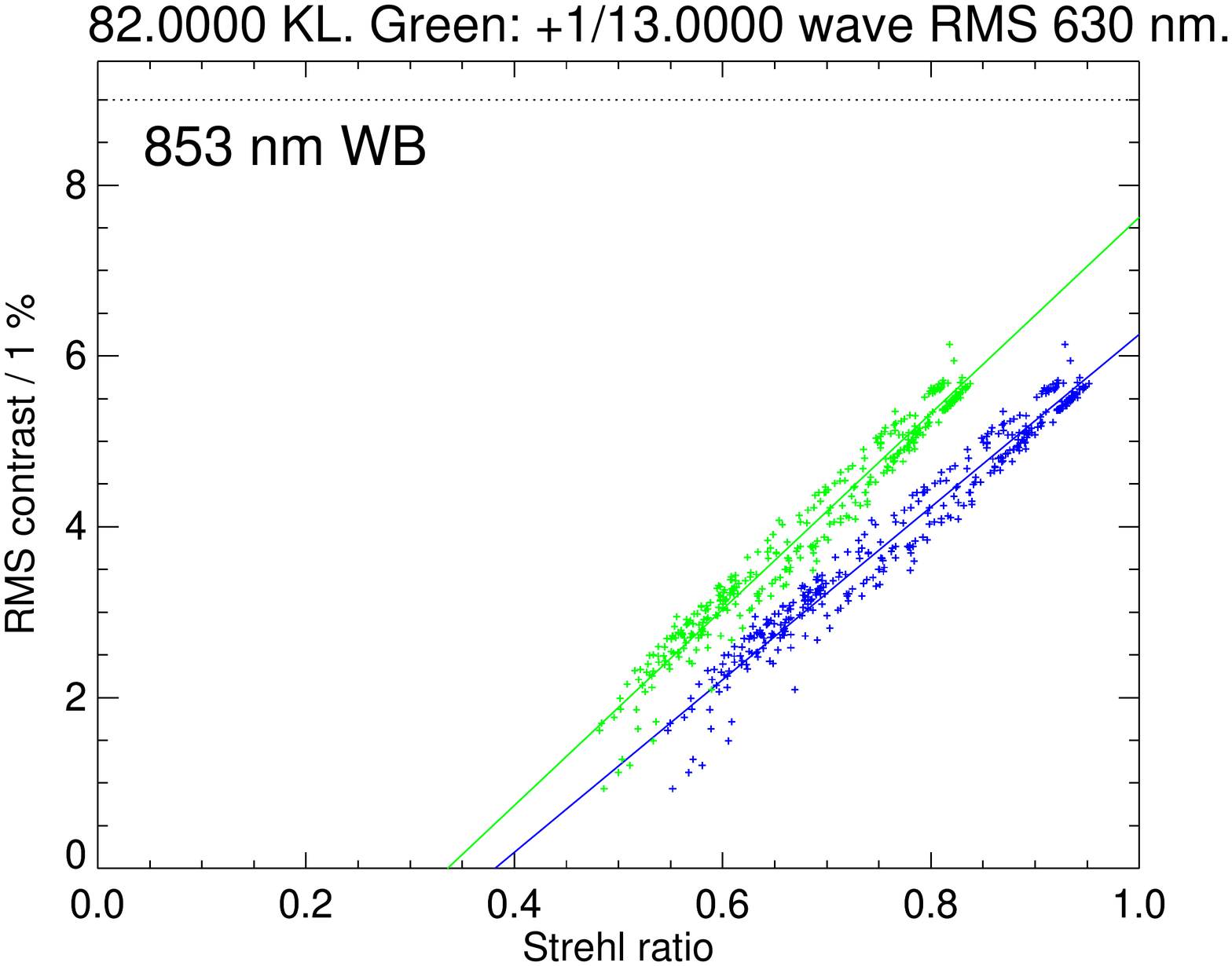}\quad
\includegraphics[viewport=73 46 690 530, width=0.32\textwidth,clip]{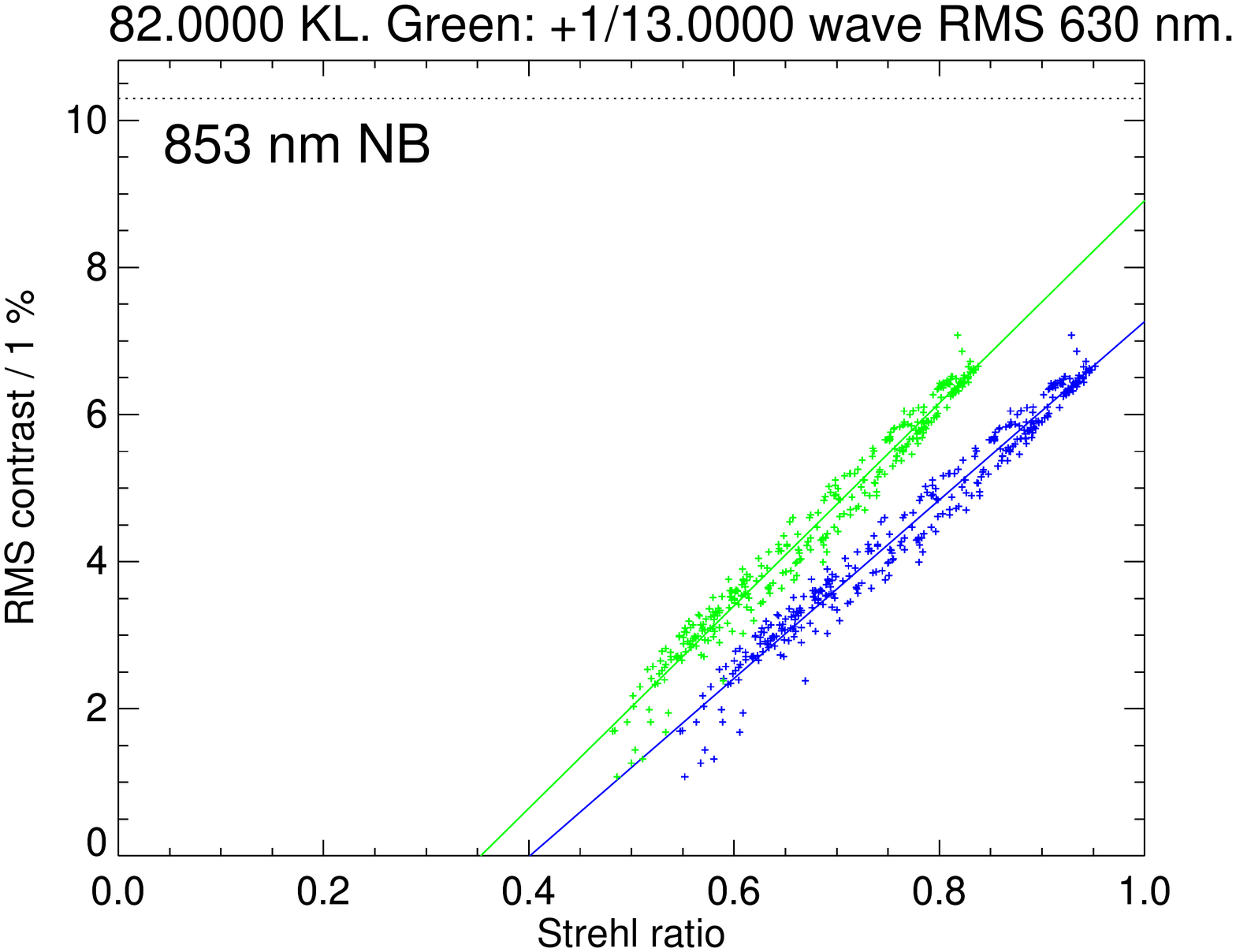}

 \caption{
RMS contrast of granulation images recorded in different seeing conditions characterized by the calculated Strehl value based on the residual wavefront RMS from $r_0$ and assuming that the AO system provides perfect  correction of 81 modes. The blue dots refer to RMS contrasts of images that are compensated only for the diffraction limited point spread function, and correspond  to the blue dots in Fig.~\ref{fig:CRISP2}. The green dots corresponds to the same contrast measurements but with the Strehl value calculated on the basis of an additional wavefront variance corresponding to $1/13$~wave RMS at 630 nm, and scaled at other wavelengths as the inverse of the wavelength. The straight lines have been fitted to data points for which the Strehl is larger than 0.15.
}
\label{fig:CRISP3}
\end{figure*}

To test this conjecture, we plot in Fig.~\ref{fig:CRISP3} the observed granulation contrast versus the Strehl value obtained from $r_0$ and an assumed number of perfectly corrected (independent) aberration parameters $N$ by the AO system\footnote{The formula given by \citet{1998PASP..110..837R} includes the inconsequential piston mode as $N=1$, whereas we refer to $N=1$ and $N=2$ as the tip-tilt modes}. The wavefront variance is estimated for zonal correction as \citep{1998PASP..110..837R}
\begin{equation}
  \sigma_N^2 = 0.34\, (D/r_0)^{5/3} (N+1)^{-5/6},
  \label{eq:9}
\end{equation}
where $\sigma_N$ is in radians and $D$ is the telescope diameter. The Strehl ratio $S$ can be estimated as
\begin{equation}
  S = S_N = \exp(-\sigma_N^2).
  \label{eq:10}
\end{equation}

Assuming almost perfect correction by the SST 85-electrode AO system, $N=81$, we show in Fig.~\ref{fig:CRISP3} the measured granulation contrast versus the inferred Strehl value. In this Figure, the blue and green dots correspond to the same measured granulation contrasts as shown in Fig.~\ref{fig:CRISP2} and the blue dots to the Strehl values calculated as described previously. The relation between the Strehl and the contrast is almost perfectly linear. From a fit of a straight line to the data, which ignores all Strehl values below 0.15, we can extrapolate to find the ``seeing-free'' contrast for $S=1$. It is evident that the extrapolated values are well below the values expected from theoretical simulations, shown as dotted lines. The green dots in Fig.~\ref{fig:CRISP3} corresponds to the Strehl ratios obtained by adding an ad-hoc wavefront variance corresponding to 48~nm RMS (1/13 wave at 630~nm), assumed to be from fixed aberrations or short-comings in the AO system. This corresponds to rewriting Eq.~(\ref{eq:10})  as
\begin{equation}
  S = \exp(-(\sigma_s^2+\sigma_N^2)) = \exp(-\sigma_s^2) ~S_{N} = S_s ~S_N.
  \label{eq:11}
\end{equation}
such that the effect of additional static aberrations is to multiply the Strehl values associated with residual seeing by a factor $S_s$.

\begin{figure*}
\center
\includegraphics[viewport=73 46 690 530, width=0.32\textwidth,clip]{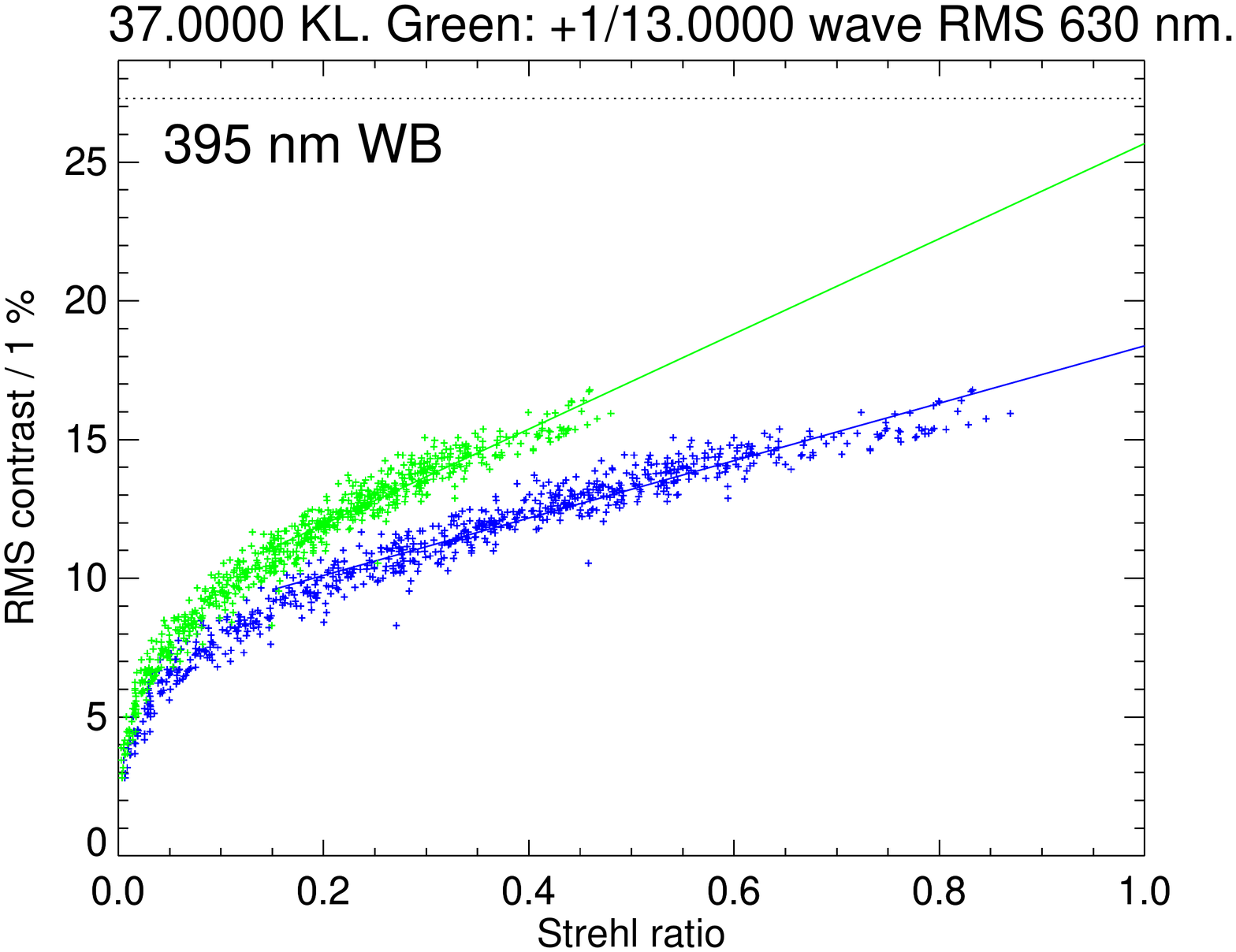}\quad
\includegraphics[viewport=73 46 690 530, width=0.32\textwidth,clip]{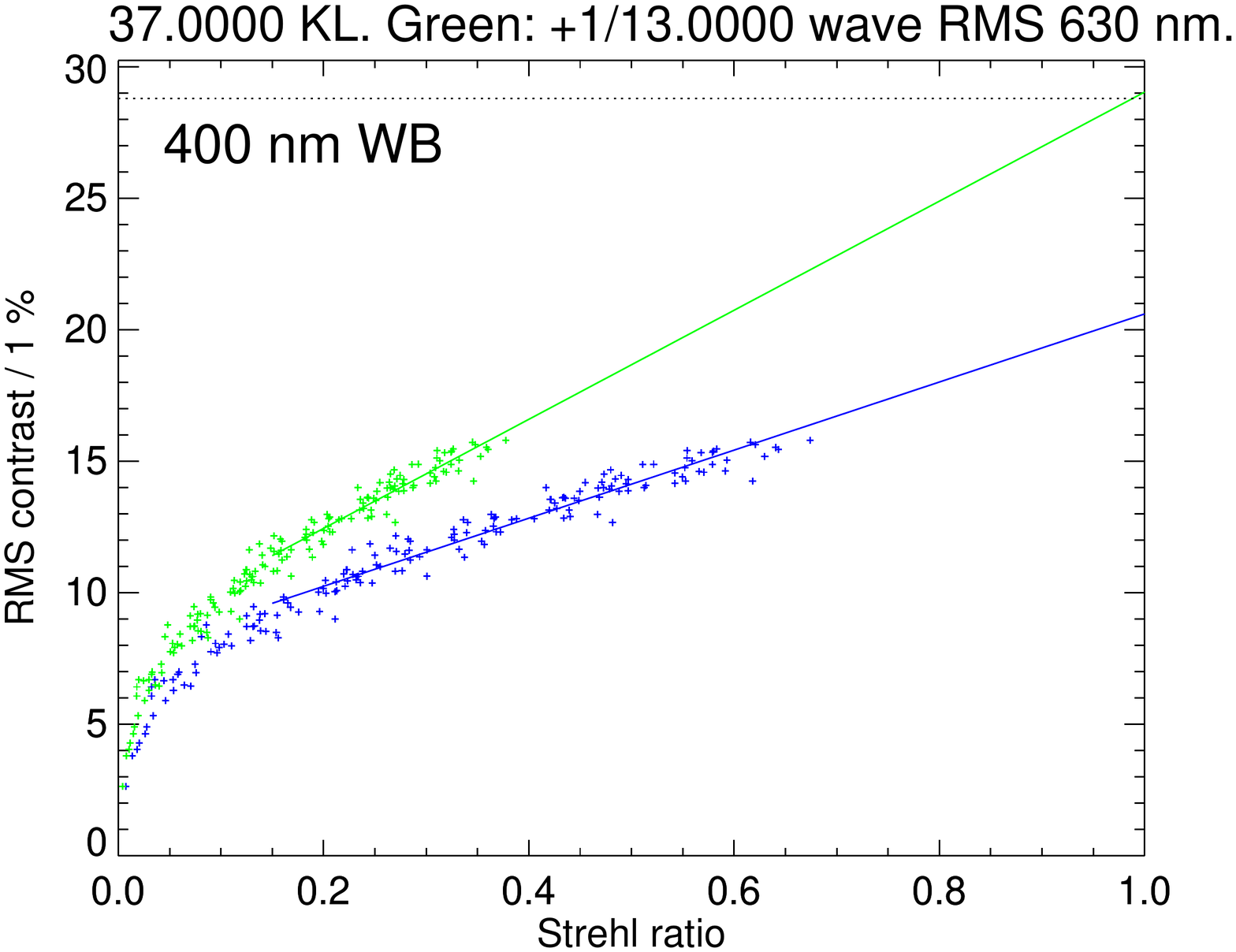}\quad
\includegraphics[viewport=73 46 690 530,  width=0.32\textwidth,clip]{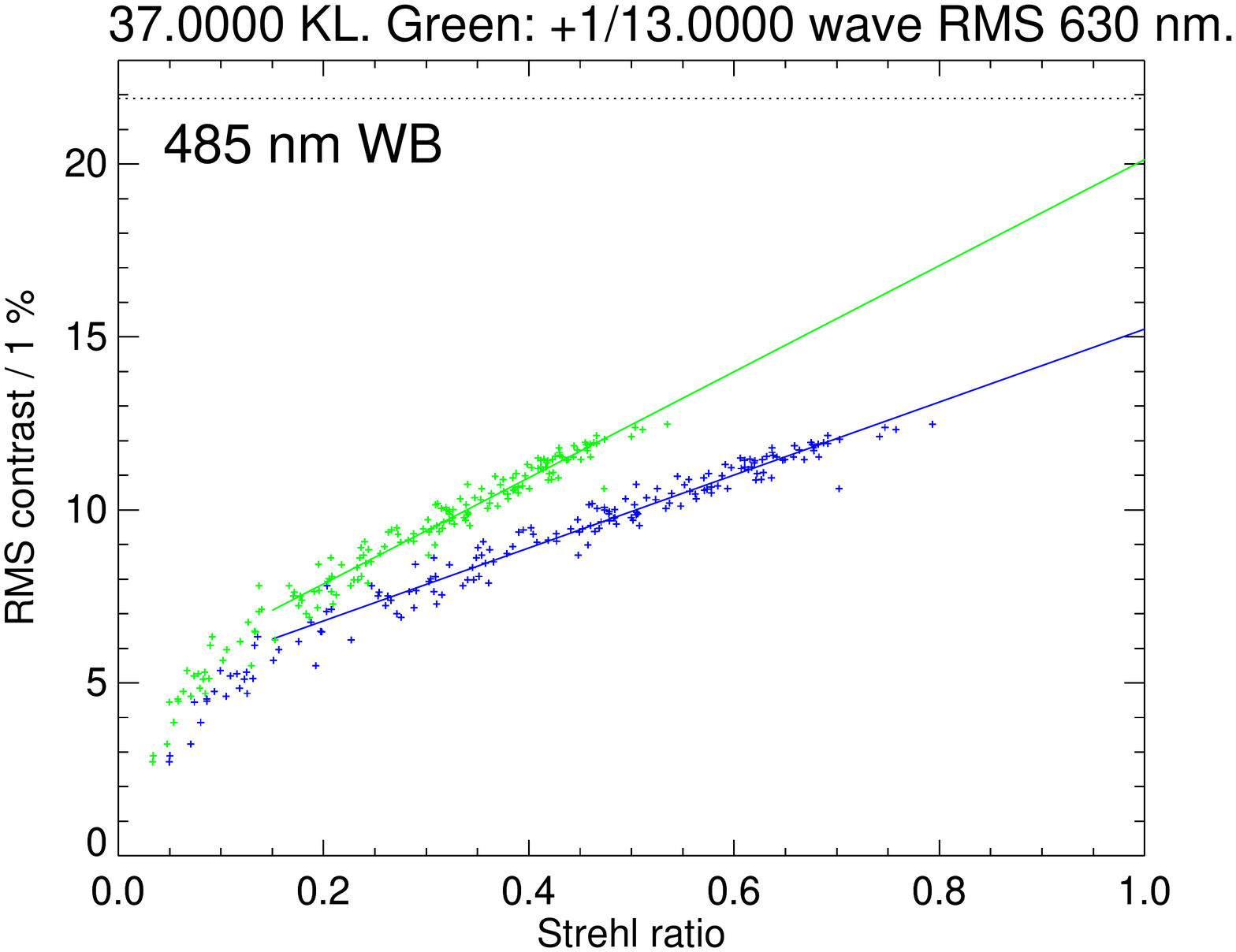}\\[1.5mm]
\includegraphics[viewport=73 46 690 530,  width=0.32\textwidth,clip]{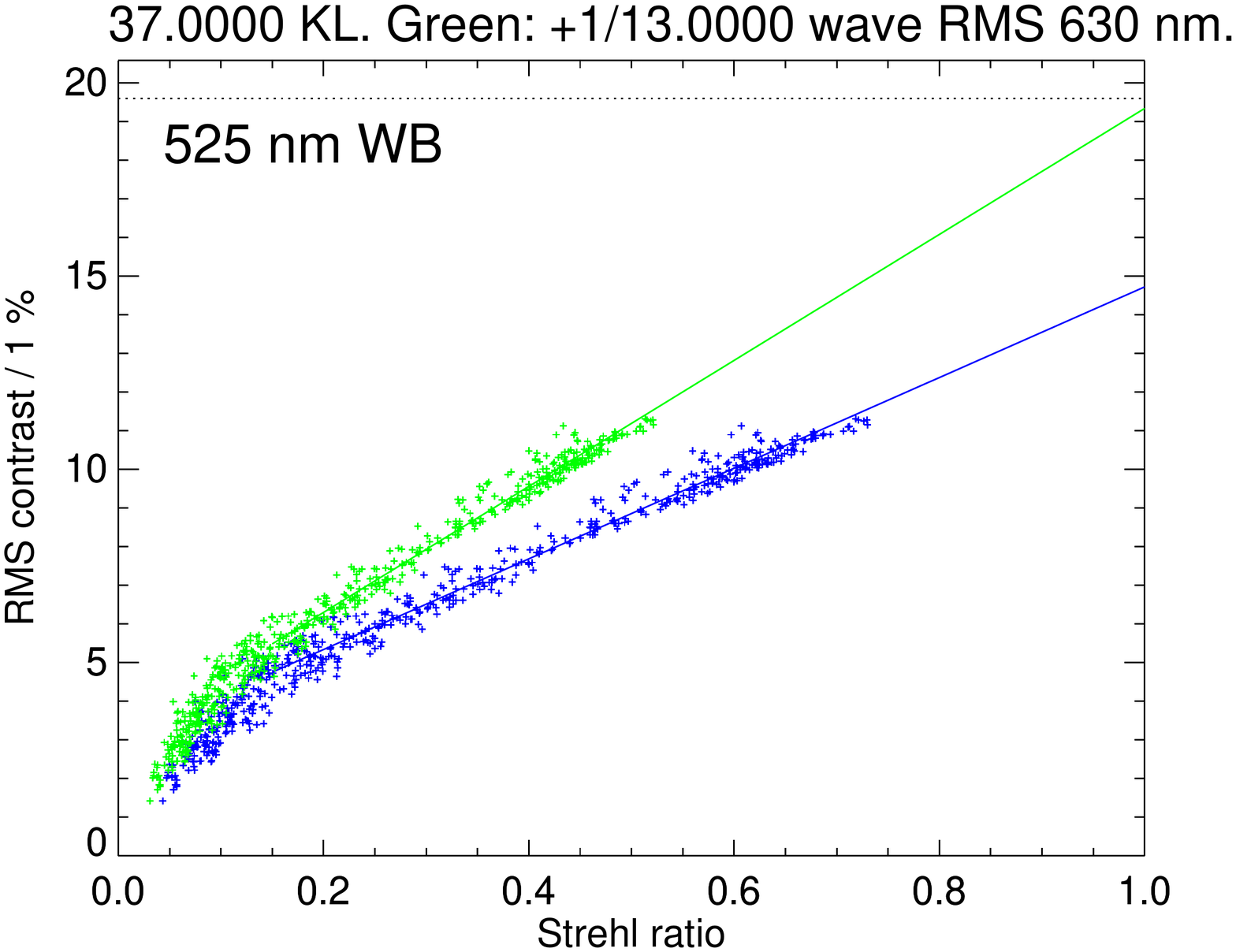}\quad
\includegraphics[viewport=73 46 690 530,  width=0.32\textwidth,clip]{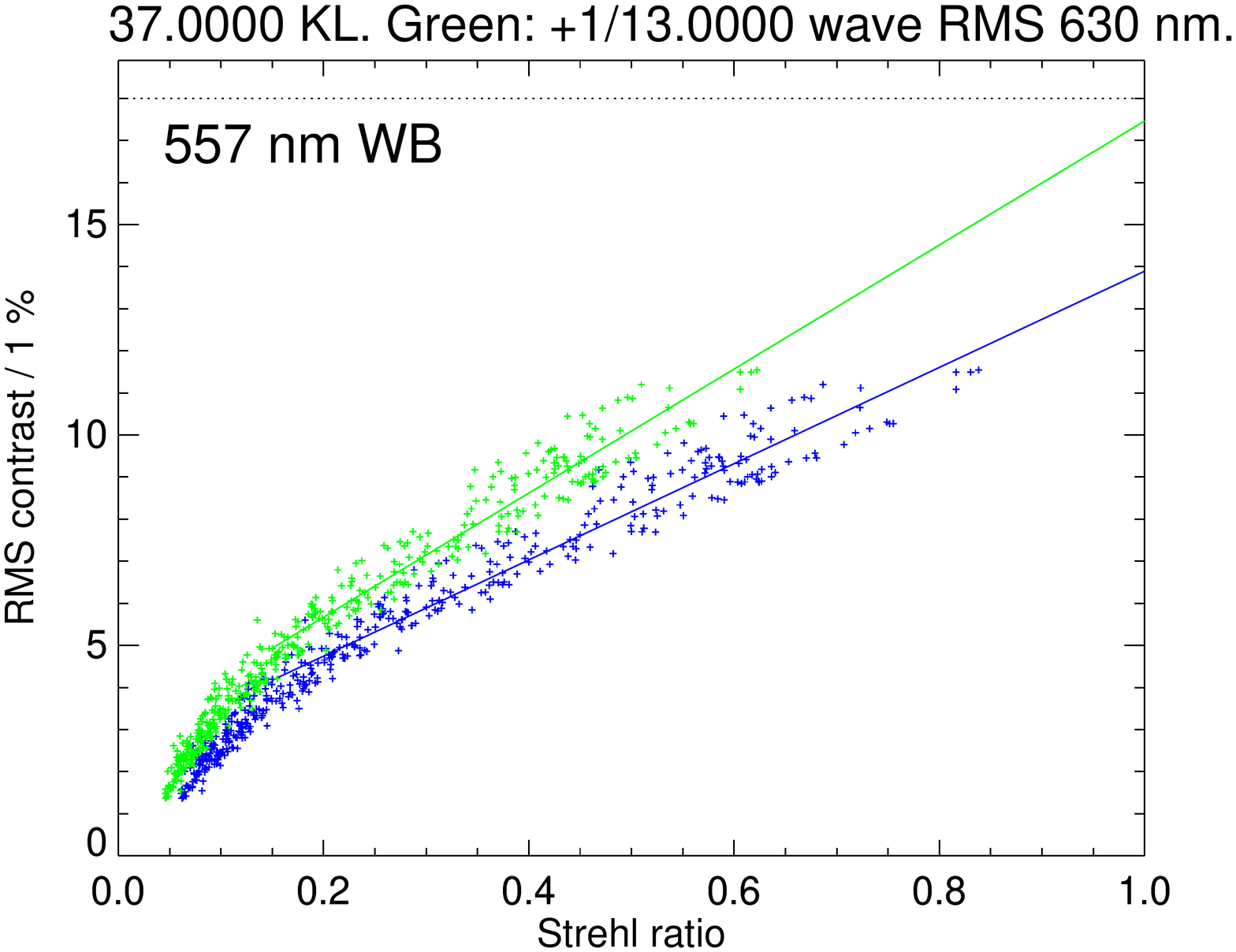}\quad
\includegraphics[viewport=73 46 690 530,  width=0.32\textwidth,clip]{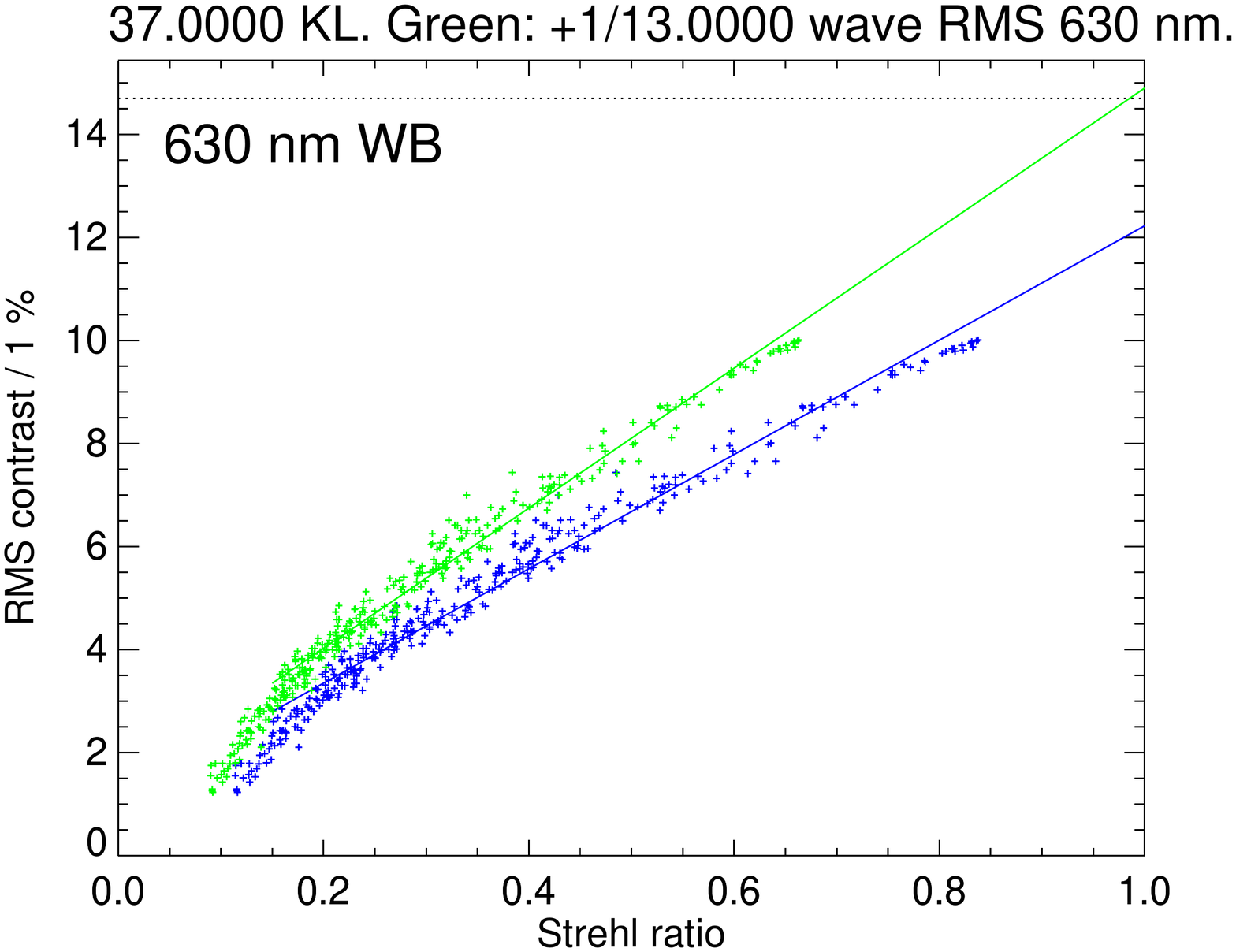}\\[1.5mm]
\includegraphics[viewport=73 46 690 530,  width=0.32\textwidth,clip]{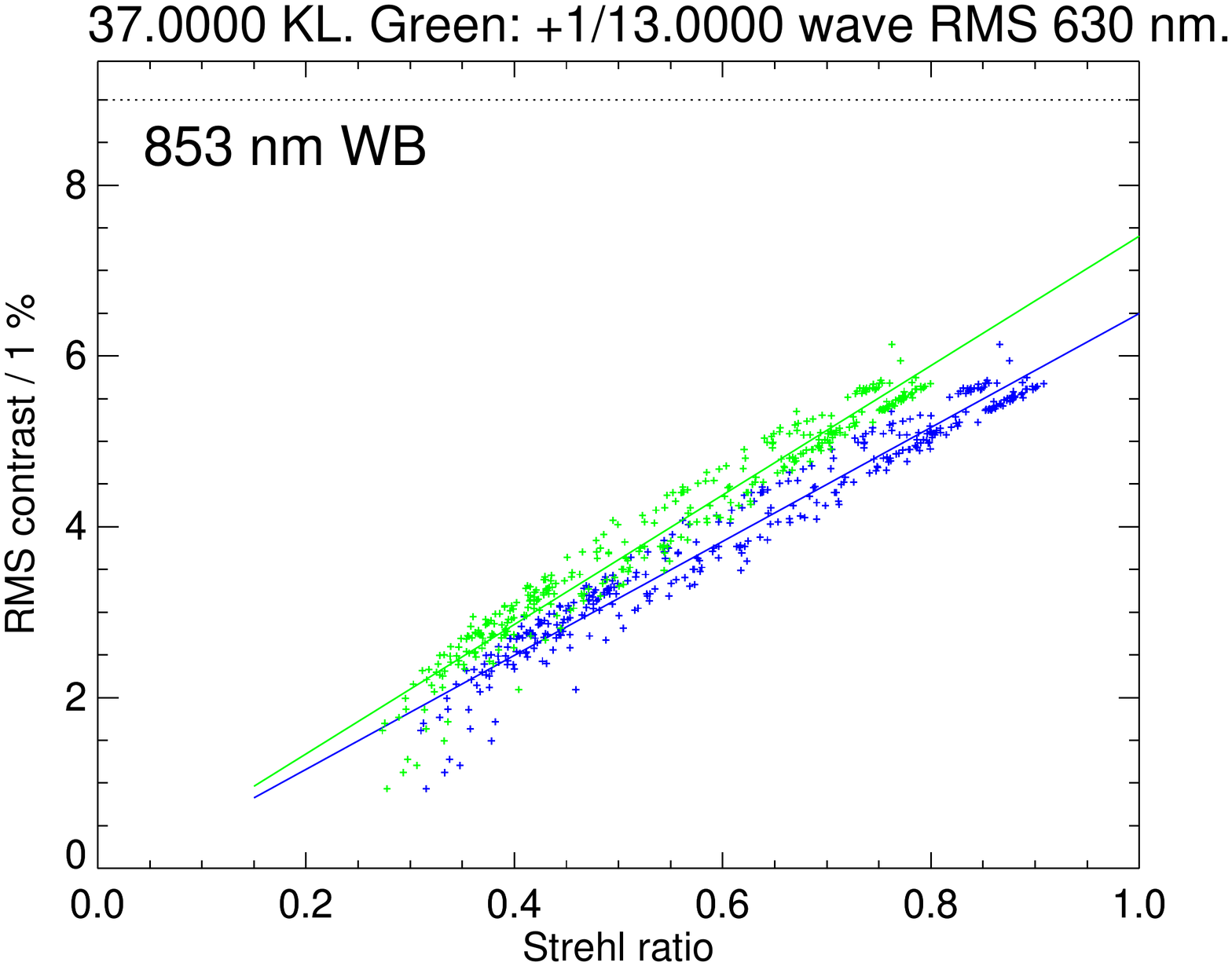}\quad
\includegraphics[viewport=73 46 690 530,  width=0.32\textwidth,clip]{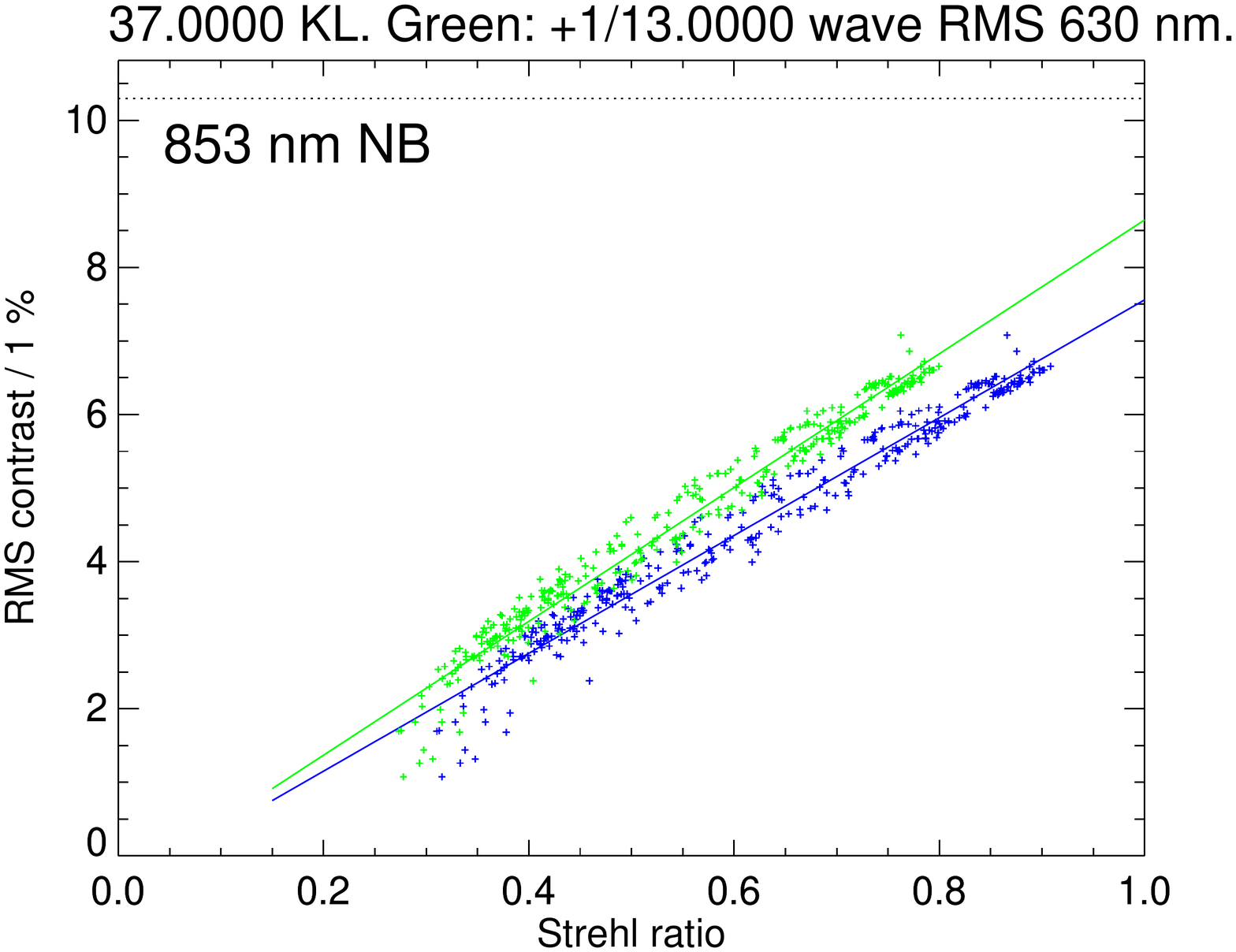}
 \caption{The same as for Fig.~\ref{fig:CRISP3} but assuming that the AO system provides perfect correction of 36 modes.
}
\label{fig:CRISP4}
\end{figure*}

The assumption of a nearly perfect AO system with $N=81$ is of course unrealistic. According to the compilation of \citet{1998PASP..110..837R}, perhaps the best we may hope for is an efficiency of around 50\%, corresponding to $N=42$. In Fig.~\ref{fig:CRISP4}, we show the relations between the Strehl ratios and the granulation contrast for $N=36$, also with straight lines fitted to all data points with Strehl values above 0.15, and with the same added static wavefront variance as in Fig.~\ref{fig:CRISP3}. Again, we find nearly linear relations for $S>0.15$ and that a static wavefront error corresponding to about $1/13$~wave RMS can explain why the ``seeing-free'' data do not seem to reach the expected granulation contrasts. We investigate these empirical relations with theoretical simulations in the following Section.

We should mention that establishing a relation between the RMS contrast and Strehl similar to that described above is not possible for the images processed and restored to compensate for the most significant 36 or 100 KL modes. This is because the restored images are combined in a way that scales the contributions from different images with the squares of their calculated transfer functions. This weighting favours contributions from individual images recorded during moments of excellent seeing but also excludes the use of the simple analysis used with the tip-tilt only compensated images.

\subsection{Simulated AO compensated point spread functions}

We simulated atmospheric turbulence with 10 random wavefronts by scaling randomized coefficients for the 1004 first Karhunen--Lo\`eve (KL) modes to the values given by Kolmogorov statistics. We assumed 100\% AO correction of low-order modes up to KL modes $N\in\{2, 18, 36, 81\}$, respectively (including piston, so removing 3 KL modes in practice corresponds to removing the tip and tilt dominated KL modes).

The remaining random wavefronts were then scaled to different Strehl ratios $S_N \in \{0.01,0.02,\ldots,1.00\}$. With these wavefronts, we calculated optical transfer functions (OTFs) and formed averages over the different atmospheric realizations. We generated PSFs based on the above averages and calculated the encircled energy for a grid of radii with a step size of 0\farcs033. The radii corresponding to 80\% encircled energy were then found by linear interpolation. 

\begin{figure*}
\center
\includegraphics[viewport=53 24 708 555, angle=0, width=0.33\textwidth,clip]{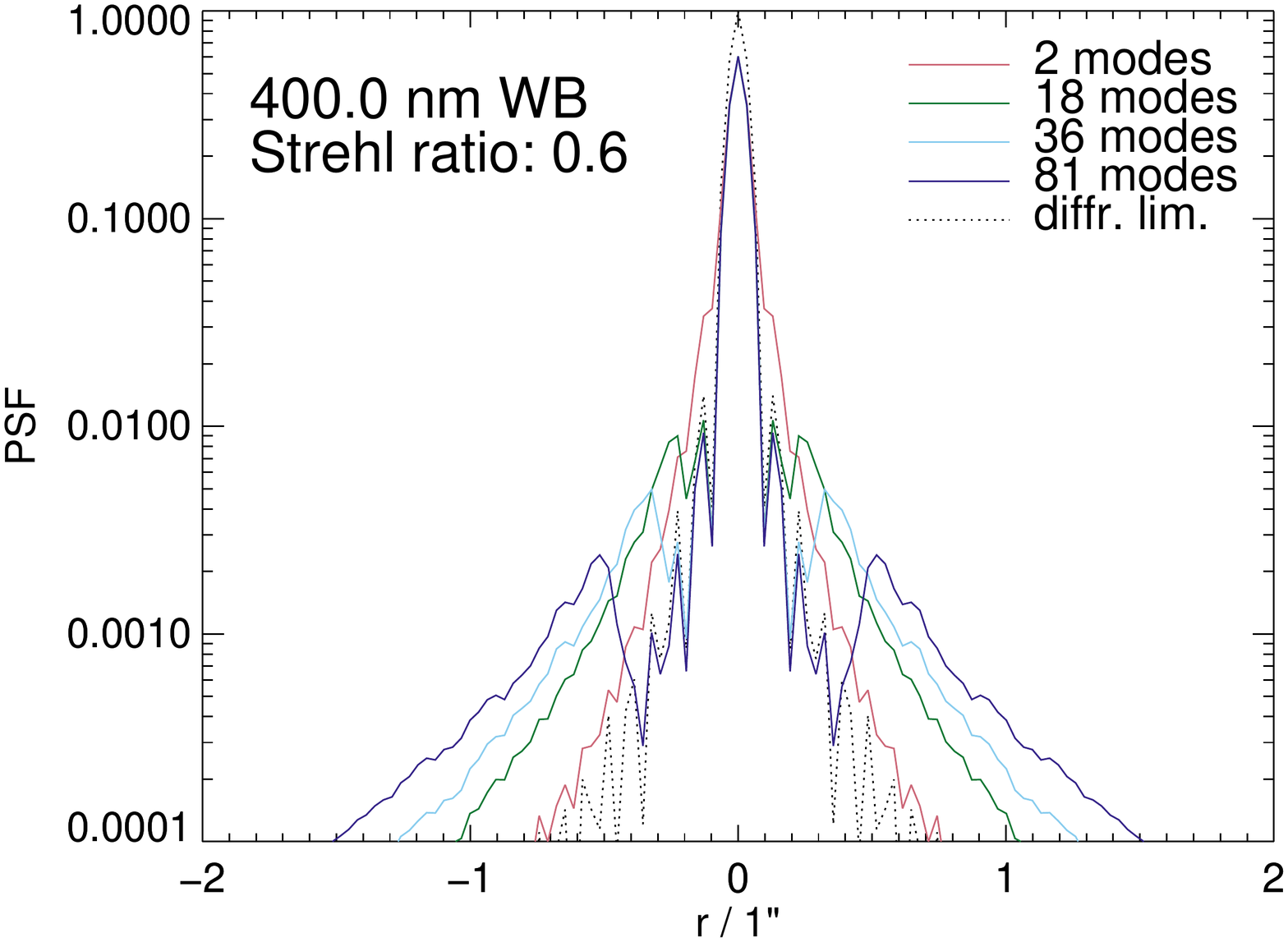}
\includegraphics[viewport=53 24 708 555, angle=0, width=0.33\textwidth,clip]{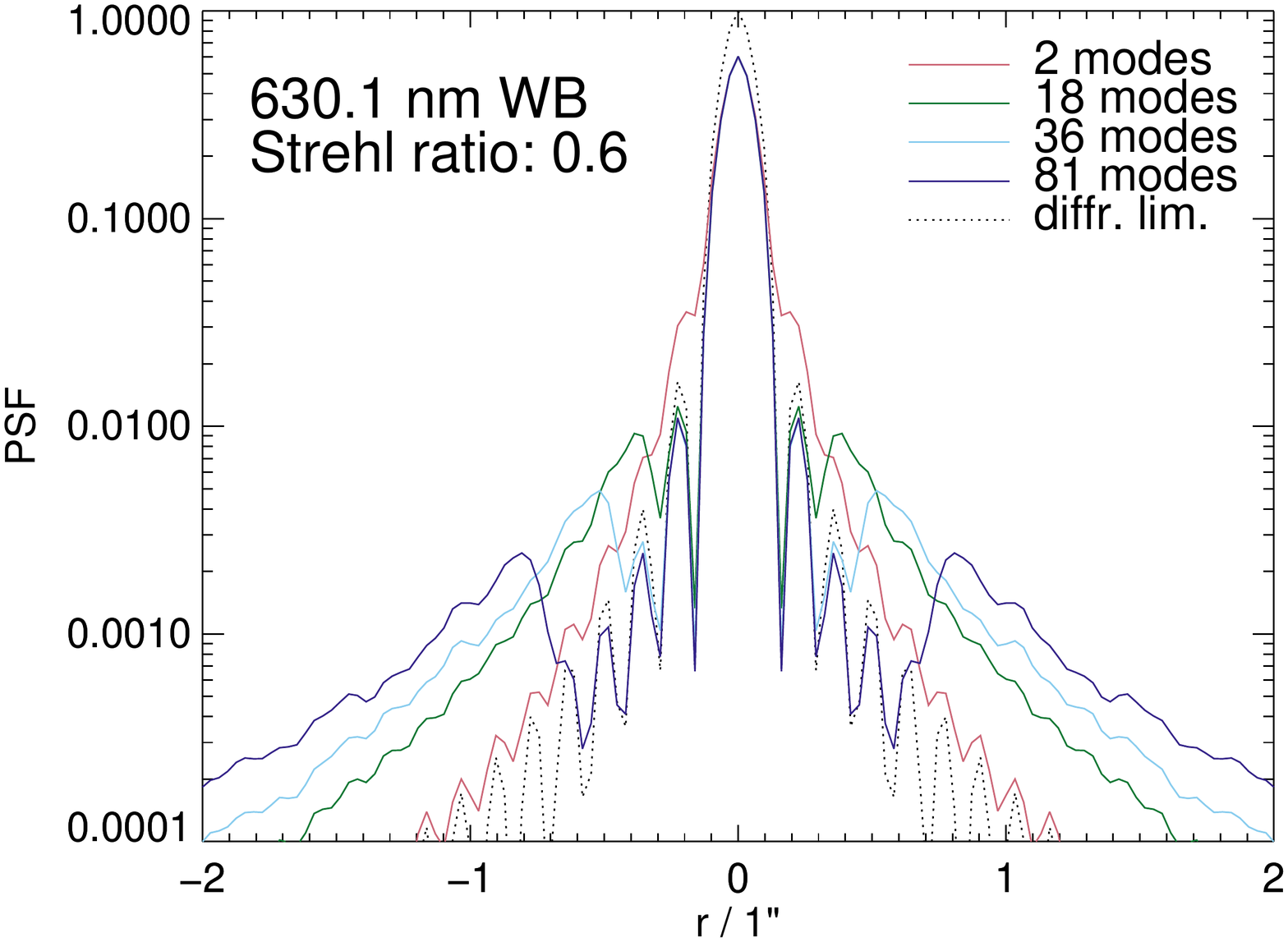}
\includegraphics[viewport=53 24 708 555, angle=0, width=0.33\textwidth,clip]{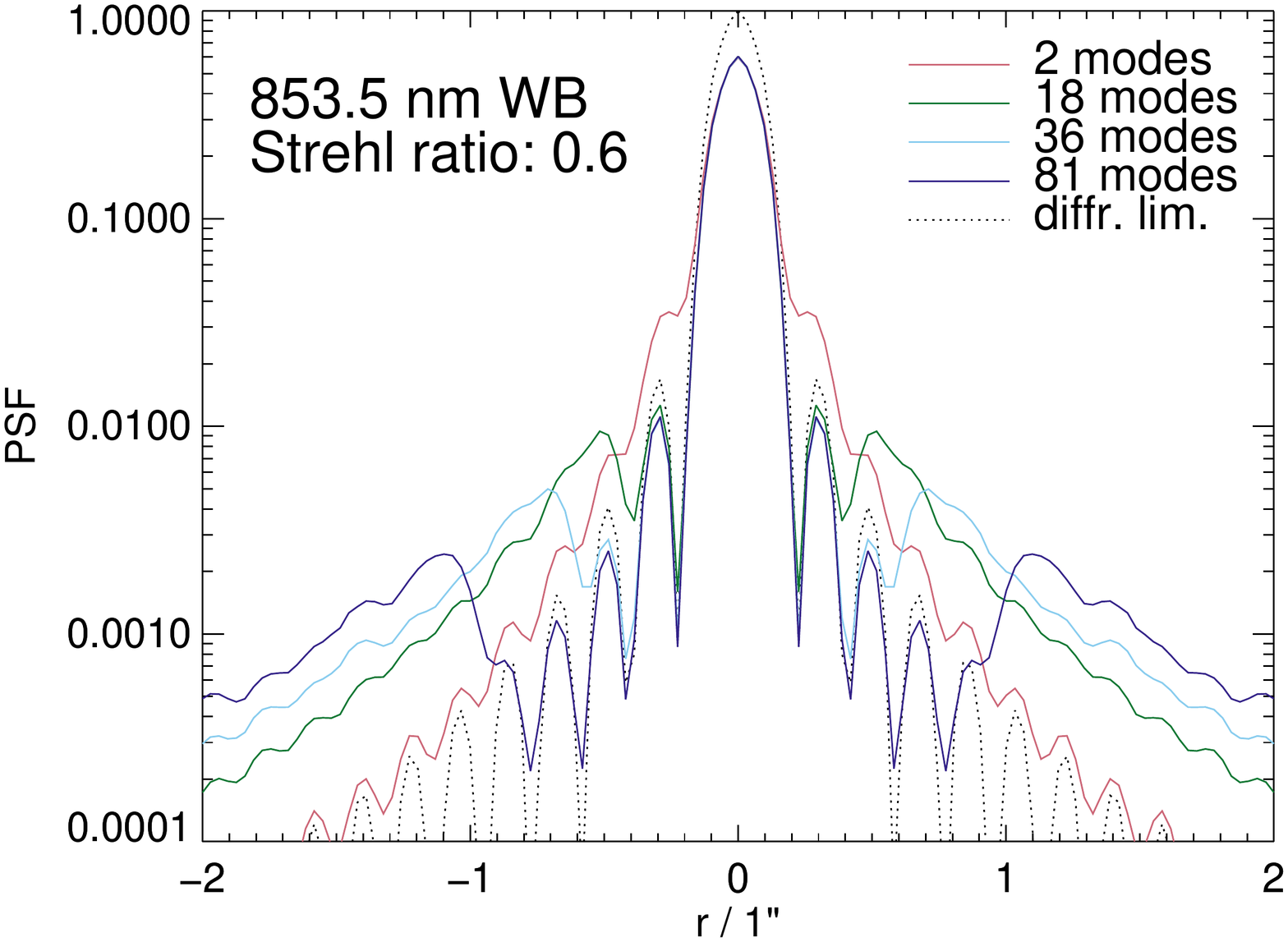}
\includegraphics[viewport=53 24 708 555, angle=0, width=0.33\textwidth,clip]{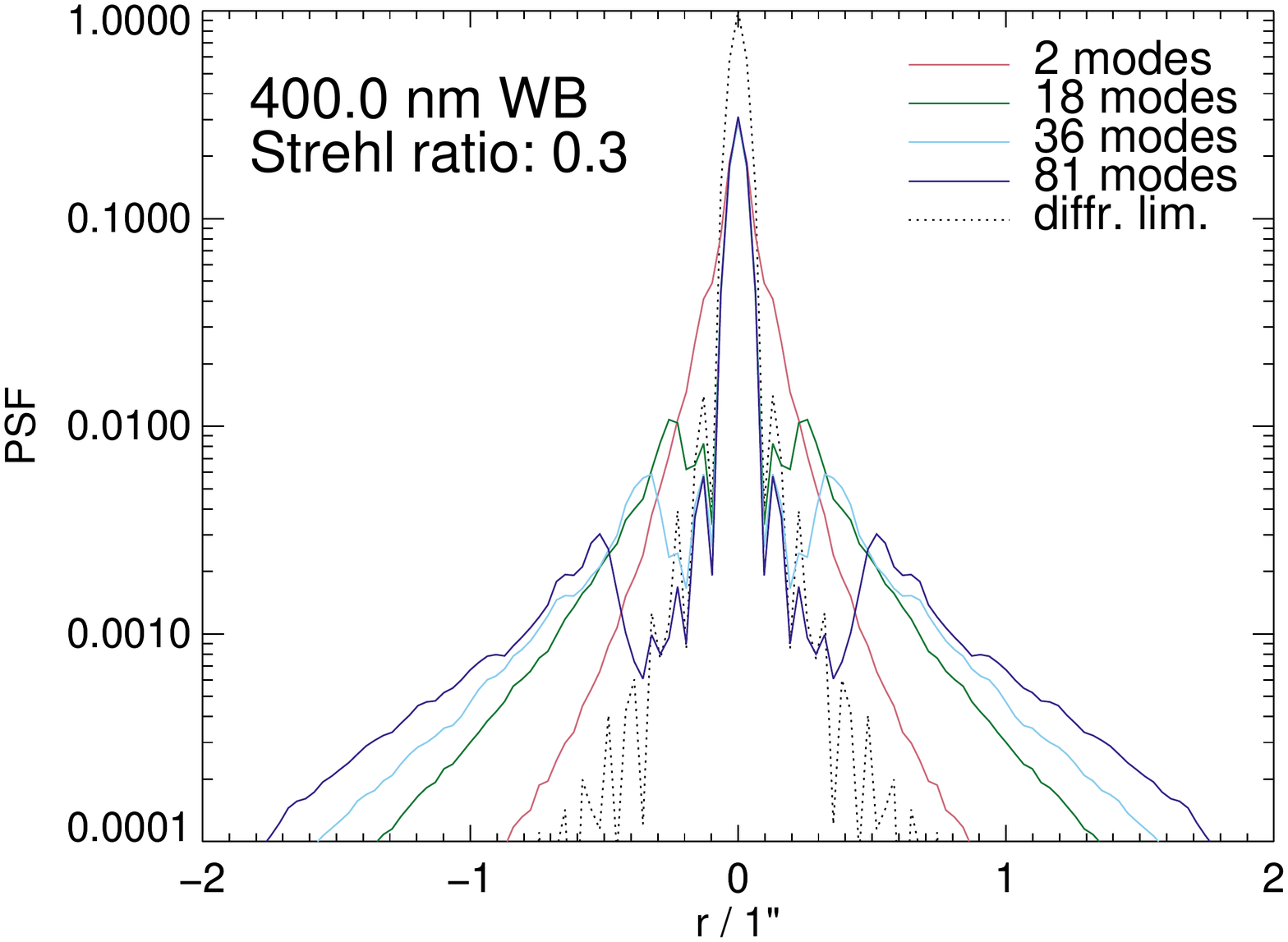}
\includegraphics[viewport=53 24 708 555, angle=0, width=0.33\textwidth,clip]{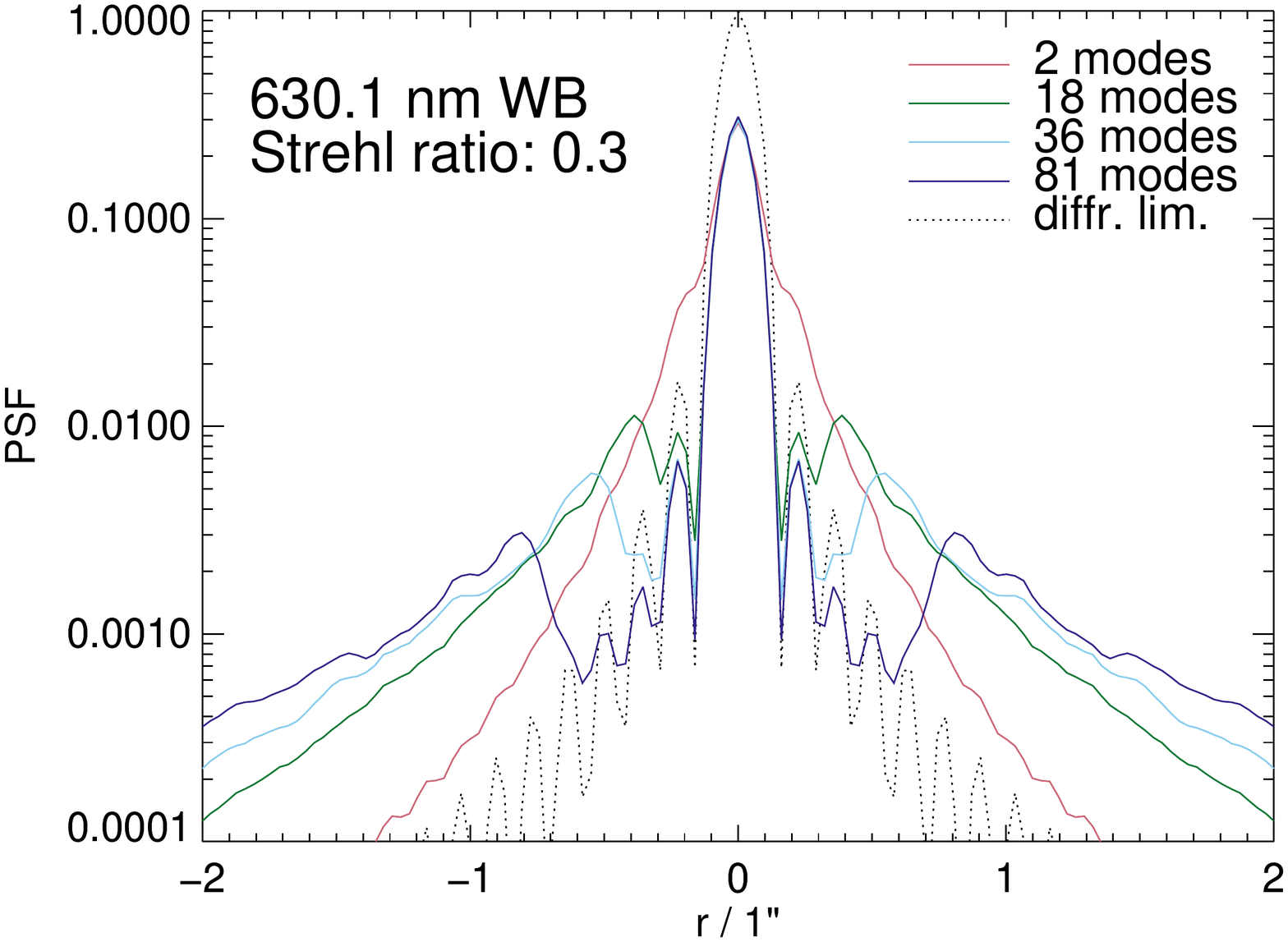}
\includegraphics[viewport=53 24 708 555, angle=0, width=0.33\textwidth,clip]{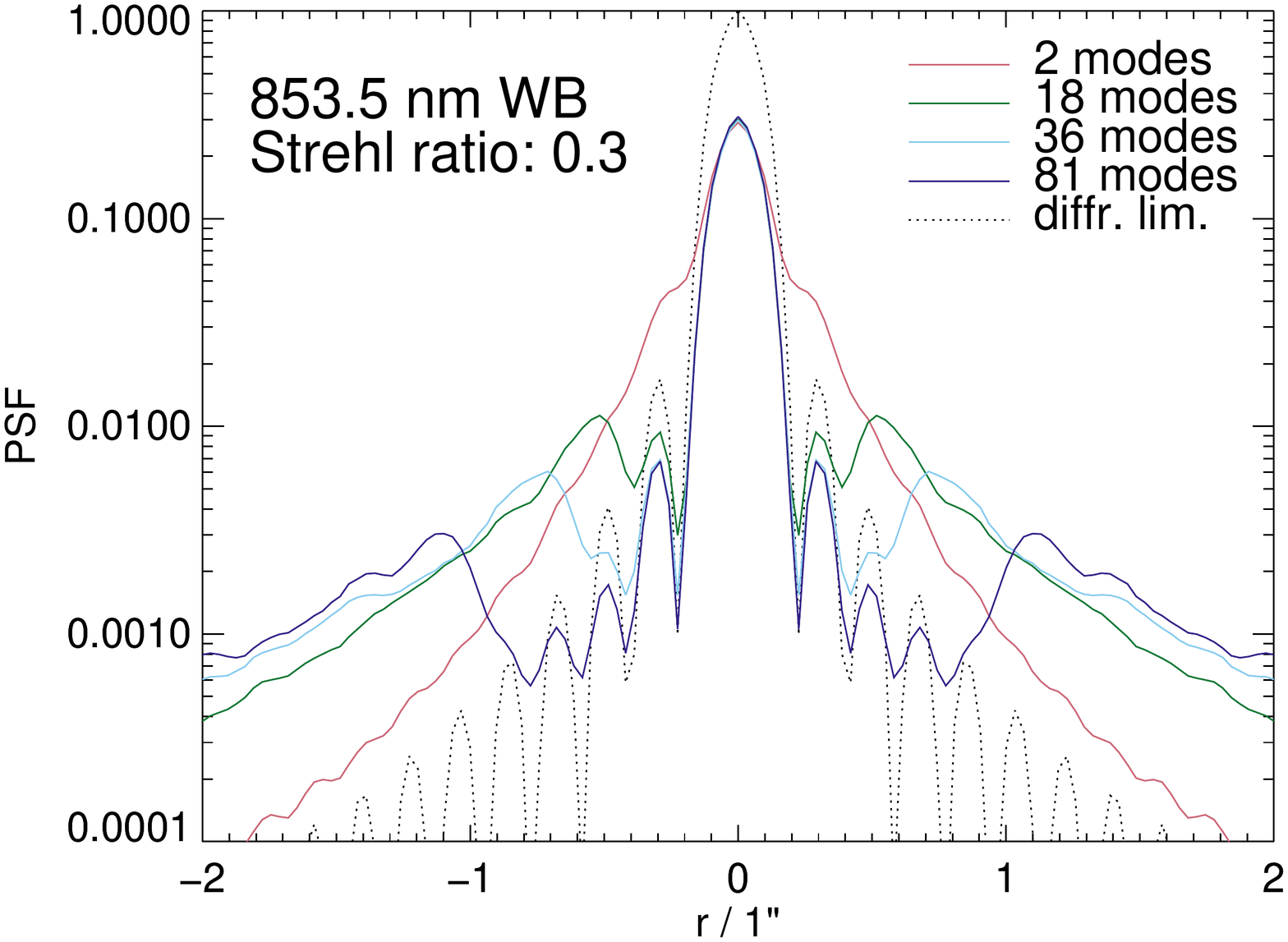}
\includegraphics[viewport=53 24 708 555, angle=0, width=0.33\textwidth,clip]{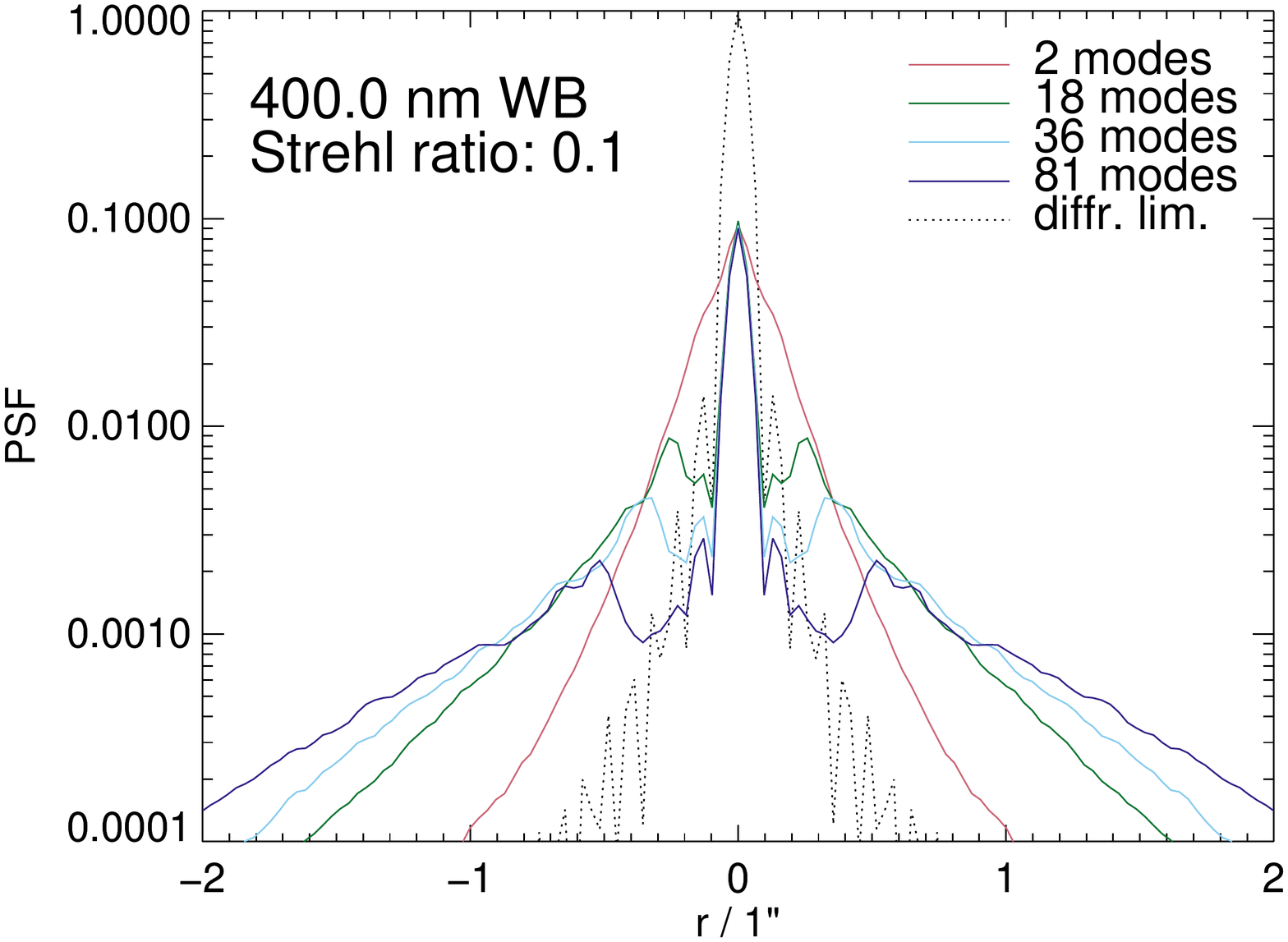}
\includegraphics[viewport=53 24 708 555, angle=0, width=0.33\textwidth,clip]{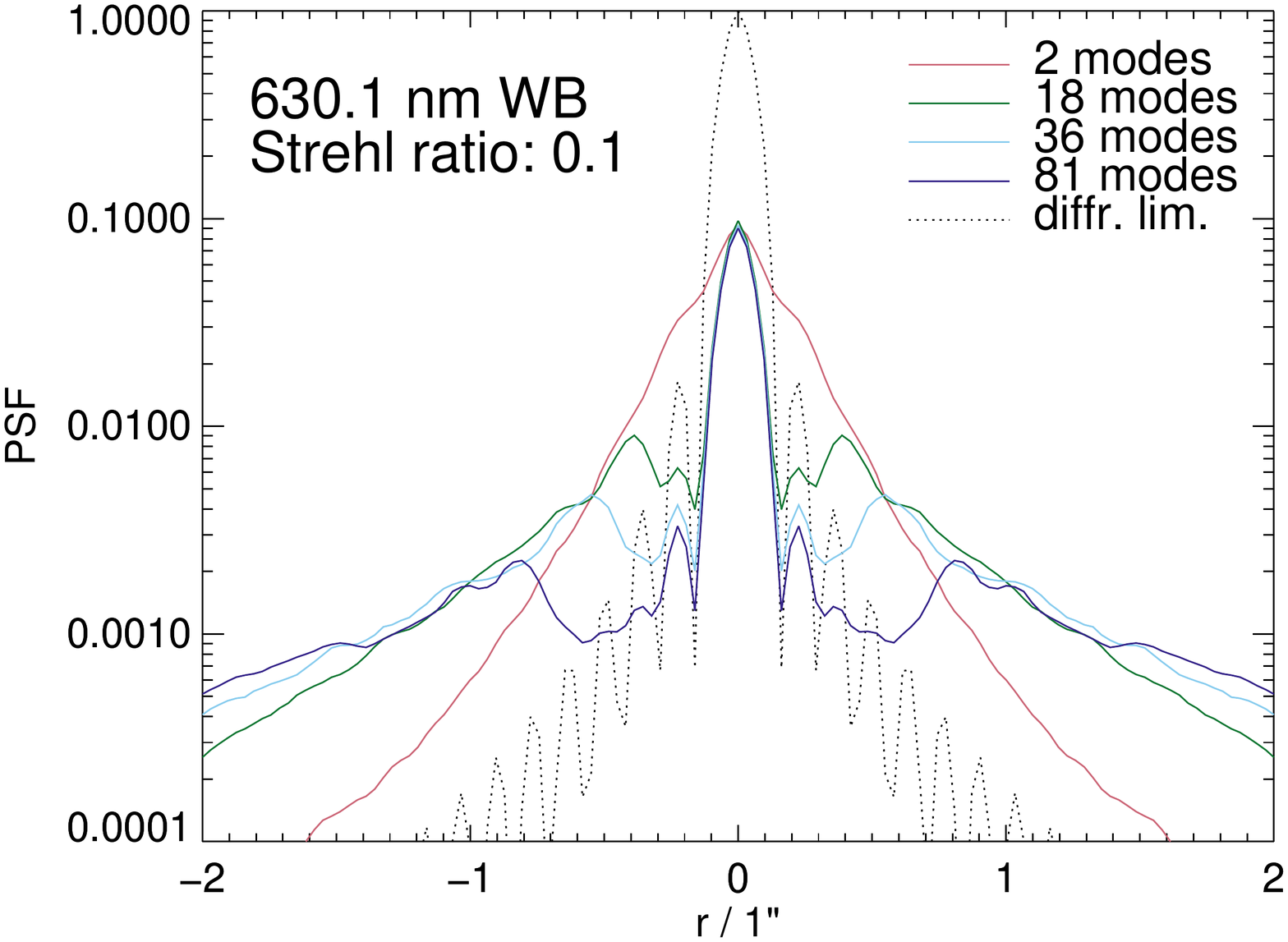}
\includegraphics[viewport=53 24 708 555, angle=0, width=0.33\textwidth,clip]{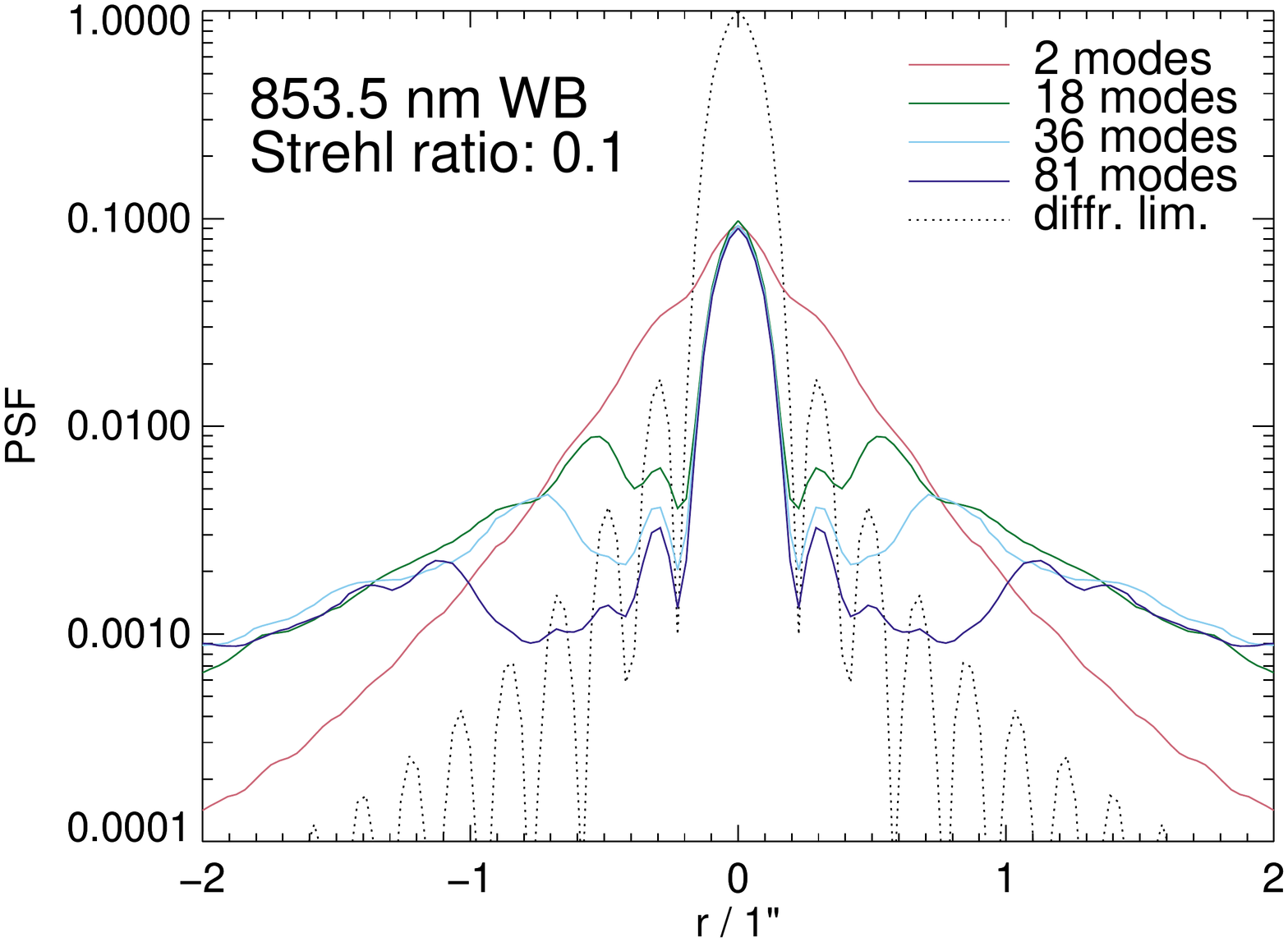}
\includegraphics[viewport=53 24 708 555, angle=0, width=0.33\textwidth,clip]{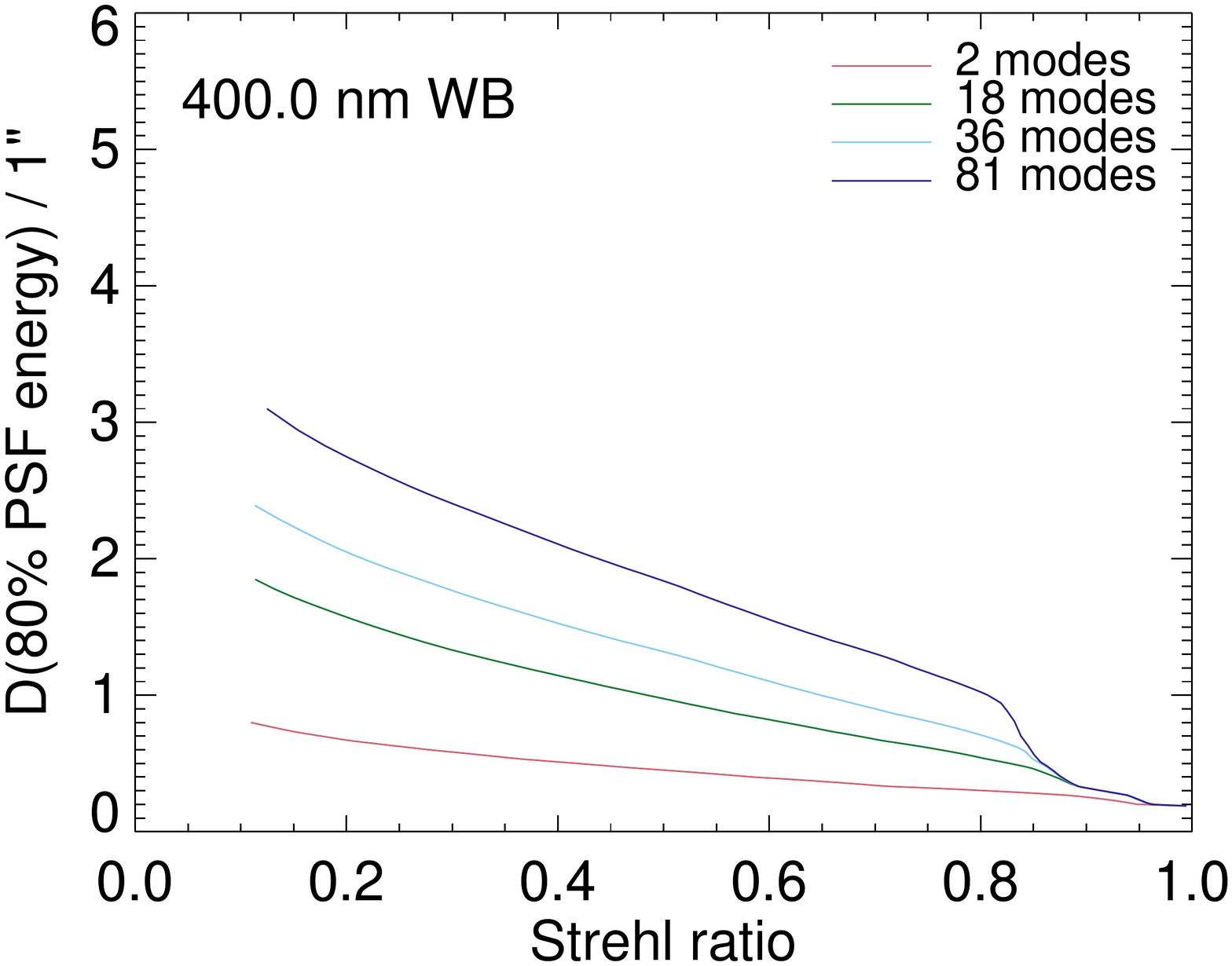}
\includegraphics[viewport=53 24 708 555, angle=0, width=0.33\textwidth,clip]{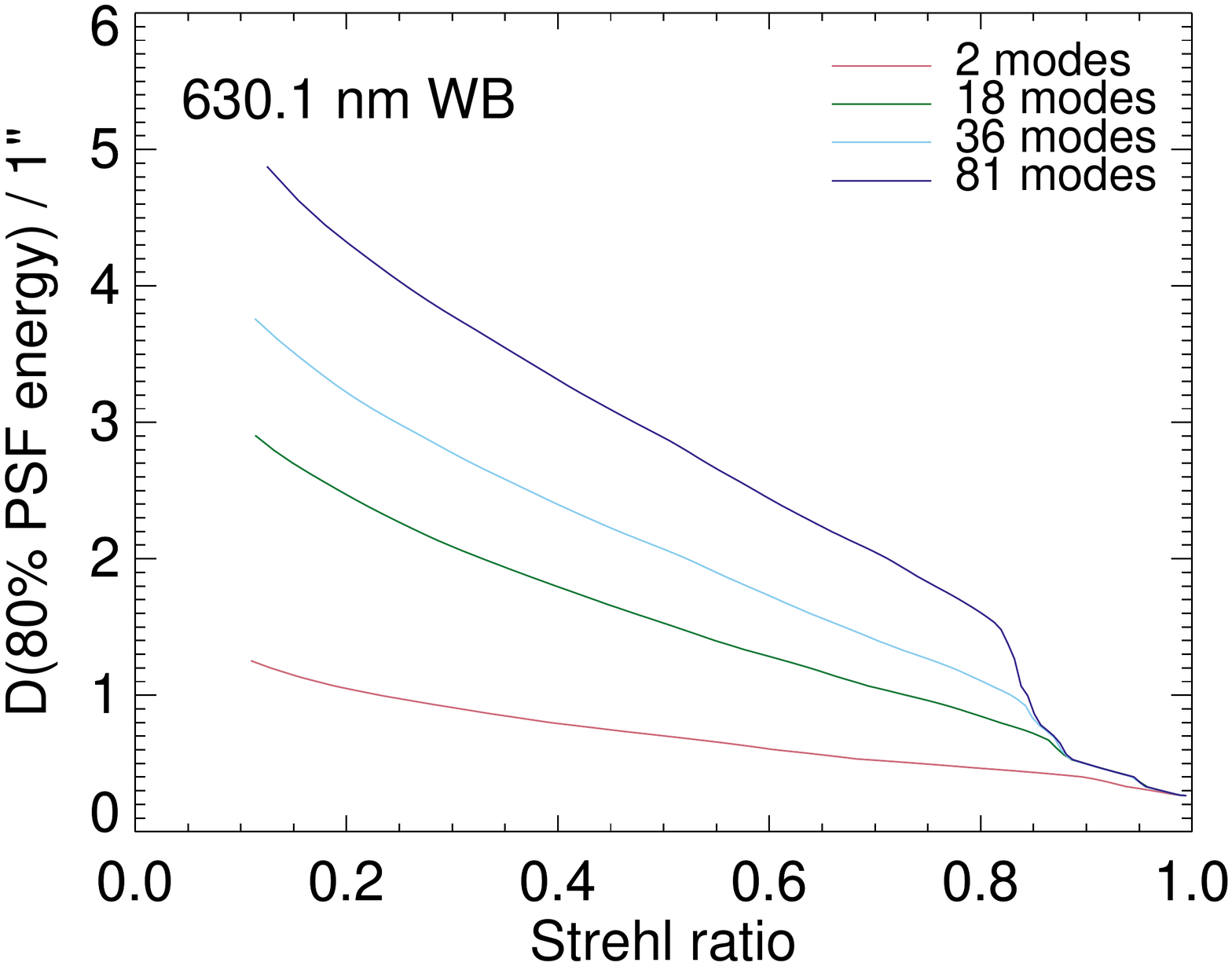}
\includegraphics[viewport=53 24 708 555, angle=0, width=0.33\textwidth,clip]{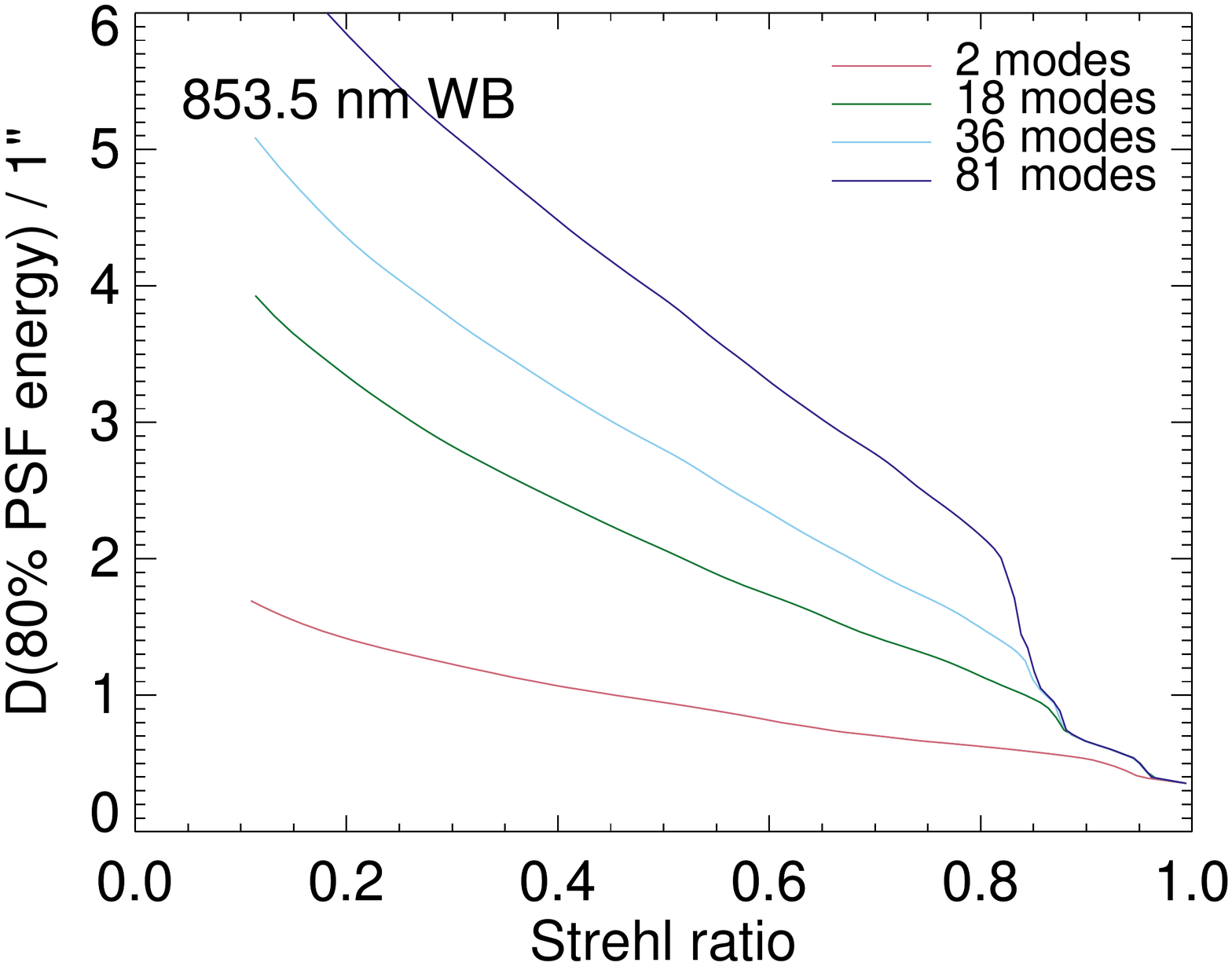}
 \caption{
The top 9 panels show the variation of the shape of the point spread function (PSF) with the Strehl ratio after perfect compensation of the first 2 (tip-tilt only), 18, 36 and 81 aberration modes. The lowermost 3 panels show the  variation with Strehl of the diameter  of the PSF containing 80\% of the encircled energy for the same number of compensated modes. 
}
\label{fig:PSF}
\end{figure*}

In Fig.~\ref{fig:PSF} is shown in the top nine panels the PSFs at 400, 630 and 854~nm from residual seeing corresponding to Strehl ratios of 0.6, 0.3 and 0.1 after removing the first 2 (tip-tilt only), 18, 36 and 81 KL modes. With only the tip-tilt modes removed, a strongly reduced Strehl is associated primarily with a broadening of the core of the PSF. For higher-order compensation, the PSF core is mostly unaffected but of reduced strength by a factor $S$ and with a characteristic rapid transition from the diffraction limited core to the surrounding ``halo''. In the lowermost three panels in the same Figure is shown the diameter of the PSF corresponding to 80\% encircled energy. This diameter is less than 1\arcsec{} at high Strehl and 400~nm wavelength, increasing to 4\arcsec{} at 50\% Strehl and 854~nm. We note that the radius of the peak of the halo is set by the number of compensated KL modes and is essentially independent of the Strehl. When the Strehl changes, the structure of the halo remains more or less the same but its strength varies as $1-S$. 

\subsubsection{Simulated granulation data and contrast calculations} \label{MHD_simulations}
To calibrate the observed contrasts through the different CRISP and CHROMIS filters, we calculated synthetic spectra using a snapshot from a 3D radiation hydrodynamic simulation of solar granulation \citep{1998ApJ...499..914S}. The simulation covers a horizontal physical domain of 6$\times$6~Mm. In the vertical direction it extends from $-453$~km to 575~km above the continuum formation layer in the photosphere. The snapshot is given in a grid with $254\times254$ horizontal points and 82 vertical points. The vertical sampling is 12.7 km/grid cell and the horizontal resolution is 23.7 km/grid cell. The same snapshot was used by \citet{2010A&A...521A..68S}. The spectra were calculated with a LTE synthesis module included in the STiC code \citep{2016ApJ...830L..30D, 2019A&A...623A..74D}. The ionization balance and chemical equilibrium were solved using an LTE equation of state, including molecules and accurate partition functions \citep{2017A&A...597A..16P}.
All line data were extracted from the VALD-3 database \citep{2015PhyS...90e4005R,1995A&AS..112..525P}. The van der Waals damping parameter was calculated using cross-sections provided by \citet{1998MNRAS.300..863B,2000A&AS..142..467B}, for all lines for which the collisional cross-sections were available. The formal solution of the transfer equation was computed using a third order Bezier interpolant \citep{2003ASPC..288....3A, 2013ApJ...764...33D}.

The so-obtained simulated spectra were multiplied with the transmission profiles of the CRISP and CHROMIS pre-filters provided  by the manufacturers, and summarised in Table~\ref{obs_data}. These spectra were then integrated over wavelength to provide the WB contrasts referred to in Table~\ref{obs_data}.

To reduce wrap-around effects from the extended wings of the PSF, we repeated the periodic FOV in order to perform the required convolutions on $512\times512$-pixel images. These synthetic images were then convolved with the PSFs described above and deconvolved with the diffraction limited PSF of SST. So, these data correspond to observed data that have been corrected for the theoretical PSF of the telescope or processed with higher order modes using MFBD. The RMS contrasts were calculated over the original FOV, avoiding the apodized edges.

\subsubsection{Results}

\begin{figure*}
\center
\includegraphics[viewport=53 24 708 555, angle=0, width=0.33\textwidth,clip]{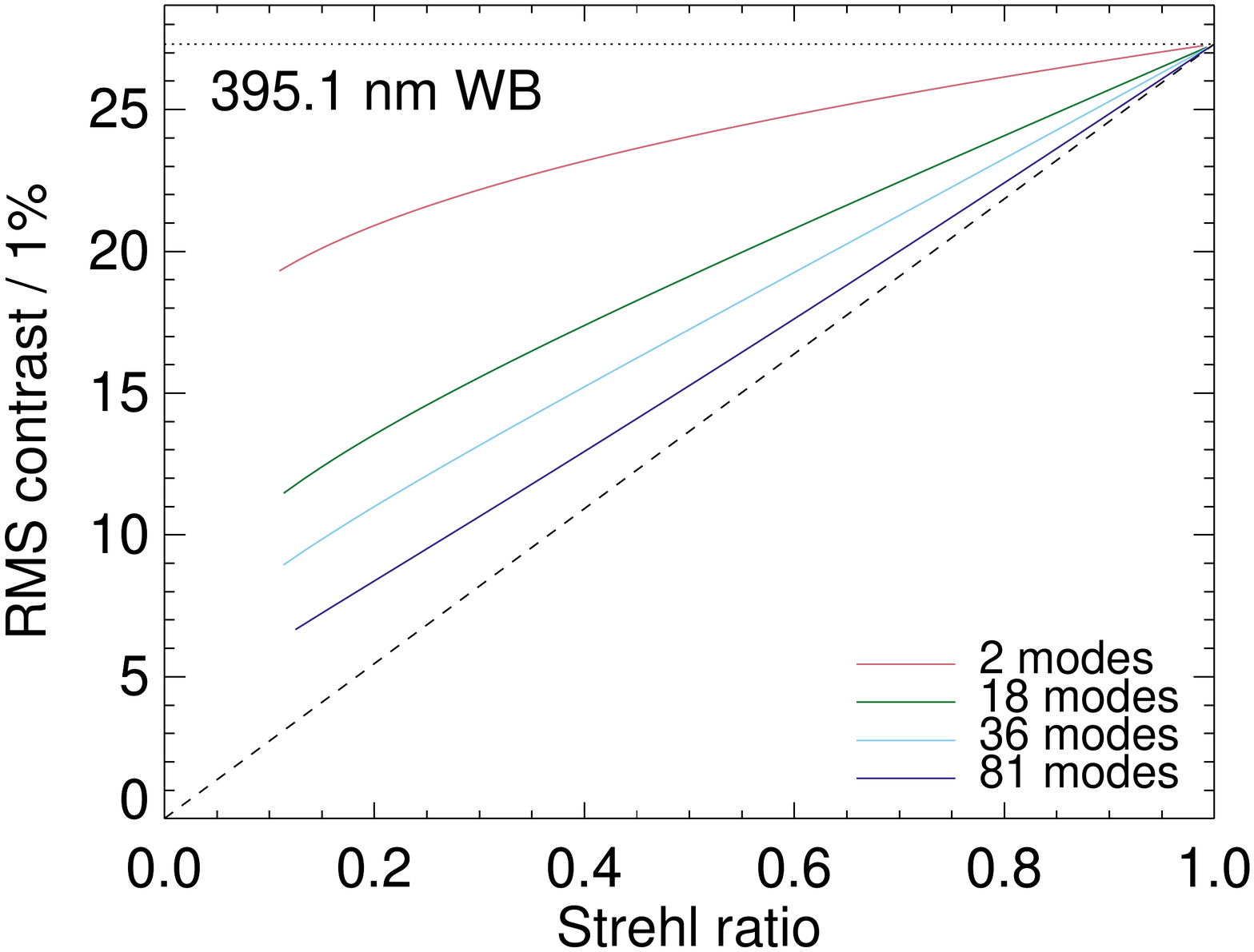}
\includegraphics[viewport=53 24 708 555, angle=0, width=0.33\textwidth,clip]{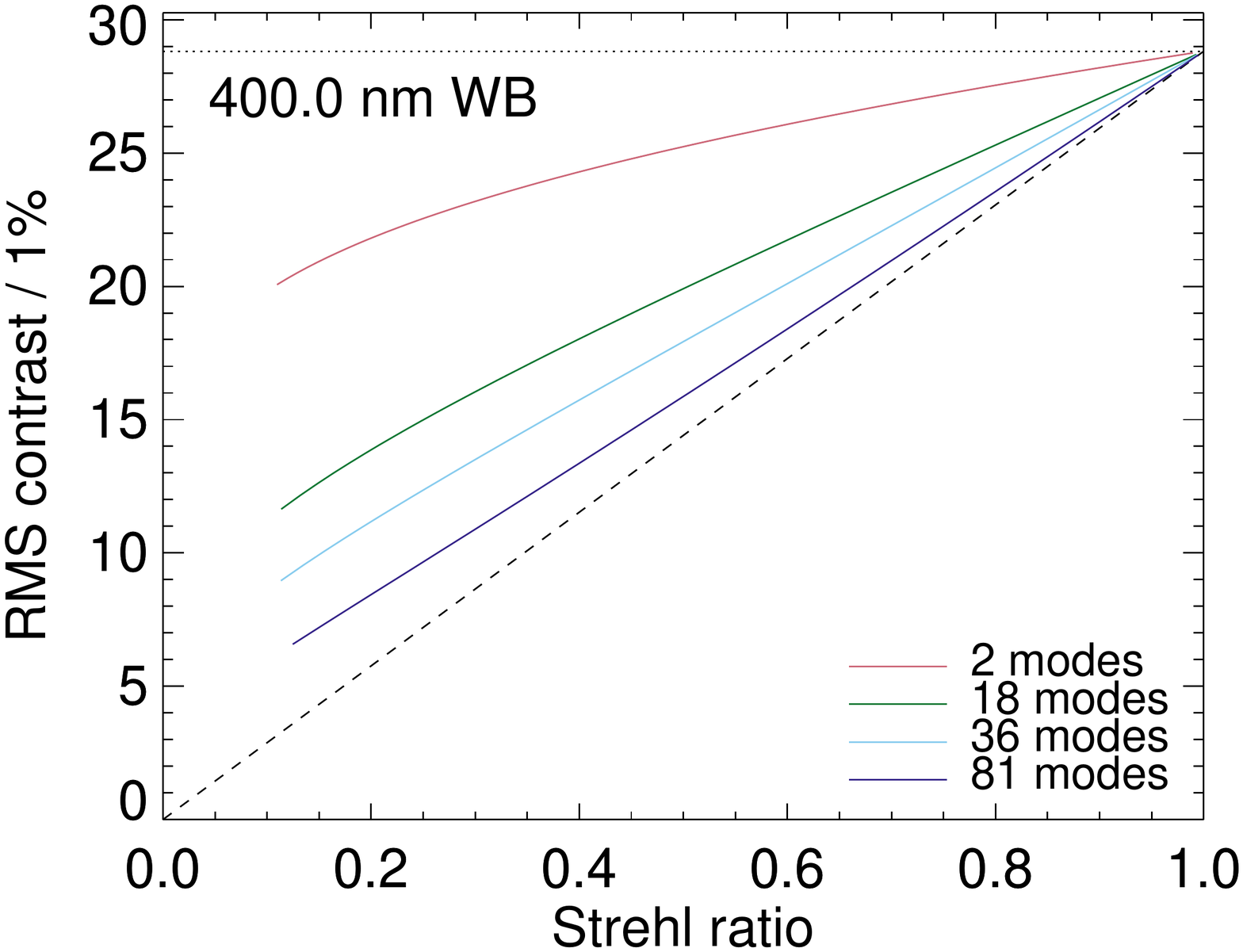}
\includegraphics[viewport=53 24 708 555, angle=0, width=0.33\textwidth,clip]{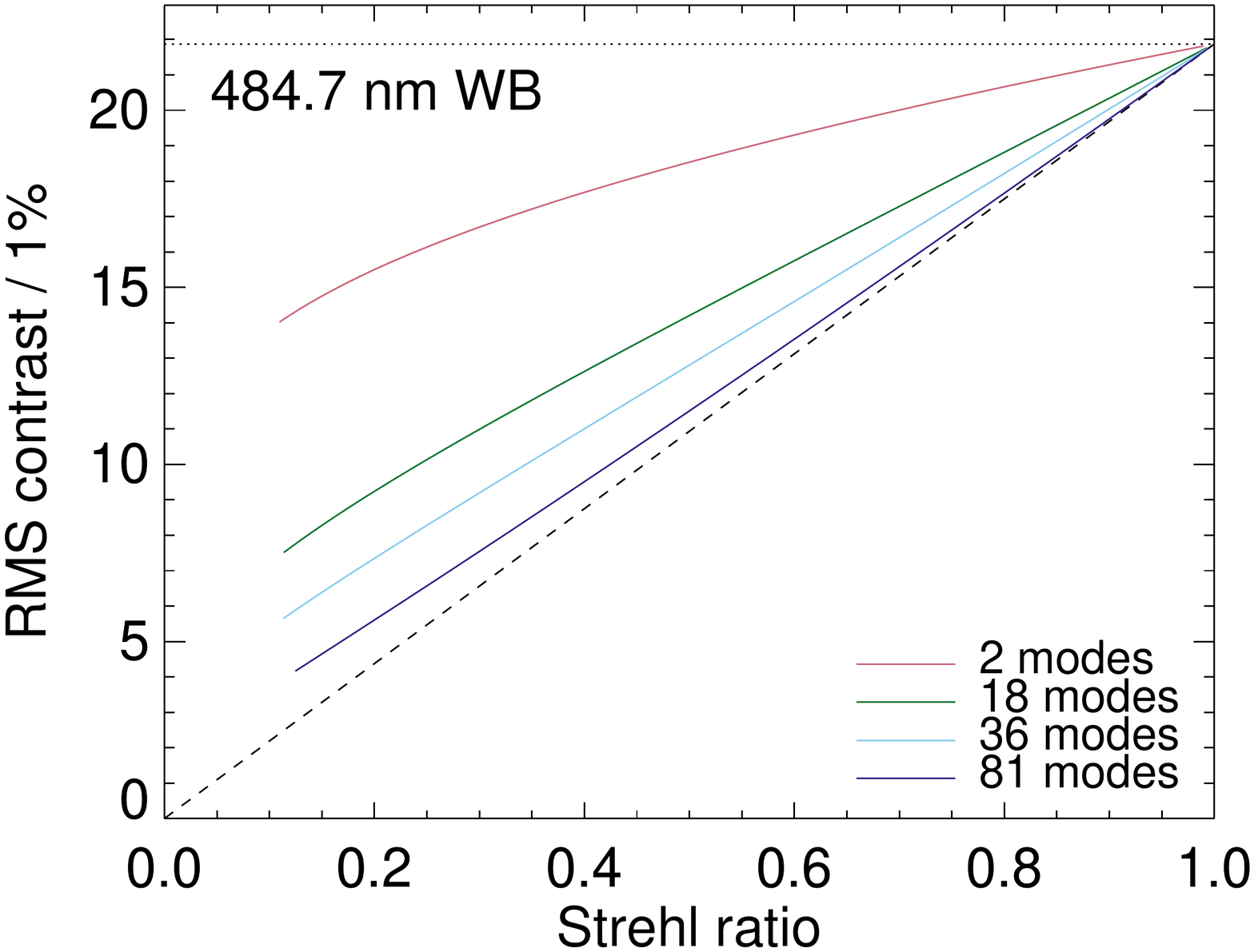}
\includegraphics[viewport=53 24 708 555, angle=0, width=0.33\textwidth,clip]{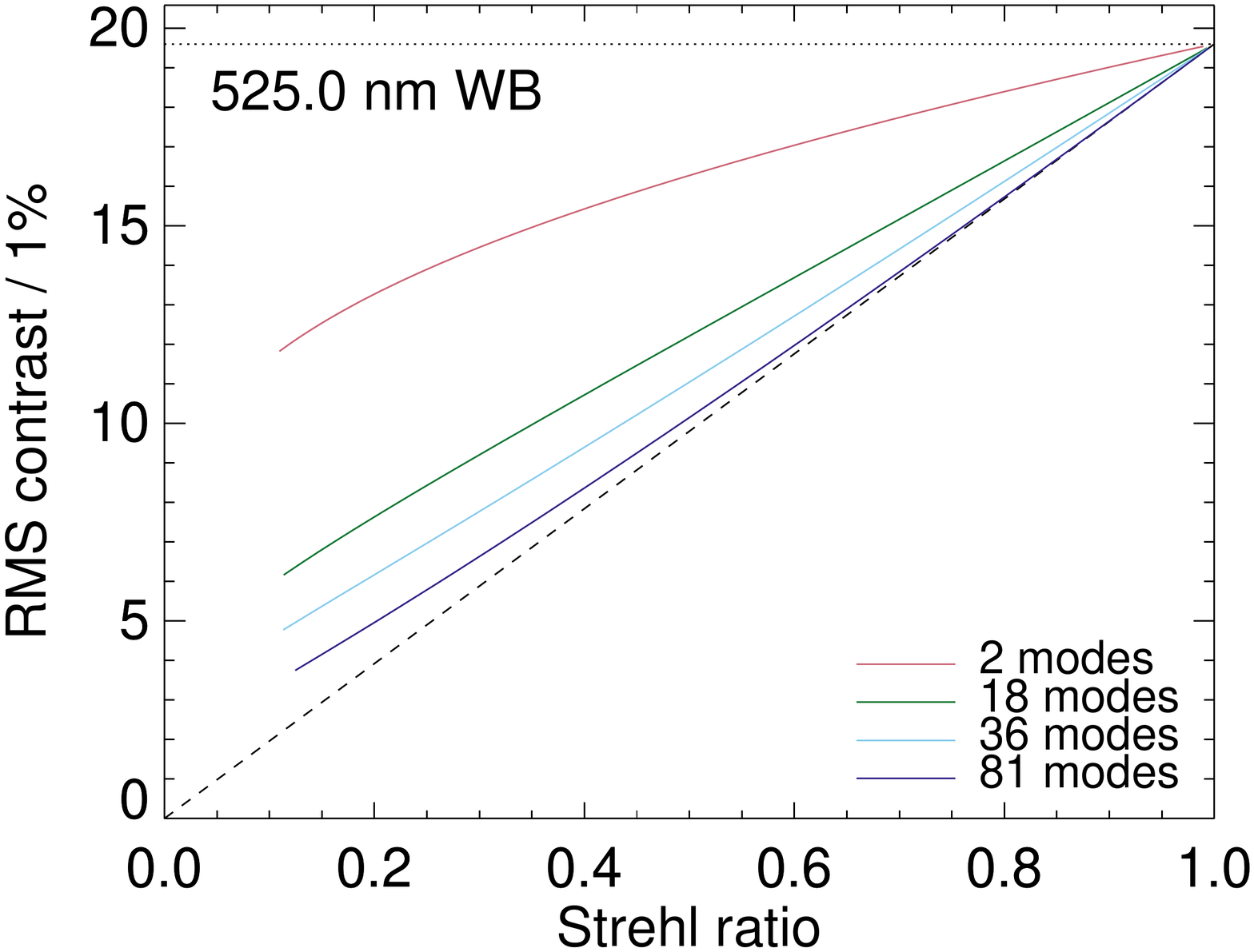}
\includegraphics[viewport=53 24 708 555, angle=0, width=0.33\textwidth,clip]{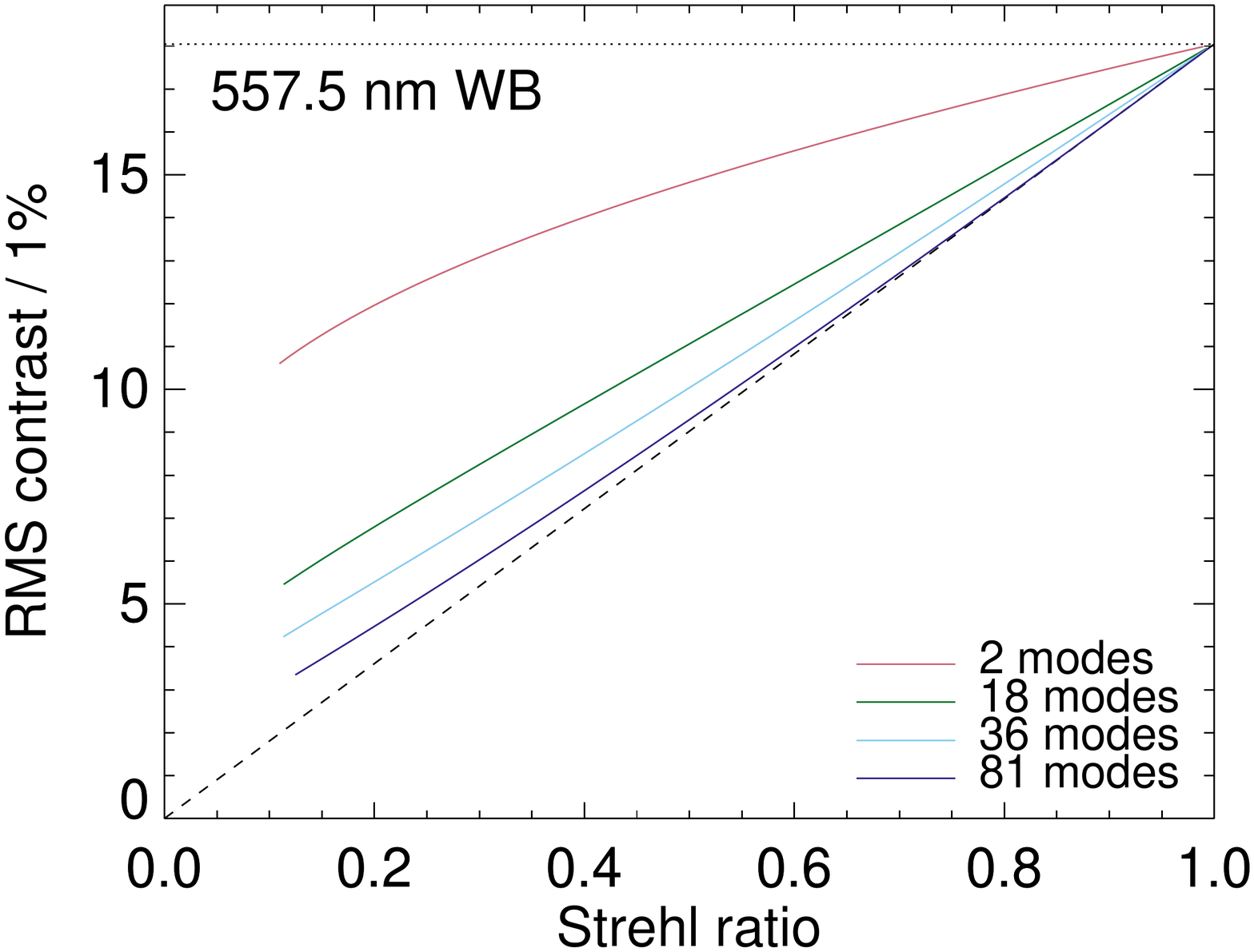}
\includegraphics[viewport=53 24 708 555, angle=0, width=0.33\textwidth,clip]{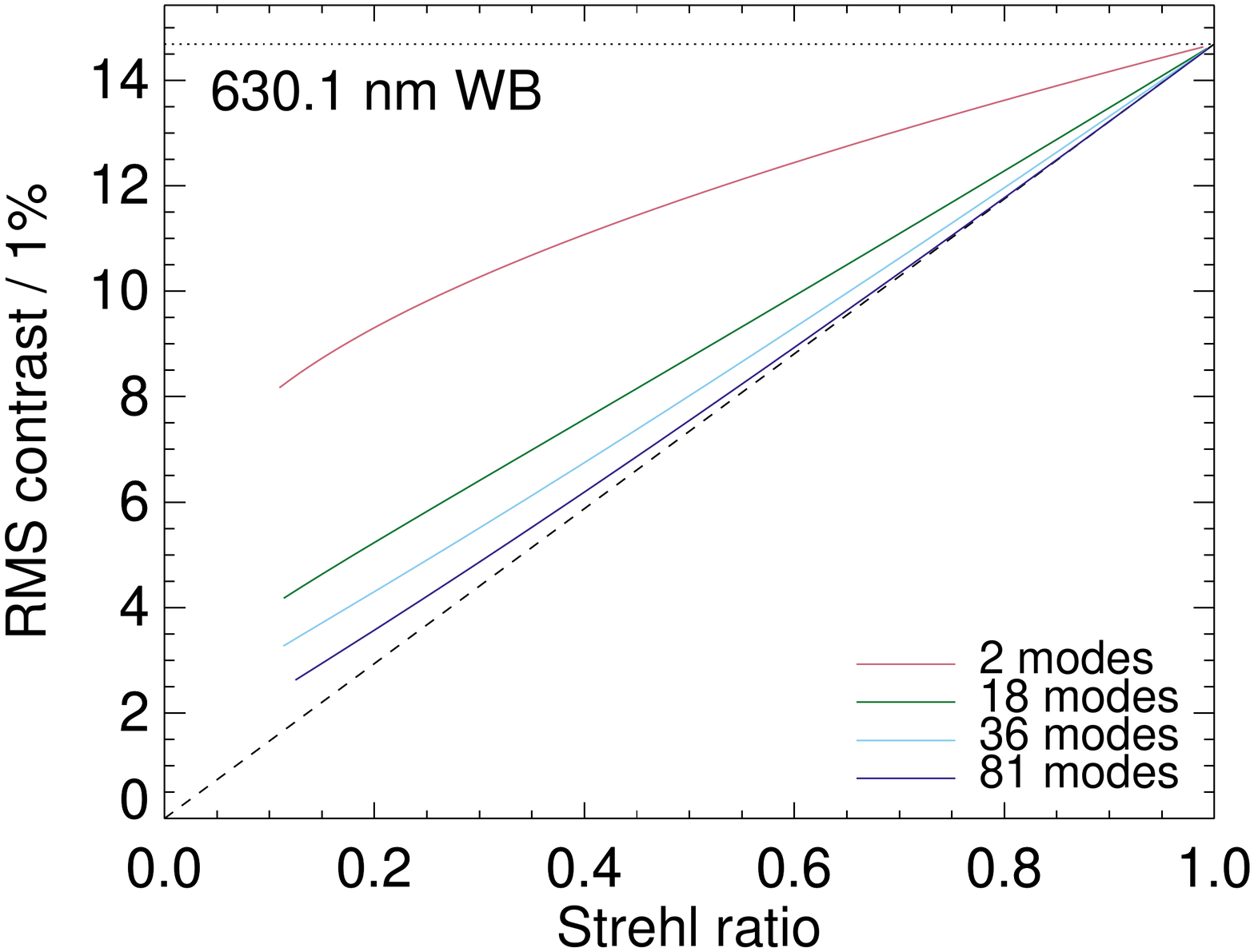}
\includegraphics[viewport=53 24 708 555, angle=0, width=0.33\textwidth,clip]{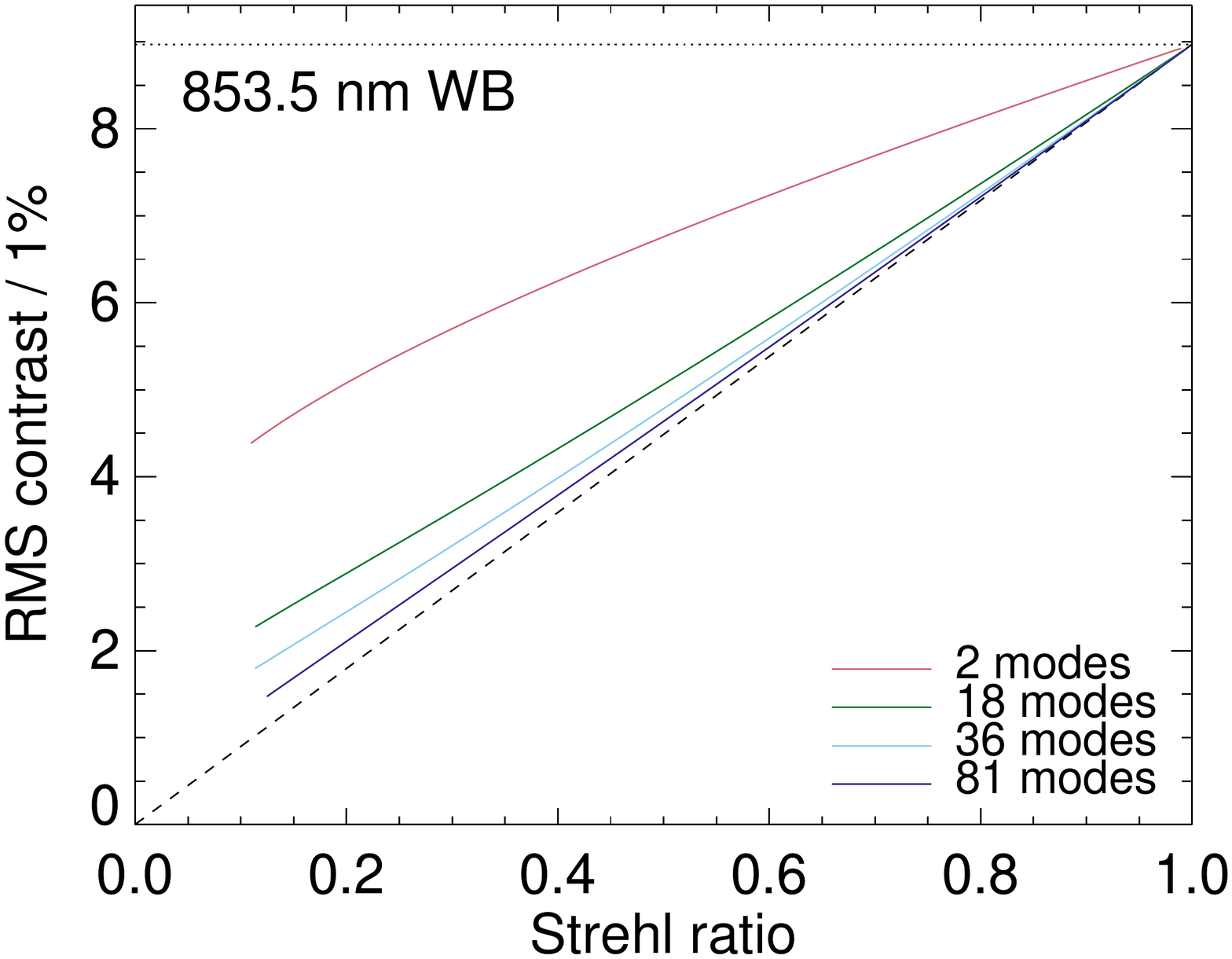}
\includegraphics[viewport=53 24 708 555, angle=0, width=0.33\textwidth,clip]{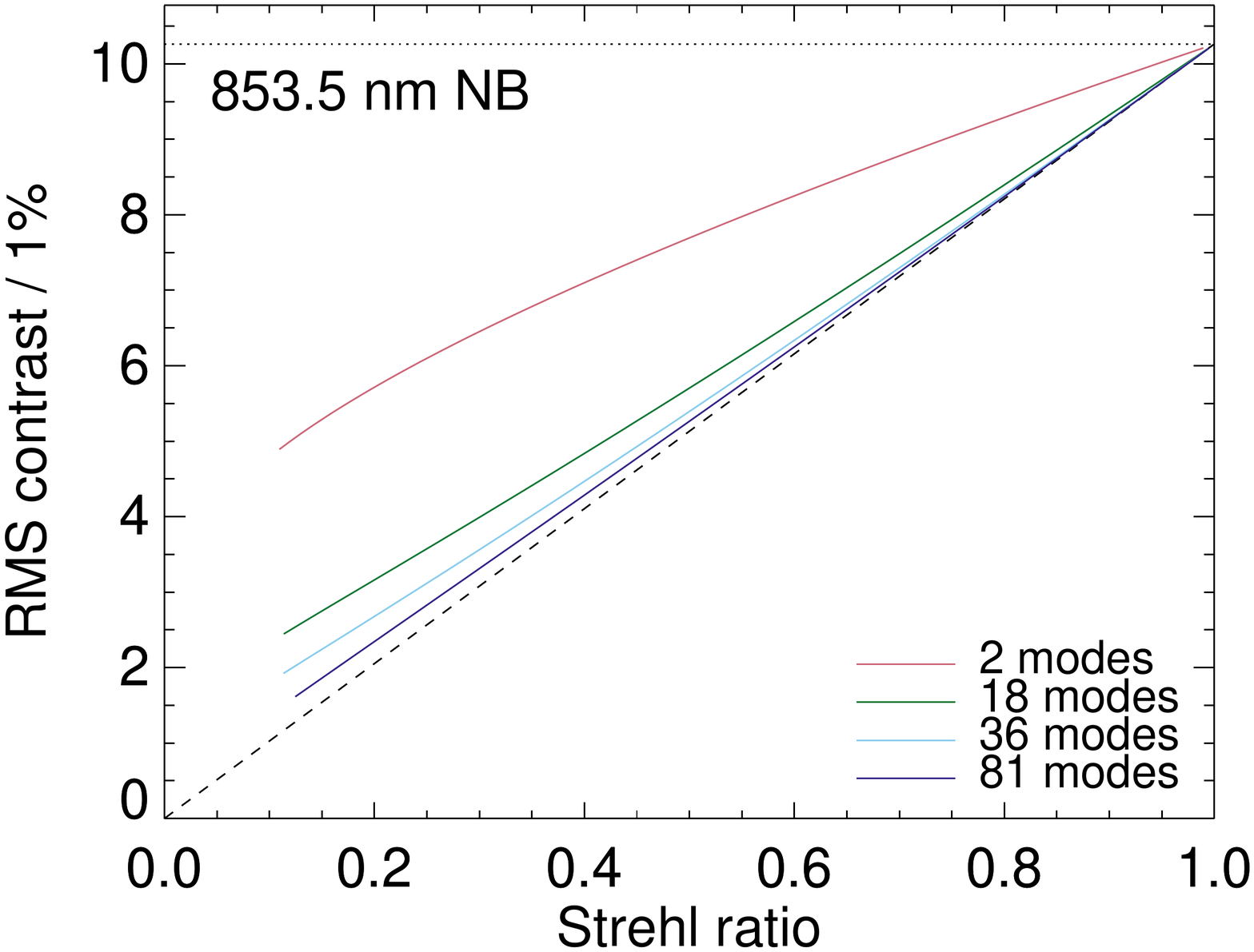}
 \caption{
Theoretically calculated RMS granulation contrast for different Strehl values, based on residual wavefront errors after perfect low-order modal compensation of seeing degraded images recorded with a 1-m telescope.  When approximately 50 modes or more are compensated, the relation between the Strehl ratio and the granulation contrast is almost linear, and independent of the number of modes compensated. This is because the wings of the PSF are then sufficiently wide (see Fig.~\ref{fig:PSF}) that more than one granule is included under the PSF. With such wide wings, making the PSF wider does not increase the amount of straylight. At short wavelengths, a higher number of compensated modes than at long wavelengths is required for this simple relation between Strehl and granulation contrast to be true.
}
\label{fig:c_vs_strehl}
\end{figure*}

In Fig.~\ref{fig:c_vs_strehl} is shown the results of the theoretically calculated granulation contrasts for different Strehl values in the presence of seeing and AO compensation of the first 2, 18, 36 and 81 modes. {\it In the limit of very large number of corrected modes, the RMS contrast is directly proportional to, and thus a direct measure of,  the Strehl value.} For a smaller number of corrected modes in the range of 18 to 81, the relation is almost linear for Strehl values above 0.1, but the {\it theoretical} curves do not extrapolate to zero contrast at zero Strehl. This is because the FWHM remains narrow enough with a relatively small number of corrected modes, such that the granulation pattern is not washed out entirely but instead the granulation contrast reaches a limiting value set by the shape of the halo of the PSF, $P_h$.  This  implies that the extrapolated granulation contrast to a Strehl of zero provides valuable information about $P_h$.

Comparing the simulations in Fig.~\ref{fig:c_vs_strehl} to the empirical results in Fig.~\ref{fig:CRISP3}, it is obvious that the assumption of $N=81$ produces unreasonable results. For the data corresponding to wavelengths at 525~nm and longer, the {\it empirical} curves extrapolate to zero contrast for Strehls in the range 0.1--0.4 whereas the extrapolated {\it theoretical} curves never reach zero contrast even at zero Strehl. Figure~\ref{fig:CRISP4}, based on the assumption that $N=36$, shows much more reasonable results in this regard, except for $\lambda=853$~nm. Plots made with $N=18$ (not shown) demonstrate that this inconsistency can be removed and suggests that the efficiency of the SST AO system perhaps corresponds to less than 36 modes.

\subsection{Discussion}
A problem with the interpretation of the 853~nm data is that CRISP uses back illuminated CCD cameras with a silicon layer that is partially transparent for light at wavelengths longer than about 700~nm. The partially transmitted light apparently is scattered back to the image plane in a way that requires a complex compensation procedure \citep{2010PhDT.......219D, 2013A&A...556A.115D} and that involves the determination of a PSF that extends to a radius of at least 6\arcsec{} from its origin. The back scatter evidently reduces the contrast of the images and although the present procedure for its compensation is successful in removing artifacts that are obvious with conventional flat-fielding, it does not yet seem clear just how accurate the back scatter compensation procedure is. We can at present not exclude that the deviations from theoretical predictions seen for the 853~nm data are associated with this compensation, although we do not have any direct evidence to suggest that this is the case.

Another limitation in our analysis is our assumptions made about the performance of the AO system. Although its efficiency can be described in terms of an effective number of corrected modes, as proposed by \citet{1998PASP..110..837R}, this is not sufficient for explaining the detailed shape of the PSF. It seems highly likely that many modes provide partial compensation only, and that this is the main reason for the limited performance of the system. To provide more quantitative information about the performance of the AO system, we need log files with more information such as the AO telemetry proposed by \citet{1997JOSAA..14.3057V} and implemented for solar telescopes by \citet{2007PhDT........41M,  Marino:10, 2010ApOpt..49.1818W}, and discussed also by  \citet{2011LRSP....8....2R}. We also need to verify that the time averaging interval of 20--30~sec., used to provide a reference for the differential image motion measurements, is long enough to not lead to overestimates of $r_0$. 

With these caveats in mind, we tentatively conclude that the variation of granulation contrast with $r_0$ can be reasonably well explained with an AO system that has around 40-45\% efficiency if we assume an additional wavefront error of about 48~nm RMS (1/13 wave RMS at 630 nm). The origin of that wavefront error is not clear. There is compelling evidence from images including both granulation and a dark sunspot that the PSF cannot have extended wings of the strength needed to explain the reduced granulation contrast, because that would make the corrected umbra intensity negative. This suggests that the 48~nm wavefront error comes from small-scale fixed aberrations in the optics of SST and its re-imaging systems. However, that interpretation is not without problems. The manufacturer of the main optical system, Opteon Oy, claims small-scale wavefront errors of the two 1.4~m flats to be in the range 8--9~nm RMS each, when used at 45$\degr$ angle of incidence, and that of the 1-meter lens about 10~nm RMS \citep{scharmer03new}. Adding up the corresponding variances delivers a total wavefront error of about 16~nm RMS. Similarly, the small-scale wavefront errors of the Schupmann corrector (lens plus mirror) were found to be about 8~nm RMS but since the lens is used in double pass, that wavefront error could possibly double. All in all, the primary optical system of SST should provide at most half of the required 48~nm RMS wavefront error. 

Another possible  explanation is that there are non-common optical path differences between CRISP and CHROMIS on the one hand, and the AO WFS on the other hand. The WFS is fed by a beam splitter cube that deflects 10\% of the light to CRISP downwards, and then this light is deflected horizontally by a right-angle prism\footnote{The reason for this arrangement is to cancel out the polarizing effect of the beam splitter that deflects 10\% of the light horizontally to the CRISP wideband system.}. The beam splitter cube is located approximately 30~cm from the focal plane in an F/46 beam, so the optical footprint on the beam splitter is about 6.5~mm for any point in the focal plane. If there are significant aberrations in the reflected beam of the beam splitter and/or right-angle prism within a circle with this diameter, then this could explain the missing 48~nm wavefront error. Whether any of the above explanations, or a combination of the two, can explain the ``missing'' wavefront error, remains an open question that deserves further attention. On the other hand, we have no indication that e.g. the beam splitter cube that divides light between the narrowband and wideband systems of CRISP introduces significant aberration differences between the two beams.

What seems clear, however, is the very significant improvement in image quality associated with the replacement of the previous 37-electrode adaptive mirror \citep{2003SPIE.4853..370S} with the new 85-electrode adaptive mirror from CILAS (Scharmer et al. in prep.) and the new tip-tilt mirror from ICOS. Previous measurements of granulation contrast correlated with $r_0$ measurements from a wide-field wavefront sensor \citep{2010A&A...513A..25S} showed ``raw'' values in the range of 6.5\% at 630~nm to 7.5\% at 538~nm close to disk center \citep{2010A&A...521A..68S}, compared to the raw values reported on in Table\ref{tab:table_obsRMS}: 9.2\% at 630~nm and 10.5--10.9\% at 525~nm. This improvement is consistent with the evaluation of aberrations in the former adaptive optics and re-imaging system of SST using phase diversity methods, which provided clear evidence for wavefront errors on the order of 42~nm RMS \citep{2012A&A...537A..80L}. The significantly increased granulation contrast observed after replacing the tip-tilt and adaptive mirrors with optics of much higher quality gives a clear indication of the importance of minimizing small-scale wavefront errors in the optics. 

We also note that the presently observed granulation contrasts compare favorably with those of any other solar telescope, whether in space, on a balloon or from the ground. For example, SOT on Hinode delivered a contrast of about 7\% at 630~nm \citep{2008A&A...484L..17D} and IMaX on Sunrise I a raw contrast in the range 8--8.5\% at 525~nm \citep{2011SoPh..268...57M}. The latter relatively low contrast was attributed to a combination of wavefront errors in the telescope and instrumentation that was estimated at $\lambda$/5.4 RMS when in operation. In the case of Hinode, the reduction in contrast is primarily from the relatively large central obscuration and spider \citep{2008A&A...484L..17D}, whereas aberrations in the optical systems are small.

\section{Conclusions}
We have described the SST primary imaging and adaptive optics (AO) systems and the telecentric dual Fabry--P\'erot systems CRISP and CHROMIS. We have attempted an evaluation of the resulting image quality through the CRISP narrowband and wideband re-imaging systems and the wideband re-imaging system of CHROMIS (at the time of recording the present data, the narrowband system of CHROMIS had not yet been installed). We have used measurements of differential image motion between four of the 85 subapertures of the SST AO system to measure the seeing quality, as quantified by the Fried parameter $r_0$, using a small FOV of only 4\arcsec$\times$4\arcsec{} to capture also the contributions from high-altitude seeing. These measurements are made during overlapping 2~sec. periods such that we obtain one $r_0$ measurement per second and capture the intermittency of the day time seeing. Simultaneously, we have recorded science images through the CRISP and CHROMIS wideband systems and processed these data in bursts of images that correspond to the time intervals of the seeing measurements. The empirical relations found between $r_0$ and the granulation contrast are discussed in terms of the expected shape of the PSF and the Strehl value of residual aberrations after partial compensation of the seeing by the AO system. We also compare the measured granulation contrasts to the values expected from numerical simulations.

Our WFS measurements of the seeing quality made with a very small FOV of only 4\arcsec$\times$4\arcsec shows excellent correlation with measured granulation contrasts and, as far as we can tell, takes into account contributions from high-altitude seeing in a reasonable manner. In our opinion, a larger FOV than this is not useful for assessing seeing and data quality.

The observed images demonstrate the outstanding image quality through the Fabry--P\'erot based system CRISP, which when compared to simultaneously obtained images through its wideband system fails to reveal any indication of image degradation by the two etalons. We attribute this to the telecentric design of CRISP, which produces very small optical footprints on the lenses and etalons for each field point. We also note that the telecentric design together with a slight tilt of the low resolution etalon, allows efficient elimination of straylight from ghost images by means of a pupil stop on the exit side of CRISP.

We find, as expected, that the measured granulation contrast varies linearly with, but is not simply proportional to, the inferred Strehl value. Reasonable agreement between the measurements and numerical calculations can be obtained if we assume the SST AO system to have an efficiency of about 40--50\%. However, to explain the discrepancy of the extrapolated ``seeing-free'' granulation contrast compared with values obtained from 3D MHD simulations, we need to introduce an additional (seeing unrelated) wavefront error of about 48~nm RMS. The origin of this may be from small-scale aberrations in the SST optics, or from noise or other shortcomings in the wavefront sensor, as discussed by \citet{2011LRSP....8....2R} and the previous sections.  

Our measurements suggest that even for a solar telescope of a relatively simple optical design with only a few high-quality optical surfaces, there are limitations in image quality that are not set by seeing in the Earths atmosphere, but rather in the optics or the AO wavefront sensor. We conjecture that for more complex solar telescopes this may be a serious problem and that the attainable image quality may well be limited  by the number or quality of the optical surfaces. In this case ``the sky is not the limit'' but rather the telescope itself. Thoroughly understanding these limitations is of crucial importance for the design and manufacture of future solar telescopes, such as the European Solar Telescope \citep[EST; e.g.,][]{2019AdSpR..63.1389J}.
\begin{acknowledgements}
  The Swedish 1-m Solar Telescope is operated on the island of La Palma by the Institute for Solar Physics of Stockholm University in
  the Spanish Observatorio del Roque de los Muchachos of the Instituto de Astrof\'isica de Canarias. The Institute for Solar Physics is
  supported by a grant for research infrastructures of national importance from the Swedish Research Council (registration number
  2017-00625). JdlCR is supported by grants from the Swedish Research Council (2015-03994) and the Swedish National Space Board (128/15). The anonymous referee is thanked for valuable suggestions.
\end{acknowledgements}


\begin{thebibliography}{47}
\expandafter\ifx\csname natexlab\endcsname\relax\def\natexlab#1{#1}\fi

\bibitem[{{Auer}(2003)}]{2003ASPC..288....3A}
{Auer}, L. 2003, in ASP Conf. Ser., Vol. 288, Stellar Atmosphere Modeling, ed.
  I.~{Hubeny}, D.~{Mihalas}, \& K.~{Werner}, 3

\bibitem[{{Barklem} \& {O'Mara}(1998)}]{1998MNRAS.300..863B}
{Barklem}, P.~S. \& {O'Mara}, B.~J. 1998, \mnras, 300, 863

\bibitem[{{Barklem} {et~al.}(2000){Barklem}, {Piskunov}, \&
  {O'Mara}}]{2000A&AS..142..467B}
{Barklem}, P.~S., {Piskunov}, N., \& {O'Mara}, B.~J. 2000, \aaps, 142, 467

\bibitem[{{Beckers} \& {Parnell}(1969)}]{1969SoPh....9...39B}
{Beckers}, J.~M. \& {Parnell}, R.~L. 1969, \solphys, 9, 39

\bibitem[{{Britton}(2006)}]{2006PASP..118..885B}
{Britton}, M.~C. 2006, \pasp, 118, 885

\bibitem[{{Conan} {et~al.}(1992){Conan}, {Madec}, \&
  {Rousset}}]{1992ESOC...42..475C}
{Conan}, J.~M., {Madec}, P.~Y., \& {Rousset}, G. 1992, in ESO Conference and
  Workshop Proceedings, Vol.~42, Progress in Telescope and Instrumentation
  Technologies, ed. {M.-H.~Ulrich}, 475

\bibitem[{{Danilovic} {et~al.}(2008){Danilovic}, {Gandorfer}, {Lagg},
  {Sch{\"u}ssler}, {Solanki}, {V{\"o}gler}, {Katsukawa}, \&
  {Tsuneta}}]{2008A&A...484L..17D}
{Danilovic}, S., {Gandorfer}, A., {Lagg}, A., {et~al.} 2008, \aap, 484, L17

\bibitem[{{de la Cruz Rodriguez}(2010)}]{2010PhDT.......219D}
{de la Cruz Rodriguez}, J. 2010, PhD thesis, Stockholm University

\bibitem[{{de la Cruz Rodr{\'{\i}}guez} {et~al.}(2016){de la Cruz
  Rodr{\'{\i}}guez}, {Leenaarts}, \& {Asensio Ramos}}]{2016ApJ...830L..30D}
{de la Cruz Rodr{\'{\i}}guez}, J., {Leenaarts}, J., \& {Asensio Ramos}, A.
  2016, \apjl, 830, L30

\bibitem[{{de la Cruz Rodr{\'{\i}}guez} {et~al.}(2019){de la Cruz
  Rodr{\'{\i}}guez}, {Leenaarts}, {Danilovic}, \&
  {Uitenbroek}}]{2019A&A...623A..74D}
{de la Cruz Rodr{\'{\i}}guez}, J., {Leenaarts}, J., {Danilovic}, S., \&
  {Uitenbroek}, H. 2019, \aap, 623, A74

\bibitem[{{de la Cruz Rodr{\'{\i}}guez} \&
  {Piskunov}(2013)}]{2013ApJ...764...33D}
{de la Cruz Rodr{\'{\i}}guez}, J. \& {Piskunov}, N. 2013, \apj, 764, 33

\bibitem[{{de la Cruz Rodr{\'{\i}}guez} {et~al.}(2013){de la Cruz
  Rodr{\'{\i}}guez}, {Rouppe van der Voort}, {Socas-Navarro}, \& {van
  Noort}}]{2013A&A...556A.115D}
{de la Cruz Rodr{\'{\i}}guez}, J., {Rouppe van der Voort}, L., {Socas-Navarro},
  H., \& {van Noort}, M. 2013, \aap, 556, A115

\bibitem[{{de la Cruz Rodr{\'{\i}}guez} \& {van
  Noort}(2017)}]{2017SSRv..210..109D}
{de la Cruz Rodr{\'{\i}}guez}, J. \& {van Noort}, M. 2017, \ssr, 210, 109

\bibitem[{{del Toro Iniesta} \& {Ruiz Cobo}(2016)}]{2016LRSP...13....4D}
{del Toro Iniesta}, J.~C. \& {Ruiz Cobo}, B. 2016, Living Reviews in Solar
  Physics, 13, 4

\bibitem[{{Hardy}(1998)}]{1998aoat.book.....H}
{Hardy}, J.~W. 1998, {Adaptive Optics for Astronomical Telescopes}, 198

\bibitem[{{Jur{\v c}{\'a}k} {et~al.}(2019){Jur{\v c}{\'a}k}, {Collados},
  {Leenaarts}, {van Noort}, \& {Schlichenmaier}}]{2019AdSpR..63.1389J}
{Jur{\v c}{\'a}k}, J., {Collados}, M., {Leenaarts}, J., {van Noort}, M., \&
  {Schlichenmaier}, R. 2019, Advances in Space Research, 63, 1389

\bibitem[{{Levy}(1971)}]{1971A&A....14...15L}
{Levy}, M. 1971, \aap, 14, 15

\bibitem[{{Lites} {et~al.}(2001){Lites}, {Elmore}, \&
  {Streander}}]{2001ASPC..236...33L}
{Lites}, B.~W., {Elmore}, D.~F., \& {Streander}, K.~V. 2001, in Astronomical
  Society of the Pacific Conference Series, Vol. 236, Advanced Solar
  Polarimetry -- Theory, Observation, and Instrumentation, ed. M.~{Sigwarth},
  33

\bibitem[{L{\"o}fdahl(2002)}]{lofdahl02multi-frame}
L{\"o}fdahl, M.~G. 2002, in Proc. SPIE, Vol. 4792, Image Reconstruction from
  Incomplete Data II, ed. P.~J. Bones, M.~A. Fiddy, \& R.~P. Millane, 146--155

\bibitem[{{L{\"o}fdahl}(2010)}]{2010A&A...524A..90L}
{L{\"o}fdahl}, M.~G. 2010, \aap, 524, A90

\bibitem[{{L{\"o}fdahl}(2016)}]{2016A&A...585A.140L}
{L{\"o}fdahl}, M.~G. 2016, \aap, 585, A140

\bibitem[{L{\"o}fdahl \& Scharmer(1994)}]{lofdahl94wavefront}
L{\"o}fdahl, M.~G. \& Scharmer, G.~B. 1994, \aaps, 107, 243

\bibitem[{{L{\"o}fdahl} \& {Scharmer}(2012)}]{2012A&A...537A..80L}
{L{\"o}fdahl}, M.~G. \& {Scharmer}, G.~B. 2012, \aap, 537, A80

\bibitem[{{Marino}(2007)}]{2007PhDT........41M}
{Marino}, J. 2007, PhD thesis, New Jersey Institute of Technology

\bibitem[{Marino \& Rimmele(2010)}]{Marino:10}
Marino, J. \& Rimmele, T. 2010, Appl. Opt., 49, G95

\bibitem[{{Mart{\'{\i}}nez Pillet} {et~al.}(2011){Mart{\'{\i}}nez Pillet}, {Del
  Toro Iniesta}, {{\'A}lvarez-Herrero}, {Domingo}, {Bonet}, {Gonz{\'a}lez
  Fern{\'a}ndez}, {L{\'o}pez Jim{\'e}nez}, {Pastor}, {Gasent Blesa}, {Mellado},
  {Piqueras}, {Aparicio}, {Balaguer}, {Ballesteros}, {Belenguer}, {Bellot
  Rubio}, {Berkefeld}, {Collados}, {Deutsch}, {Feller}, {Girela}, {Grauf},
  {Heredero}, {Herranz}, {Jer{\'o}nimo}, {Laguna}, {Meller}, {Men{\'e}ndez},
  {Morales}, {Orozco Su{\'a}rez}, {Ramos}, {Reina}, {Ramos},
  {Rodr{\'{\i}}guez}, {S{\'a}nchez}, {Uribe-Patarroyo}, {Barthol}, {Gandorfer},
  {Knoelker}, {Schmidt}, {Solanki}, \& {Vargas
  Dom{\'{\i}}nguez}}]{2011SoPh..268...57M}
{Mart{\'{\i}}nez Pillet}, V., {Del Toro Iniesta}, J.~C., {{\'A}lvarez-Herrero},
  A., {et~al.} 2011, \solphys, 268, 57

\bibitem[{{Mathew} {et~al.}(2009){Mathew}, {Zakharov}, \&
  {Solanki}}]{2009A&A...501L..19M}
{Mathew}, S.~K., {Zakharov}, V., \& {Solanki}, S.~K. 2009, \aap, 501, L19

\bibitem[{{Piskunov} \& {Valenti}(2017)}]{2017A&A...597A..16P}
{Piskunov}, N. \& {Valenti}, J.~A. 2017, \aap, 597, A16

\bibitem[{{Piskunov} {et~al.}(1995){Piskunov}, {Kupka}, {Ryabchikova}, {Weiss},
  \& {Jeffery}}]{1995A&AS..112..525P}
{Piskunov}, N.~E., {Kupka}, F., {Ryabchikova}, T.~A., {Weiss}, W.~W., \&
  {Jeffery}, C.~S. 1995, \aaps, 112, 525

\bibitem[{{Rimmele} \& {Marino}(2011)}]{2011LRSP....8....2R}
{Rimmele}, T.~R. \& {Marino}, J. 2011, Living Reviews in Solar Physics, 8, 2

\bibitem[{{Roddier}(1998)}]{1998PASP..110..837R}
{Roddier}, F. 1998, \pasp, 110, 837

\bibitem[{{Ryabchikova} {et~al.}(2015){Ryabchikova}, {Piskunov}, {Kurucz},
  {Stempels}, {Heiter}, {Pakhomov}, \& {Barklem}}]{2015PhyS...90e4005R}
{Ryabchikova}, T., {Piskunov}, N., {Kurucz}, R.~L., {et~al.} 2015, \physscr,
  90, 054005

\bibitem[{{S{\'a}nchez Cuberes} {et~al.}(2000){S{\'a}nchez Cuberes}, {Bonet},
  {V{\'a}zquez}, \& {Wittmann}}]{2000ApJ...538..940S}
{S{\'a}nchez Cuberes}, M., {Bonet}, J.~A., {V{\'a}zquez}, M., \& {Wittmann},
  A.~D. 2000, \apj, 538, 940

  \bibitem[{{Sarazin} \& {Roddier}(1990)}]{1990A&A...227..294S}
{Sarazin}, M. \& {Roddier}, F. 1990, \aap, 227, 294

\bibitem[{Scharmer(2006)}]{scharmer06comments}
Scharmer, G.~B. 2006, \aap, 447, 1111

\bibitem[{Scharmer {et~al.}(2003)Scharmer, Bjelksj{\"o}, Korhonen, Lindberg, \&
  Pettersson}]{scharmer03new}
Scharmer, G.~B., Bjelksj{\"o}, K., Korhonen, T.~K., Lindberg, B., \&
  Pettersson, B. 2003, in Proc. SPIE, Vol. 4853, Innovative Telescopes and
  Instrumentation for Solar Astrophysics, ed. S.~Keil \& S.~Avakyan, 341--350

\bibitem[{{Scharmer} {et~al.}(2003){Scharmer}, {Dettori}, {L{\"o}fdahl}, \&
  {Shand}}]{2003SPIE.4853..370S}
{Scharmer}, G.~B., {Dettori}, P.~M., {L{\"o}fdahl}, M.~G., \& {Shand}, M. 2003,
  in Society of Photo-Optical Instrumentation Engineers (SPIE) Conference
  Series, Vol. 4853, Society of Photo-Optical Instrumentation Engineers (SPIE)
  Conference Series, ed. {S.~L.~Keil \& S.~V.~Avakyan}, 370--380

\bibitem[{{Scharmer} {et~al.}(2011){Scharmer}, {Henriques}, {Kiselman}, \& {de
  la Cruz Rodr{\'{\i}}guez}}]{2011Sci...333..316S}
{Scharmer}, G.~B., {Henriques}, V.~M.~J., {Kiselman}, D., \& {de la Cruz
  Rodr{\'{\i}}guez}, J. 2011, Science, 333, 316

\bibitem[{{Scharmer} {et~al.}(2010){Scharmer}, {L{\"o}fdahl}, {van Werkhoven},
  \& {de la Cruz Rodr{\'{\i}}guez}}]{2010A&A...521A..68S}
{Scharmer}, G.~B., {L{\"o}fdahl}, M.~G., {van Werkhoven}, T.~I.~M., \& {de la
  Cruz Rodr{\'{\i}}guez}, J. 2010, \aap, 521, A68+

\bibitem[{{Scharmer} \& {van Werkhoven}(2010)}]{2010A&A...513A..25S}
{Scharmer}, G.~B. \& {van Werkhoven}, T.~I.~M. 2010, \aap, 513, A25

\bibitem[{{Stein} \& {Nordlund}(1998)}]{1998ApJ...499..914S}
{Stein}, R.~F. \& {Nordlund}, A. 1998, \apj, 499, 914

\bibitem[{{Tsuneta} {et~al.}(2008){Tsuneta}, {Ichimoto}, {Katsukawa}, {Nagata},
  {Otsubo}, {Shimizu}, {Suematsu}, {Nakagiri}, {Noguchi}, {Tarbell}, {Title},
  {Shine}, {Rosenberg}, {Hoffmann}, {Jurcevich}, {Kushner}, {Levay}, {Lites},
  {Elmore}, {Matsushita}, {Kawaguchi}, {Saito}, {Mikami}, {Hill}, \&
  {Owens}}]{2008SoPh..249..167T}
{Tsuneta}, S., {Ichimoto}, K., {Katsukawa}, Y., {et~al.} 2008, \solphys, 249,
  167

\bibitem[{{van Noort} {et~al.}(2005){van Noort}, {Rouppe van der Voort}, \&
  {L{\"o}fdahl}}]{2005SoPh..228..191V}
{van Noort}, M., {Rouppe van der Voort}, L., \& {L{\"o}fdahl}, M.~G. 2005,
  \solphys, 228, 191

\bibitem[{{Veran} {et~al.}(1997){Veran}, {Rigaut}, {Maitre}, \&
  {Rouan}}]{1997JOSAA..14.3057V}
{Veran}, J.-P., {Rigaut}, F., {Maitre}, H., \& {Rouan}, D. 1997, Journal of the
  Optical Society of America A, 14, 3057

\bibitem[{{Wedemeyer-B{\"o}hm}(2008)}]{2008A&A...487..399W}
{Wedemeyer-B{\"o}hm}, S. 2008, \aap, 487, 399

\bibitem[{{Wedemeyer-B{\"o}hm} \& {Rouppe van der
  Voort}(2009)}]{2009A&A...503..225W}
{Wedemeyer-B{\"o}hm}, S. \& {Rouppe van der Voort}, L. 2009, \aap, 503, 225

\bibitem[{{W{\"o}ger}(2010)}]{2010ApOpt..49.1818W}
{W{\"o}ger}, F. 2010, \ao, 49, 1818

\end{thebibliography}
\end{document}